\documentclass[aps,pre,reprint,superscriptaddress]{revtex4-1}

\usepackage{subfigure}
\usepackage{amsfonts, amssymb, amsmath}
\usepackage{bm}
\usepackage{graphics}
\usepackage{graphicx}
\usepackage{wasysym}

\usepackage[utf8x]{inputenc}
\usepackage[T1]{fontenc}

\usepackage{color}
\definecolor{darkblue}{rgb}{0,0,0.6}
\definecolor{darkred}{rgb}{0.6,0,0}

\usepackage[colorlinks=true,urlcolor=darkblue,citecolor=darkblue,linkcolor=darkred,hyperfootnotes=false]{hyperref}

\newcommand{\argc}[1]{\left[#1\right]}
\newcommand{\arga}[1]{\left\lbrace #1\right\rbrace }
\newcommand{\argp}[1]{\left(#1\right)}
\newcommand{\valabs}[1]{\vert #1\vert}
\newcommand{\moy}[1]{\left\langle  #1 \right\rangle }

\newcommand{\clap}[1]{\hbox to 0pt{\hss#1\hss}}
\newcommand{\mathclapinternal}[2]{\clap{$\mathsurround=0pt#1{#2}$}}
\def\mathclap{\mathpalette\mathclapinternal}
\newcommand{\Wbar}{ \mathclap{\phantom{W}\overline{\phantom{I}}} W}

\newcommand{\tho}{{\text{\thorn}} }
\newcommand{\Erf}{\operatorname{Erf}}

\begin{document}

\title{
Static fluctuations of a thick 1D interface in the 1+1 Directed Polymer formulation}

\author{Elisabeth Agoritsas}
\email[]{Elisabeth.Agoritsas@unige.ch}
\affiliation{DPMC-MaNEP, University of Geneva, 24 Quai Ernest-Ansermet, 1211 Geneva 4, Switzerland}
\author{Vivien Lecomte}
\affiliation{DPMC-MaNEP, University of Geneva, 24 Quai Ernest-Ansermet, 1211 Geneva 4, Switzerland}
\affiliation{Laboratoire Probabilit\'es et Mod\`eles Al\'eatoires (CNRS UMR 7599), Universit\'es Paris VI \& Paris VII, B\^atiment Sophie Germain \\ Avenue de France, 75013 Paris, France}
\author{Thierry Giamarchi}
\affiliation{DPMC-MaNEP, University of Geneva, 24 Quai Ernest-Ansermet, 1211 Geneva 4, Switzerland}

\date{\today}


\begin{abstract}

Experimental realizations of a 1D interface always exhibit a finite microscopic width $\xi>0$;
its influence is erased by thermal fluctuations at sufficiently high temperatures,
but turns out to be a crucial ingredient for the description of the interface fluctuations below a characteristic temperature $T_c(\xi)$. 
Exploiting the exact mapping between the static 1D interface and a 1+1 Directed Polymer (DP) growing in a continuous space,
we study analytically both the free-energy and geometrical fluctuations of a DP,
at finite temperature $T$, with a short-range elasticity and submitted to a quenched random-bond Gaussian disorder of \textit{finite} correlation length $\xi$.

We derive the exact `time'-evolution equations of the disorder free-energy ${\bar{F}(t,y)}$ --~which encodes the microscopic disorder integrated by the DP up to a growing `time'~$t$ and an endpoint position~$y$~-- its derivative~${\eta (t,y)}$, and their respective two-point correlators ${\bar{C}(t,y)}$ and ${\bar{R}(t,y)}$.
%
We compute the exact solution of its linearized evolution~${\bar{R}^{\text{lin}}(t,y)}$, and we combine its qualitative behavior and the asymptotic properties known for an uncorrelated disorder (${\xi=0}$), to justify the construction of a `toymodel' leading to a simple description of the DP properties.
This model is characterized by Gaussian Brownian-like free-energy fluctuations,
correlated at small ~${\valabs{y} \lesssim \xi}$,
and of amplitude~${\widetilde{D}_{\infty}(T,\xi)}$.
We present an extended scaling analysis of the roughness, supported by saddle-point arguments on its path-integral representation, which predicts ${\widetilde{D}_{\infty} \sim 1/T}$ at high-temperatures and  ${\widetilde{D}_{\infty} \sim 1/T_c(\xi)}$ at low-temperatures.
We identify the connection between the temperature-induced crossover of~${\widetilde{D}_{\infty}(T,\xi)}$ and the full replica-symmetry breaking (full-RSB) in previous Gaussian Variational Method (GVM) computations.
In order to refine our toymodel with respect to finite-`time' geometrical fluctuations, we propose an effective `time'-dependent amplitude~${\widetilde{D}_t}$.

Finally we discuss the consequences of the low-temperature regime for two experimental realizations of KPZ interfaces, namely the static and quasistatic behavior of magnetic domain walls and the high-velocity steady-state dynamics of interfaces in liquid crystals.

\end{abstract}


\maketitle

\tableofcontents


\section{Introduction} \label{section-intro}

Effective one-dimensional (1D) interfaces can be spotted in various experimental contexts,
encompassing domain walls (DWs) in ferromagnetic \cite{lemerle_1998_PhysRevLett80_849,repain_2004_EurPhysLett68_460,metaxas_2007_PhysRevLett99_217208}
or ferroic \cite{tybell_2002_PhysRevLett89_097601,paruch_2005_PhysRevLett94_197601,pertsev_2011_JApplPhys110_052001}
thin films,
fractures in brittle materials \cite{santucci_2007_PhysRevE75_016104}
or paper \cite{alava_2006_RepProgPhys69_669},
contact line in wetting experiments \cite{alava_2004_AdvPhys53_83,santucci_2011_EurPhysLett94_46005}.
%
The generic framework of the \textit{disordered elastic systems} (DES) \cite{agoritsas_2012_ECRYS2011} has been proven to provide a quite successful modelling for such systems,
describing them as point-like elastic strings living in a two-dimensional disordered energy landscape.
%
The competition between the elasticity -- the tendency to minimize their distortions -- and the disorder -- the inhomogeneities of the underlying medium --, blurred by thermal fluctuations at finite temperature, accounts for the resulting metastability and the consequent glassy properties observed in such systems.
%
Moreover the value of the roughness exponent $\zeta$, which characterizes the scaling properties of a self-affine manifold, is fully determined for a given DES once the dimensionality, the type of elasticity and of disorder are chosen, thus promoting the value of $\zeta$ to a reliable signature of the disorder universality class to which a given system might belong.

The specific case of a 1D interface with a short-range elasticity and a random-bond (RB) quenched Gaussian disorder can actually be mapped on other statistical-physics models in the Kardar-Parisi-Zhang (KPZ) universality class \cite{kardar_1986_originalKPZ_PhysRevLett56_889,krug_1997_AdvPhys_46_139,corwin_2011_arXiv:1106.1596}, including in particular the so-called `1+1 Directed Polymer' (DP) which has stimulated an increased activity lately, among both statistical physicists \cite{spohn_2006_physicaA369_71,kriecherbauer_krug_2010_JPhysA43_403001,sasamoto_spohn_2010_JStatMech2010_P11013}
and mathematicians
\cite{amir_arXiv:1003.0443,borodin_corwin_2012_arXiv:1204.1024}.
A large variety of results emphasizes the deep connection which exists between the descriptions of a wide range of systems up to random matrices \cite{johansson_2000_CommMathPhys209_437,praehofer_spohn_2000_PhysRevLett84_4882},
such as the Burgers equation in hydrodynamics~\cite{forster_nelson_stephen_1977_PhysRevA16_732},
roughening phenomena and stochastic growth~\cite{halpin_zhang_1995_PhysRep254},
last-passage percolation~\cite{krug-spohn_1991_Godreche_BegRohu},
dynamics of cold atoms~\cite{kulkarni-lamacraft_2012_arXiv:1201.6363},
and vicious walkers~\cite{spohn_2006_physicaA369_71,rambeau-schehr_2010_EurPhysLett91_60006,forrester-majumdar-schehr_2011_NuclPhysB844-500}.
%
A shared feature between those related models is the well-known KPZ exponent $\zeta_{\text{KPZ}}=2/3$, which characterizes the exact scaling at \textit{asymptotically} large lengthscales or `times',
generated by the nonlinear KPZ evolution equation and assuming an \textit{uncorrelated} disorder \cite{kardar_1987_NuclPhysB290_582,huse_henley_fisher_1985_PhysRevLett55_2924,johansson_2000_CommMathPhys209_437,balazs_arXiv:0909.4816}.

Although of interest regarding the whole KPZ class problems, there are two additional issues which turn out to be relevant especially for the study of experimental interfaces:
on one hand, the characterization of the scaling properties \textit{at finite lengthscales}, with possibly different regimes and crossover lengthscales regarding both the roughness exponent and the amplitude of the geometrical fluctuations;
on the other hand, the consequences of the interplay \textit{at finite temperature} between thermal fluctuations and disorder.
%
However, in order to have then a complete realistic description, an additional physical ingredient must be included in the DES model: an experimental realization of interface always exhibits a finite microscopic width $\xi>0$, which translates equivalently for a point-like interface into \textit{a finite disorder correlation length}.
%
Above a characteristic temperature $T_c(\xi)>0$, thermal fluctuations simply erase the existence of such a microscopic width, whereas at sufficiently low temperature it becomes relevant even for the macroscopic properties of the interface.
%
Those two temperature regimes can be hinted by simple scaling arguments \cite{agoritsas_2012_ECRYS2011},
which are reflected in the two opposite Functional-Renormalization-Group (FRG) regimes of high-temperature \cite{bustingorry_2010_PhysRevB82_140201} versus zero-temperature fixed-point \cite{balents-fisher_1993_PhysRevB48_5949,chauve_2000_ThesePC_PhysRevB62_6241}.
Their connection has already been addressed analytically in a single computation in a Gaussian-Variational-Method (GVM) approximation \cite{agoritsas_2010_PhysRevB_82_184207,agoritsas_2012_ECRYS2011}.
%
Its predictions for the low-temperature regime turned out to be potentially accessible and thus crucially relevant for ferromagnetic DWs in ultra-thin films \cite{lemerle_1998_PhysRevLett80_849,repain_2004_EurPhysLett68_460,metaxas_2007_PhysRevLett99_217208};
these boundaries between regions of homogeneous magnetization are believed to be the experimental realization of precisely the 1D DES considered here,
and actually exhibit temperatures ${T_c(\xi)}$
--~extracted from their dynamical response to an external magnetic field~--
which are well above room-temperature~\cite{bustingorry_kolton_2012_PhysRevB85_214416}.

Unfortunately the GVM computation does not allow to grasp directly the correct asymptotic fluctuations of the 1D interface,  
as it predicts ${\zeta=3/5}$ instead of $\zeta_{\text{KPZ}}$, thus jeopardizing its predictions for the scaling in temperature of the roughness~\cite{agoritsas_2010_PhysRevB_82_184207,agoritsas_2012_ECRYS2011}.
In order to circumvent this known GVM artefact, we have actually performed in~Ref.~\citep{agoritsas_2010_PhysRevB_82_184207} a GVM computation on an effective `toymodel' of the interface free-energy in a 1+1 DP formulation.
Following Mézard and Parisi footsteps~\cite{mezard_parisi_1992_JPhysI02_2231},
we essentially assumed Gaussian fluctuations of the DP free-energy --~as of a Brownian-walk type~-- but in addition including explicitly a finite correlation length~${\tilde{\xi} \approx \xi}$.
A central and physically meaningful quantity in this model is the adjustable amplitude of the free-energy fluctuations, denoted~$\widetilde{D}_{\infty}$, which turns out to control also the amplitude of the geometrical fluctuations, along with its characteristic crossover lengthscales such as its Larkin length~\cite{Larkin_model_1970-SovPhysJETP31_784}.
At high temperatures (or equivalently ${\xi=0}$) it is known that ${\widetilde{D}_{\infty}\sim 1/T}$ \cite{huse_henley_fisher_1985_PhysRevLett55_2924},
whereas at low temperatures we expect by scaling arguments~\cite{agoritsas_2012_ECRYS2011} a saturation to ${\widetilde{D}_{\infty}\sim 1/T_c(\xi)}$ that cures what would otherwise have been an unphysical divergence in the zero-temperature limit.
However,
a proper justification of our DP `toymodel' assumptions was needed in order to assess the validity of its GVM predictions for the roughness~\cite{agoritsas_2010_PhysRevB_82_184207,agoritsas_2012_ECRYS2011}.
Moreover, an analytical prediction for the full temperature-induced crossover of~${\widetilde{D}_{\infty}(T,\xi)}$ itself, although crucially relevant, was still missing, and has thus been our focus in this study.

%
In this paper,
using the exact mapping between the static 1D interface and a 1+1 DP growing in a continuous 2D space,
we study analytically the temperature-dependence of the free-energy fluctuations \textit{in a spatially-correlated random potential}, as a function of lengthscale or DP growing `time'~$t$, and its consequences on the geometrical fluctuations.
In order to dissociate the effects due to disorder from the pure thermal ones, which hide them at small lengthscales and actually blur the physical picture,
we focus on the \textit{disorder free-energy} ${\bar{F}(t,y)}$ of the DP endpoint,
a quantity that integrates all the microscopic disorder explored by the DP up to its endpoint position $y$ at a fixed `time'.
For an uncorrelated disorder~(${\xi=0}$) the universal distribution of its fluctuations has recently been completely elucidated at all `times' \cite{sasamoto_2010_NuclPhysB_834_523,calabrese_2010_EPL90_20002,dotsenko_2010_EPL90_20003,amir_arXiv:1003.0443}
whereas for a correlated disorder~(${\xi>0}$) such a universal distribution is believed to be jeopardized by the specificity of the microscopic disorder correlation.
As a first step, we have addressed in~Ref.~\cite{agoritsas-2012-FHHpenta} a generalized correspondence between the geometrical and free-energy fluctuations at large~$y$, via their respective two-point correlators and an adjustable amplitude assimilable to~$\widetilde{D}_{\infty}$.
Here we complete this study by focusing on the fluctuations of~${\partial_y \bar{F}(t,y)}$,
whose two-point correlator at fixed~$t$ and small~$y$ allows us to follow, in the KPZ language, how the interplay between the disorder correlation and the feedback of the KPZ non-linearity controls the universal scaling in temperature of the amplitude~${\widetilde{D}_{\infty}(T,\xi)}$.

The plan of the paper is as follows.
In~Sec.~\ref{section-DPformulation} we define the full model of the static 1D interface in the 1+1 DP formulation, along with the quantities of interest for the characterization of its geometrical and free-energy fluctuations at a given lengthscale $r$ of the 1D interface or growing `time' $t$ of the DP.
Then in~Sec.~\ref{section-DPtoymodel} we recall the exact properties of the model at asymptotically large `times' or in its `linearized' version --~obtained by neglecting the KPZ non-linearity~--, and use them to  justify the construction of our DP `toymodel'.
In~Sec.~\ref{section-scaling-saddle}, extensive scaling arguments are given in order to tackle the opposite low- \textit{versus} high-temperature regimes and their connection,
and the underlying scaling assumptions are actually made explicit using saddle-point arguments;
these arguments allow to reinterpret previous GVM computations with full Replica-Symmetry-Breaking (full-RSB) as a quantitative interpolation of ${\widetilde{D}_{\infty}(T,\xi)}$ between these two opposite asymptotics.
In~Sec.~\ref{section-synthetic-outlook} we combine our analytical arguments in a synthetic outlook and we derive from it in~Sec.~\ref{section-Dtildeinfty-T-xi} an analytical prediction for an effective `time'-dependent amplitude $\widetilde{D}_t$, as a refinement of our DP `toymodel'.
We finally discuss in~Sec.~\ref{section-discussion-exp} our results with respect to two experimental systems, namely the domain walls in ultrathin magnetic films and interfaces in liquid crystals,
and we conclude in~Sec.~\ref{section-conclusion}.
%

For completeness, most of the technical details of the paper have been gathered in the appendices.
For the convenience of the reader interested in a specific issue, we list thereafter the content of the different appendices.
Associated to the definition of the full model of the static 1D interface of~Sec.~\ref{section-DPformulation},
Appendix~\ref{A-GVM-PRB2010} first recalls briefly previous GVM predictions for the corresponding roughness of this model, predictions that will be revisited and reinterpreted in regards of our actual understanding of the physics at stake;
Appendix~\ref{A-STS-bythebook} is devoted to the STS, central to the definition of the disorder free-energy;
Appendix~\ref{A-FHHderivation} gives the starting point of the Feynman-Kac `time'-evolution equations of the free-energy, namely, the stochastic heat equation with a careful treatment of its normalization issues;
Appendix~\ref{A-flow_CbarRbar} finally details the derivation via the  It\=o formula of the `time'-evolution of averaged quantities such as the two-point correlators ${\bar{C}(t,y)}$ and ${\bar{R}(t,y)}$.
Associated to the construction of the DP toymodel in~Sec.~\ref{section-DPtoymodel}, the exact two-point correlators for the linearized dynamics of~${\bar{F}_V(t,y)}$ are derived in~Appendix~\ref{A-short-time-dynamics-Fbar-generic} and the steady-state solution of the Fokker-Planck equation for the disorder free-energy is examined in~Appendix~\ref{A-FokkerPlanck-equation}.
Finally, Appendix~\ref{A-saddle_scalings-maths} discusses the specific case of a temperature-dependent elasticity, a convention widely considered in the Mathematics litterature since it is equivalent to taking a temperature-independent Wiener measure for the DP trajectories.


\section{1+1 Directed-Polymer formulation of the static 1D interface} \label{section-DPformulation}

\subsection{DES model of a 1D interface} \label{section-def-DES}

\begin{figure}
 \includegraphics[width=0.8\columnwidth]{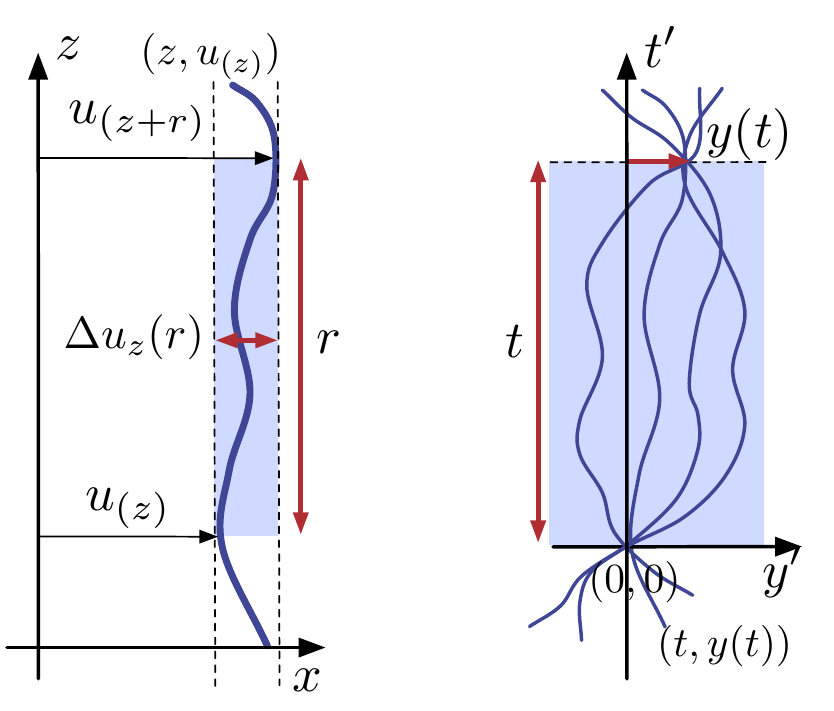}
 \caption{
 \textit{Left:} 1D interface configuration of displacement field $u(z)$ with respect to the $z$-axis; definition of its relative displacement ${\Delta u_z(r)}$ at a lengthscale $r$ with respect to the internal coordinate $z$.
 \textit{Right:} Focus on all the segments of the 1D interface starting from $(0,0)$ and ending at $(t,y(t))$; definition of the DP's end-point position $y(t)$ after a growing `time' $t$.
 The translation table between those two representations is given by
 ${(z,x) \leftrightarrow (t',y')}$ for the coordinates,
 ${u(z) \leftrightarrow y(t')}$ for the trajectory,
 ${\mathcal{P}\argp{\Delta u(r)} \leftrightarrow \mathcal{P}(t,y)}$ for the geometrical PDF
 and ${B(r) = \overline{\left< \Delta u(r)^2\right>} \leftrightarrow \overline{\left< y(t)^2\right>} = B(t)}$ for the roughness function.
 \label{fig:1D-DP-roughness}
 }
\end{figure}

We consider a 1D interface, living in an infinite and continuous 2D space of respectively internal and transverse coordinates $(z,x)\in \mathbb{R}^2$.
Restricting the model to the case where the interface has no bubbles nor overhangs, each possible configuration is described by a univalued displacement field ${u(z) \in \mathbb{R}}$ with respect to a flat configuration defined by the $z$-axis (cf. Fig.~\ref{fig:1D-DP-roughness} left).

In the elastic limit of small distortions and for a short-range elasticity, the energetic cost of elastic distortions are given by the elastic Hamiltonian ${\mathcal{H}_{\text{el}} \argc{u} = \frac{c}{2} \int_{\mathbb{R}} dz \, \argp{\nabla_z u(z)}^2}$
with $c$ the elastic constant.

Assuming that we have a quenched disorder, accounting typically for a weak collective pinning of the interface by many impurities, the microscopic disorder is described by a random potential $V(z,x)$ with the corresponding energy ${\mathcal{H}_{\text{dis}} \argc{u,V} = \int_{\mathbb{R}} dz \, V(z,u(z))}$.
The disorder average $\overline{\mathcal{O}}$ of an observable at fixed disorder $\mathcal{O}_V$ is then defined with respect to the probability distribution of the disorder configurations ${\bar{\mathcal{P}}\argc{V}}$, which is assumed to be Gaussian, \textit{i.e.} fully defined by its mean and its two-point disorder correlator:
\begin{equation}
\begin{split}
 & \overline{V(z,x)} = 0 \\
 & \overline{V(z,x)V(z',x')} = D \cdot \delta(z-z') \cdot R_{\xi}(x-x')
 \end{split}
 \label{eq-def-moydis}
\end{equation}
with $D$ the strength of disorder, which quantifies the typical amplitude of the random potential.
%
The disorder should be statistically translational-invariant in space, and it is actually assumed to be uncorrelated along its internal direction $z$ and correlated on a typical length $\xi>0$ along its transverse direction $x$.
Finally, we consider the specific case of a random-bond (RB) disorder, \textit{i.e.} with a symmetric function ${R_{\xi}(x)}$ decreasing sufficiently fast to encode a short-range disorder and with the chosen normalization ${\int_{\mathbb{R}} dx \,R_{\xi}(x) \equiv 1}$.

At equilibrium and for a given disorder configuration,
the statistical average over thermal fluctuations $\moy{\mathcal{O}}_V$ is then defined
with respect to the normalized Boltzmann weight ${\mathcal{P}_V \argc{u} \propto e^{-\mathcal{H}\argc{u,V}/T}}$ of Hamiltonian
${\mathcal{H}\argc{u,V}=\mathcal{H}_{\text{el}} \argc{u}+\mathcal{H}_{\text{dis}}\argc{u,V}}$
(the Boltzmann constant is fixed once and for all at $k_B=1$ so that the temperature has the dimensions of an energy).
For a self-averaging disorder, a given observable must be averaged analytically first over thermal fluctuations and secondly over disorder $\overline{\moy{\mathcal{O}}}$, recovering in particular a translational invariance in space.

The choice of those different assumptions is explained in detail in Ref.~\cite{agoritsas_2012_ECRYS2011}.
In order to compute the GVM roughness of such a static 1D interface, $R_{\xi}(x)$ was chosen in Ref.~\cite{agoritsas_2010_PhysRevB_82_184207} to be a normalized Gaussian function of variance $2\xi^2$,
encoding thus the typical width $\xi$ as the single feature of this correlator function (cf. Appendix~\ref{A-GVM-PRB2010}).

\subsection{Mapping of the 1D interface on the 1+1 Directed Polymer}  \label{section-mapping-2models}

The characterization of the geometrical fluctuations of the \textit{static} 1D interface goes through the determination of the probability distribution function (PDF) of the relative displacements $\mathcal{P}(\Delta u(r))$ at a given lengthscale $r$, with ${\Delta u_z(r) \equiv u(z+r)-u(z)}$.
The contribution of the combined PDF of thermal fluctuations $\mathcal{P}_V \argc{u}$ and of disorder $\bar{\mathcal{P}}\argc{V}$ can be disconnected by focusing directly on the fluctuations of segments of length $r$ on the interface.
As defined in Fig.~\ref{fig:1D-DP-roughness}, such a segment can be mapped on the trajectory of a directed polymer starting from $(0,0)$ and growing in `time' $t$ in the 2D disordered energy landscape described by the random potential $V(t,y)$.
The fluctuations of the DP end-point $y(t)$ at a `time' $t$, of PDF $\mathcal{P}(t,y)$, encode thus precisely the translational-invariant $\mathcal{P}(\Delta u(r))$ at the lengthscale $r$.

The energy of a segment of lengthscale $r \leftrightarrow t_1$, of trajectory $y(t)$ connecting $(0,0)$ to $(t_1,y_1)$, is given by the partial Hamiltonian:
\begin{equation}
 \mathcal{H} \argc{y,V;t_1}
 = \int_0^{t_1} dt \, \argc{\frac{c}{2} \argp{\partial_t y(t)}^2 + V(t,y(t))}
 \label{eq-partial-Hamiltonian-DP}
\end{equation}
with the disorder distribution defined by \eqref{eq-def-moydis}.
Integrating over the thermal fluctuations at fixed disorder $V$,
the \textit{unnormalized} Boltzmann weight of a DP ending at $(t_1,y_1)$ is then given by the path-integral:
\begin{equation}
 W_V (t_1,y_1)
 = \int_{y(0)=0}^{y(t_1)=y_1} \mathcal{D}y(t) \, e^{-\mathcal{H} \argc{y,V;t_1}/T}
 \label{eq-def-unnorm-Boltzmann-Wv}
\end{equation}
with the underlying four DES parameters $\arga{c,D,T,\xi}$.
The connection between this continuous formulation of the DP, well-known among physicists, and its discretized version on a lattice with the solid-on-solid constraint has recently been properly established \cite{alberts-quastel_2012_arXiv:1202.4403v1}.
The corresponding free-energy ${F_V(t,y)}$, that will be defined in the next section with the proper normalization by~${W_V(t,y) \propto e^{-F_V(t,y)/T}}$, follows a KPZ evolution equation and thus connects our study of the static 1D interface to the broader 1D KPZ universality class, via the present mapping on the growing 1+1 DP.

We restrict our study to the case where the polymer is attached in ${y=0}$ at initial `time'. This choice corresponds to the so-called `sharp-wedge' initial conditions~\cite{sasamoto_spohn_2010_JStatMech2010_P11013} of the KPZ equation as opposed \textit{e.g.} to the `flat' ones where the initial position would be integrated upon.


\subsection{Geometrical and free-energy fluctuations} \label{section-def-fluctuations}

We start with the definition of the relevant quantities for the characterization of the geometrical and free-energy fluctuations.

With the following normalization at fixed `time' $t$:
\begin{equation}
 \Wbar_V (t)
 \equiv \int_{-\infty}^{\infty} dy \cdot  W_V (t,y)
 \label{eq-WV-normalization}
\end{equation}
we can define the PDF of the DP end-point, respectively at fixed disorder $V$ and after the disorder average:
\begin{equation}
 \mathcal{P}_V(t,y)
 \equiv \frac{W_V (t,y)}{\Wbar_V (t)}
 \, , \; 
 \mathcal{P}(t,y)
 =\overline{\mathcal{P}_V(t,y)}
 \label{eq-def-PDF-yt}
\end{equation}
and use them for the computation of averages for any observable $\mathcal{O}$ which depends on the sole DP end-point position $y(t)$ (and not on its whole trajectory $y(t')$, with $t' \in \argc{0,t}$):
\begin{eqnarray}
 \moy{\mathcal{O}\argc{y(t)}}_V
 &=& \int_{-\infty}^{\infty} dy \cdot \mathcal{O}\argc{y} \cdot \mathcal{P}_V(t,y)
 \label{eq-def-moyV-obs} \\
 \overline{\moy{\mathcal{O}\argc{y(t)}}}
 &=& \int_{-\infty}^{\infty} dy \cdot \mathcal{O}\argc{y} \cdot \mathcal{P}(t,y)
 \label{eq-def-moy-obs}
\end{eqnarray}
and in particular the different moments of the PDF \eqref{eq-def-PDF-yt}:
\begin{eqnarray}
 \moy{y(t)^k}_V
 &=& \int_{-\infty}^{\infty} dy \cdot y^k \cdot \mathcal{P}_V(t,y)
 \label{eq-def-moyV-yk} \\
 \overline{\moy{y(t)^k}}
 &=& \int_{-\infty}^{\infty} dy \cdot y^k \cdot \mathcal{P}(t,y)
 \label{eq-def-moy-yk}
\end{eqnarray}

Actually the PDF $\mathcal{P}(t,y)$ is known to be fairly Gaussian (although the study of its small non-Gaussian deviations encodes relevant physics
\cite{halpin_1991_PhysRevA44_R3415,zumofen_1992_PhysRevA45_7624,goldschmidt-blum_1993_PhysRevE47_R2979,goldschmidt-blum_1993_PhysRevE48_161,halpin_zhang_1995_PhysRep254}), in the sense that
\begin{equation}
 \mathcal{P}(t,y) \approx \frac{e^{-y^2/(2 B(t))}}{\sqrt{2 \pi B(t)}}
 \label{eq-PDF-fairGaussian}
\end{equation}
with its main feature being summarized in its second moment, namely the \textit{roughness function} $B(t)$ and its corresponding roughness exponent $\zeta(t)$:
\begin{equation}
 B(t)
 \equiv \overline{\moy{y(t)^2}}
 \, , \;
 \zeta (t)
 \equiv \frac12 \frac{\partial \log B(t)}{\partial \log t}
 \label{eq-def-roughness-zeta}
\end{equation}
a proper exponent $\zeta$ being defined only if a powerlaw can be identified on a certain range in $t$; this is typically the case at large lengthscales, the beginning of this asymptotic regime defining the so-called `Larkin length' \cite{Larkin_model_1970-SovPhysJETP31_784} $L_c$.
In absence of disorder, the DP is a Brownian random walk whose PDF $\mathcal{P}_{\text{th}}(t,y)$ is then exactly a Gaussian function of thermal roughness ${B_{\text{th}}(t)=\frac{Tt}{c}}$.
In presence of a short-range RB disorder, there is a crossover from this thermal roughness at small lengthscales to an asymptotic roughness $B_{\text{RM}}(t)\sim t^{4/3}$ in the `random-manifold' (RM) regime of large lengthscales.
A 1D interface is thus a self-affine manifold in these two lengthscales regimes, its geometrical fluctuations being characterized by the scaling ${y(t)^2 \sim A(c,D,T,\xi) \, t^{2\zeta}}$
with the \textit{diffusive} exponent $\zeta_{\text{th}}=\frac12$ at sufficiently small lengthscales (extended at all lengthscales in absence of disorder),
and the \textit{superdiffusive} exponent $\zeta_{\text{RM}}=\frac23$ at asymptotically large lengthscales (obtained in Ref.~\cite{kardar_1987_NuclPhysB290_582,huse_henley_fisher_1985_PhysRevLett55_2924,johansson_2000_CommMathPhys209_437,balazs_arXiv:0909.4816}  assuming ${\xi=0}$).
Actually the existence of a finite width $\xi>0$ strongly modifies the scaling of the prefactor $A(c,D,T,\xi)$ and the roughness crossover, with in particular a whole intermediate `Larkin-modified' lengthscale regime for temperatures below $T_c(\xi) = (\xi c D)^{1/3}$ (cf. Ref.~\cite{agoritsas_2010_PhysRevB_82_184207,agoritsas_2012_ECRYS2011} or Appendix~\ref{A-GVM-PRB2010}).

The PDF ${\mathcal{P}(t,y)}$ and its roughness $B(t)$ are precisely the quantities accessible experimentally via an analysis of a `snapshot' of an interface configuration (defined as in Fig.~\ref{fig:1D-DP-roughness}); however only a single roughness regime has been observed up to now in ferromagnetic DWs, which are believed to be the prototype of our idealized 1D interface (e.g. ${\zeta=0.69 \pm 0.07}$ in Ref.~\cite{lemerle_1998_PhysRevLett80_849}).
From an analytical point of view, additional information can be extracted from the fluctuations of the probability $\mathcal{P}_V (t,y)$ itself, or alternatively from its corresponding pseudo free-energy ${F_V(t,y)}$ defined at fixed disorder by:
\begin{eqnarray}
 & \frac{W_V (t,y)}{\Wbar_{V\equiv 0}(t)}
 \equiv \exp \argc{-\frac{1}{T} F_V (t,y)} &
 \label{eq-def-free-energy-1} \\
 & F_V (t,y)
 = F_{V \equiv 0} (t,y) + \bar{F}_V (t,y) &
 \label{eq-def-free-energy-2}
\end{eqnarray} 
with the following conventions:
\begin{eqnarray}
 F_{V \equiv 0} (t,y)
 &=& F_{\text{th}} (t,y) + T \log \Wbar_{V \equiv 0} (t)
 \label{eq-def-free-energy-3} \\  
 \frac{F_{\text{th}} (t,y)}{T} &=& \frac12 \frac{y^2}{B_{\text{th}} (t)}
 \Leftrightarrow 
 F_{\text{th}} (t,y)
 = \frac{c y^2}{2t}
 \label{eq-def-free-energy-4} \\
 \Wbar_{V \equiv 0} (t)
 &=& \sqrt{2 \pi B_{\text{th}} (t)}
 = \sqrt{2 \pi \frac{Tt}{c}}
 \label{eq-def-free-energy-5}
\end{eqnarray}
The decomposition of \eqref{eq-def-free-energy-2} defines the disorder free-energy $\bar{F}_V(t,y)$, which fully encodes the integrated disorder encountered by the DP up to a `time' $t$.
This is the central quantity that we study throughout this paper, as it allows to examine in a systematic way the role of disorder as a function of the lengthscale or growing `time'.
This contribution can be dissociated from the pure thermal free-energy $F_{\text{th}}(t,y)$ because of the statistical tilt symmetry (STS) of the model, whose different incarnations are discussed in Appendix \ref{A-STS-bythebook}.
Indeed, with the particular form of the short-range elasticity ${\frac{c}{2} \argp{\partial_t y(t)}^2}$ in \eqref{eq-partial-Hamiltonian-DP} and $y$ being a continuous variable,
the effective disorder $\bar{F}_V(t,y)$ inherits the statistical translation-invariance of the microscopic disorder ${\bar{\mathcal{P}}\argc{V(t,y)}}$ defined by \eqref{eq-def-moydis}.
Its PDF at fixed `time' (and similarly any functional of $\bar{F}_V (t,y)$) thus satisfies:
\begin{equation}
 \bar{\mathcal{P}} \argc{\bar{F}_V (t,y+Y)}
 = \bar{\mathcal{P}} \argc{\bar{F}_V (t,y)}
 \label{eq-STS-PFbarV}
\end{equation}
Note that the decomposition~\eqref{eq-def-free-energy-2} is specific to the `sharp-wedge' initial condition of the KPZ equation. It allows to work with a translation-invariant quantity, broken otherwise by the thermal contribution ${F_{V \equiv 0}(t,y)}$ (contrarily to the `flat' initial condition where we would have~${F_{V \equiv 0}(t,y)\equiv 0}$).
In order to single out the $y$-dependent additive contribution of $\bar{F}_V(t,y)$, we also define the random phase $\eta_V(t,y)$ in a kind of `random-field' formulation of the disorder free-energy:
\begin{equation}
\begin{split}
 \eta_V (t,y)
 & \equiv \partial_y \bar{F}_V (t,y) \\
 \bar{F}_V (t,y) 
 &=  \frac12 \argp{\int_{-\infty}^{y} - \int_{y}^{\infty}} dy' \, \eta_V (t,y') + \text{cte}_V (t)
\end{split}
\label{eq-def-etaV}
\end{equation}
where $\text{cte}_V (t)$ is a $y$-independent constant.
Note that the STS implies that ${\bar{\mathcal{P}} \argc{\eta_V (t,y+Y)} = \bar{\mathcal{P}} \argc{\eta_V (t,y)}}$.

We may assume that the scaling of the distribution ${\bar{\mathcal{P}} \argc{\bar{F},t}}$ and ${\bar{\mathcal{P}} \argc{\eta,t}}$ is in large part controlled by their two-point disorder correlators, on which we focus our interest:
\begin{eqnarray}
 \bar{C} (t,\valabs{y_1-y_2})
 \equiv & \overline{\argc{\bar{F}_V(t,y_1)-\bar{F}_V (t,y_2)}^2} &
 \label{eq-def-corr-FbarFbar2} \\
 \bar{R} (t,\valabs{y_1-y_2})
 \equiv & \overline{\eta_V (t,y_1) \eta_V (t,y_2)} &
 \label{eq-def-corr-etaeta}
\end{eqnarray}
which reflect explicitly the translation invariance, and are related by
\begin{equation}
 \bar{C} (t,y)
 = \int^{y}_{0} dy_1 \int^{y}_{0} dy_2 \cdot \bar{R} (t,\vert y_1-y_2 \vert)
 \label{eq-def-corr-CbarRbar-integral}
\end{equation}
or alternatively by ${\partial_y^2 \bar{C}(t,y) = 2 \bar{R}(t,y)}$ using their parity.
Note that the second moment ${\bar{C}(t,y)}$ is equal to the second \textit{cumulant} of the total free-energy ~${\overline{\argc{F_V(t,y_1)-F_V (t,y_2)}^2}}^c$, but that this direct equivalence breaks down for higher $n$-point correlation functions. It is then more transparent to focus on the fluctuations of the disorder free-energy~${\bar{F}_V(t,y)}$, whose distribution is translation-invariant as stated by~\eqref{eq-STS-PFbarV}.

The connection between the fluctuating disorder free-energy $\bar{F}_V(t,y)$ and the PDF $\mathcal{P}(t,y)$ with its moments $\overline{\moy{y(t)^k}}$ can formally be defined as:
\begin{equation}
\begin{split}
 & \overline{\moy{y(t)^k}}
 = \int \mathcal{D}V \, \bar{\mathcal{P}} \argc{V} \frac{\int dy \, y^k \, e^{-F_V(t,y)/T}}{\int dy \, e^{-F_V(t,y)/T}} \\
 &= \int \mathcal{D} \bar{F} \, \bar{\mathcal{P}} \argc{\bar{F},t} \frac{\int dy \, y^k \, e^{-(F_{\text{th}}(t,y)+\bar{F}_V (t,y) )/T}}{\int dy \, e^{-(F_{\text{th}}(t,y)+\bar{F}_V (t,y) )/T}}
\end{split}
\label{eq-def-moments-pdf-replicas-1}
\end{equation}
where only the $y$-dependent part of the pseudo free-energy, \textit{i.e.} the information encoded in the random phase $\eta_V (t,y)$ actually matters. However, even the roughness $B(t)$ cannot be computed straightforwardly through the disorder average, this would require \textit{e.g.} the introduction of replic{\ae} in a GVM framework \cite{mezard_parisi_1992_JPhysI02_2231,agoritsas_2010_PhysRevB_82_184207,agoritsas-2012-FHHpenta} (cf. Appendix~\ref{A-GVM-PRB2010}).

In order to determine the quantities introduced in this section and that characterize the 1D interface, one can either perform analytical studies, which is the object of the present paper, or numerical ones that will be the object of the separate publication Ref.~\cite{agoritsas_2012_FHHtri-numerics}.

\subsection{Feynman-Kac evolution equations} \label{section-FeynmanKac}

Since we work with a one-dimensional object (the 1D interface, likewise the DP), explicit evolution equations can be written for $W_V (t,y)$, $F_V (t,y)$, $\bar{F}_V (t,y)$ and $\eta_V (t,y)$ \cite{huse_henley_fisher_1985_PhysRevLett55_2924}.
We can thus follow the evolution with continuous `time' or lengthscale $t$ of the effective PDF-related quantities at fixed disorder, and also of the mean values $\overline{\bar{F}_V(t,y)}$ and $\overline{\eta_V(t,y)}$.

However, no such closed equation for the correlators $\bar{C}(t,y)$ and $\bar{R}(t,y)$, the normalized PDF $\mathcal{P}(t,y)$, nor the roughness $B(t)$ of course, are available.
This limitation in the lengthscale renormalization of the disorder-average quantities is conceptually similar to the fact that the FRG flow equations
\cite{balents-fisher_1993_PhysRevB48_5949,chauve_2000_ThesePC_PhysRevB62_6241,ledoussal_wiese_2005_PhysRevE72_035101,wiese-ledoussal_2007_arXiv:cond-mat/0611346}
of the disorder correlator $R_{\xi}(x)$ \eqref{eq-def-moydis} are truncated in a perturbative expansion in $\epsilon=4-d$ (with the dimension $d=1$ for the 1D interface), an exact analytical description at all lengthscales remaining thus unsolved.

At fixed microscopic disorder, in a continuous-time limit and at finite temperature, the so-called `Feynman-Kac' formula \cite{feynman_1948_RevModPhys20_367,kac_1949_TransAmerMathSoc65_1,kardar_2007_StatPhysFields} for $W_V(t,y)$,
is a continuum stochastic heat equation with multiplicative noise \cite{huse_henley_fisher_1985_PhysRevLett55_2924,bertini-cancrini_1995_JStatPhys78_1377,bertini-giacomin_1997_CommMathPhys183_571,alberts-quastel_2012_arXiv:1202.4403v1}:
\begin{equation}
 \partial_t \argc{\frac{W_V (t,y)}{\Wbar_{V\equiv 0}(t)}}
 = \argc{\frac{T}{2c} \partial_y^2 - \frac{1}{T} V(t,y) } \argc{\frac{W_V (t,y)}{\Wbar_{V\equiv 0}(t)}}
 \label{eq-FeynmanKac-Wvnorm}
\end{equation}
where the normalization ${\Wbar_{V \equiv 0} (t)}$ is usually hidden in the functional integration ${\int \mathcal{D}y(t)}$ of \eqref{eq-def-unnorm-Boltzmann-Wv}.
In order to clarify the normalization issues that arise due to the disorder, this last equation is rederived in~Appendix~\ref{A-FHHderivation} both in continuous and discretized `time'.
In absence of disorder, we recover the standard heat equation:
\begin{equation}
 \partial_t \mathcal{P}_{V \equiv 0} (t,y) = \frac{T}{2c} \partial_y^2 \mathcal{P}_{V \equiv 0} (t,y)
 \label{eq-Feynman-Kac-Wvnorm-thermal}
\end{equation}
whose solution at fixed `time' is the thermal PDF $\mathcal{P}_{V \equiv 0}(t,y) {= \mathcal{P}_{\text{th}}(t,y)}$, \textit{i.e.} a Gaussian function of zero mean and variance ${B_{\text{th}}(t)=\frac{Tt}{c}}$.

Moving in on the pseudo free-energy $F_V(t,y)$ defined by \eqref{eq-def-free-energy-1} yields a KPZ equation with an additive noise \cite{kardar_1986_originalKPZ_PhysRevLett56_889,corwin_2011_arXiv:1106.1596}:
\begin{equation}
 \partial_t F_V  (t,y)
 =	\frac{T}{2c} \partial_y^2 F_V (t,y)
	- \frac{1}{2c} \argc{\partial_y F_V (t,y)}^2
	+ V(t,y)
 \label{eq-FeynmanKac-FV}
\end{equation}
So the free-energy landscape seen by the DP end-point is a KPZ growing surface, whose disorder correlation length $\xi$ lies along the \textit{internal} direction of the surface, whereas $\xi$ has been initially defined as a microscopic disorder correlation along the \textit{transverse} direction of the 1D interface or growing DP.

As for the disorder free-energy $\bar{F}_V  (t,y)$ \eqref{eq-def-free-energy-2}, it evolves with a tilted KPZ equation:
\begin{equation}
 \begin{split}
 \partial_t \bar{F}_V  (t,y)
 =&	\frac{T}{2c} \partial_y^2 \bar{F}_V (t,y)
 	- \frac{1}{2c} \argc{\partial_y \bar{F}_V (t,y)}^2 \\
 &	- \frac{y}{t} \partial_y \bar{F}_V (t,y) + V(t,y)
 \end{split}
 \label{eq-FeynmanKac-FbarV}
\end{equation}
with the new additive term stemming from $-\frac{1}{c} \argc{\partial_y F_{\text{th}} (t,y)} \argc{\partial_y \bar{F}_V (t,y)} = -\frac{y}{t} \eta_V (t,y)$.
One advantage of focusing on the disorder free-energy is precisely the decomposition of the KPZ nonlinearity~${\frac{1}{2c} \argc{\partial_y F_V (t,y)}^2}$ into two contributions: the remaining nonlinearity~${\frac{1}{2c} \argc{\partial_y \bar{F}_V (t,y)}^2}$ and the tilt term~${\frac{y}{t} \partial_y \bar{F}_V (t,y)}$. Neglecting the nonlinearity in~\eqref{eq-FeynmanKac-FV} yields the well-studied Edward-Wilkinson (EW) equation~\cite{edwards_wilkinson_1982_ProcRSocLondA381_17}, whereas the linearized version of~\eqref{eq-FeynmanKac-FbarV} goes one step further than the EW equation: it still contains a part of the initial KPZ nonlinearity via the tilt term, while remaining solvable as we will show in~Sec.~\ref{section-DPtoymodel-A}.

Applying $\partial_y$ on~\eqref{eq-FeynmanKac-FbarV} yields finally the evolution equation of the random phase ${\eta_V  (t,y)}$ \eqref{eq-def-etaV} itself:
\begin{equation}
 \begin{split}
 \partial_t \eta_V  (t,y)
 =&	\frac{T}{2c} \partial_y^2 \eta_V (t,y)
 	- \frac{1}{2c} \partial_y \argc{\eta_V (t,y)}^2 \\
 &	- \partial_y \argc{\frac{y}{t} \, \eta_V (t,y)}
 	+ \partial_y V(t,y)
 \end{split}
 \label{eq-FeynmanKac-etaV}
\end{equation}

\begin{figure}
 \begin{center}
 \subfigure{\includegraphics[width=0.8\columnwidth]{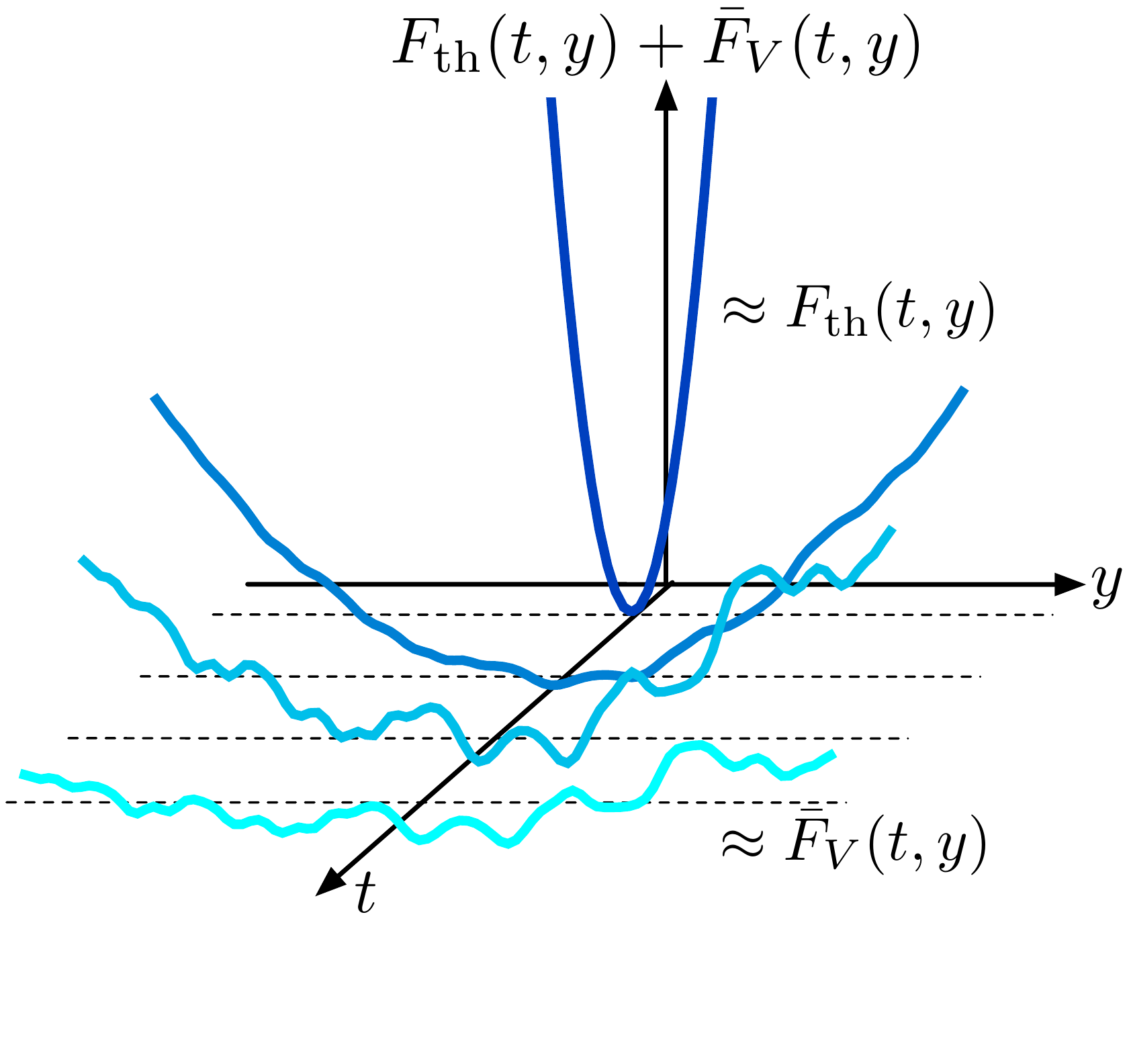}  \label{fig:graphFthFbarV-difft}}
 \subfigure{\includegraphics[width=0.9\columnwidth]{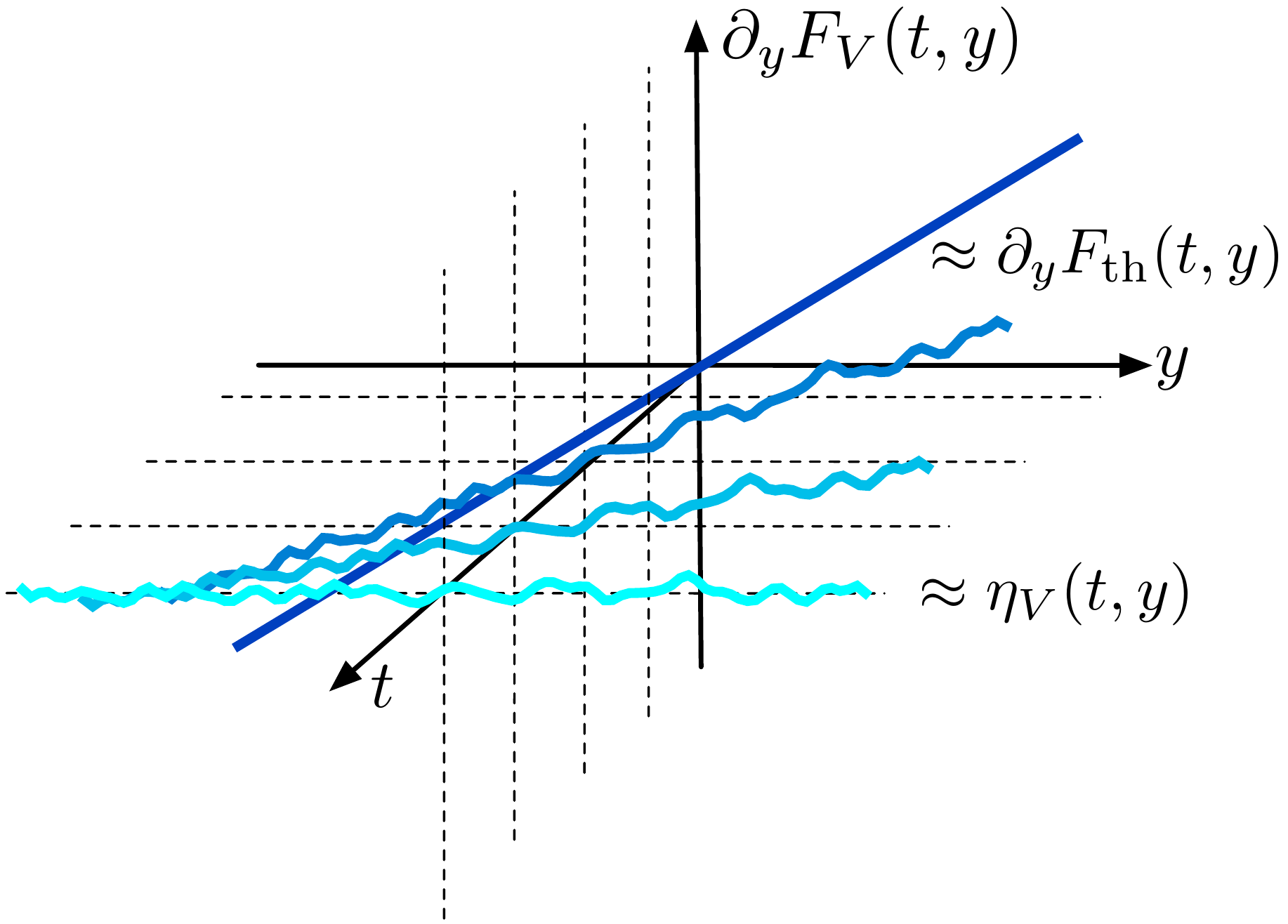}  \label{fig:graphetaV-difft}}
 \caption{
 (Color online)
 Free-energy landscape seen by the DP end-point as a function of `time' or lengthscale $t$.
 \textit{Top}:~Graph of ${F_{\text{th}}(t,y) + \bar{F}_V (t,y)}$ (imposing ${\overline{\bar{F}_V(t,y)}\equiv 0}$ for simplification): the thermal parabola ${F_{\text{th}}(t,y)=\frac{cy^2}{2t}}$ flattens and unveils the disorder fluctuations ${\bar{F}_V (t,y)}$, which sketches a KPZ surface in its steady state at asymptotically large `times'.
 \textit{Bottom}:~Alternative point of view with the graph of ${\partial_y F_V(t,y)=\frac{cy}{t} + \eta_V(t,y)}$, where the random phase is progressively revealed with increasing lengthscale.
 }
 \label{fig:graphFthFbarV-etaV-difft}
 \end{center}
\end{figure}

The disorder free-energy and its random phase encode all the information concerning the effects of disorder, so both $\bar{F}_{V \equiv 0} (t,y)$ and $\eta_{V \equiv 0} (t,y)$ are zero.
In presence of disorder those quantities are moreover completely hidden at small `times' by thermal fluctuations:
\begin{equation}
 F_V (t,y) \stackrel{t \to 0}{\approx} F_{V \equiv 0} (t,y)
 \Rightarrow
 \bar{F}_{V} (t,y) \approx 0
 , \, \;
 \eta_{V} (t,y) \approx 0
 \label{eq-FKbehavior-at-small-times}
\end{equation}
whereas they completely dominate the large-lengthscales behavior ($F_V (t,y) \approx \bar{F}_{V} (t,y) + \text{cte}(t) $), the evolution equation \eqref{eq-FeynmanKac-FV} and \eqref{eq-FeynmanKac-FbarV} thus sharing the same statistical steady-state at asymptotically large `times'.
Those disorder-induced quantities can be properly defined at all `times' (cf. Fig.~\ref{fig:graphFthFbarV-etaV-difft}), yielding in particular the following initial conditions:
\begin{eqnarray}
 \mathcal{P}_V (t=0,y) &=& \delta(y) \label{eq-initial-PV} \\
 \bar{F}_V (t=0,y) & \equiv & 0 \label{eq-initial-FbarV} \\
 \eta_V (t=0,y) & \equiv & 0 \label{eq-initial-etaV}
\end{eqnarray}
For the total free-energy this initial condition corresponds to the `sharp-wedge' ${\lim_{t \to 0} F_{V \equiv 0}(t,y)}$ --~as defined in~\eqref{eq-def-free-energy-3}-\eqref{eq-def-free-energy-4}~-- which non-trivially yields back the Dirac $\delta$ function~\eqref{eq-initial-PV} of the DP fixed endpoint.

Considering at last the evolution of the mean values $\overline{\bar{F}_V(t,y)}$ and $\overline{\eta_V(t,y)}$, at first the translation invariance by STS \eqref{eq-STS-PFbarV} trivially implies that ${\overline{\bar{F}_V(t,y)}=\text{cte}(t)}$ (and ${\overline{\moy{y(t)}}=0}$).
Exchanging the disorder average and the partial derivatives $\partial_{y,t}$, on the definition \eqref{eq-def-etaV} and on \eqref{eq-FeynmanKac-FbarV} respectively, we obtain:
\begin{eqnarray}
 & \overline{\eta_V (t,y)}
 = \partial_y \overline{\bar{F}_V(t,y)}=0 &
 \label{eq-mean-etaV} \\
 & \partial_t \overline{\bar{F}_V (t,y)}
 = -\frac{1}{2c} \overline{\argc{\eta_V (t,y)}^2}
 = -\frac{1}{2c} \cdot \bar{R}(t,y=0) &
 \label{eq-mean-FbarV}
\end{eqnarray}
whereas \eqref{eq-FeynmanKac-etaV} simply yields the consistency check ${\partial_y \bar{R}(t,y=0)=0}$.
So the evolution of the mean disorder free-energy is directly given by the sole two-point correlator of $\eta_V(t,y)$ in $y=0$ at a given `time' $t$.

As we will discuss at length in the next section, the behavior of $\bar{R}(t,y)$ at small $\vert y \vert$ corresponds to the curvature of the disorder free-energy correlator $\bar{C}(t,y)$ around ${y=0}$ which fully determines the amplitude of the geometrical fluctuations characterized by the roughness prefactor $A(c,D,T,\xi)$.
$\bar{R}(t,y)$ has essentially a symmetrical peak centered at $y=0$, whose maximum is well-defined for a finite correlation length $\xi$ but diverges in the limit ${\xi \to 0}$ (corresponding equivalently to the high-temperature regime).
The connection between this regularization at $\xi>0$ and the `time'-evolution of the peak main features, \textit{i.e.} its typical width $\tilde{\xi}_t$ and amplitude $\widetilde{D}_t$, will be the two ingredients of the DP toymodel constructed in the next section.

\subsection{`Time'-evolution equations for the two-point correlators ${\bar{R}(t,y)}$ and ${\bar{C}(t,y)}$}
\label{subsection-FK-Rbar-Cbar-Ito}

There are no closed equations for ${\partial_t \bar{R}(t,y)}$ and ${\partial_t \bar{C}(t,y)}$, but the combination of the Feynman-Kac equations \eqref{eq-FeynmanKac-FbarV}-\eqref{eq-FeynmanKac-etaV} with the It\={o}'s formula yields nevertheless, as presented in details in Appendix~\ref{A-flow_CbarRbar}:
\begin{equation}
\begin{split}
 \partial_t \bar{R}(t,y)
 =	& \frac{T}{c} \partial_y^2 \bar{R}(t,y) - \frac{1}{t} \arga{\bar{R}(t,y) + \partial_y \argc{y \bar{R}(t,y)}} \\ 
 	& - \frac{1}{c} \partial_y \bar{R}_3 (t,y) - D R_{\xi}''(y) 
\end{split}
\label{eq-noclose-Rbar}
\end{equation}
\begin{equation}
\begin{split}
 \partial_t \bar{C}(t,y)
 =	& \frac{T}{c} \partial_y^2  \argc{\bar{C}(t,y) - \bar{C}(t,0)} - \frac{y}{t} \partial_y \bar{C}(t,y) \\
	&  - \frac{1}{c} \bar{C}_3 (t,y)  - 2D \argc{R_{\xi}(y) - R_{\xi}(0)}
\end{split}
\label{eq-noclose-Cbar}
\end{equation}
which would be closed but for the presence of the three-point correlators:
\begin{eqnarray}
 \bar{R}_3 (t,y) & \equiv & \overline{\eta(t,y)^2 \eta(t,0)} \label{eq-R3-def} \\
 \bar{C}_3 (t,y) & \equiv & -2 \overline{ \argc{\bar{F}(t,y)- \bar{F}(t,0)} \argc{\partial_y \bar{F} (t,0)}^2 } \label{eq-C3-def}
\end{eqnarray}

Neglecting the non-linear KPZ term in the evolution equation \eqref{eq-FeynmanKac-FbarV} for $\bar{F}_V(t,y)$ is equivalent to neglecting those three-point contributions.
The solution for the corresponding \textit{linearized} correlator ${\bar{R}^{\text{lin}}(t,y)}$ for a generic RB disorder correlator ${R_{\xi}(y)}$ is given in the next section, and its complete derivation is detailed in Appendix~\ref{A-short-time-dynamics-Fbar-generic}.
It will be used in the next section, in order to discuss on one hand the expected qualitative behavior of the correlator ${\bar{R}(t,y)}$,
and to identify on the other hand the role of the KPZ nonlinearity in the short- \textit{versus} large-`times' and the low- \textit{versus} high-$T$ regimes.


\section{Exact properties and construction of a DP toymodel} \label{section-DPtoymodel}

The DP free-energy fluctuations for a RB uncorrelated disorder and the `sharp-wedge' initial condition~\eqref{eq-initial-PV} have been progressively elucidated,
first at infinite `time'~\cite{huse_henley_fisher_1985_PhysRevLett55_2924},
then at asymptotically large `time'~\cite{praehofer-spohn_2002_JStatPhys108_1071}
and finally recently at all `times'~\citep{calabrese_2010_EPL90_20002,dotsenko_2010_EPL90_20003,sasamoto_2010_NuclPhysB_834_523,amir_arXiv:1003.0443}, using a wide range of different mappings and techniques which strongly rely on the assumption~${\xi=0}$.

In this section we first recall the analytical results for the asymptotically large `times' DP fluctuations, exact for an \textit{uncorrelated} disorder (${\xi=0}$), and we discuss their possible generalization for a \textit{correlated} disorder ($\xi>0$): we examine in particular the connection between the KPZ nonlinearity and the non-Gaussianity of the free-energy fluctuations in the Fokker-Planck approach.
Then we present the complete solution of the linearized equation for~${\partial_t \bar{F}_V(t,y)}$~\eqref{eq-FeynmanKac-FbarV}, obtained for a generic RB disorder correlator~${R_{\xi}(y)}$ and at all `times', and we use it as a qualitative benchmark for the `time'-dependent phenomenology summarized by~Fig.~\ref{fig:graphDPtoymodel}.

Merging the intuition gained from these considerations of both the asymptotic properties and the linearized solution, we define a DP toymodel for the disorder free-energy fluctuations, valid by construction for `times' larger than a characteristic scale $t_{\text{sat}}$ --~bounded above by the Larkin length~$L_c$~-- and aimed at grasping the temperature dependence of the DP fluctuations.

\subsection{Free-energy fluctuations at asymptotically large `times' and~${\xi=0}$} \label{section-DPtoymodel-A}

At infinite `time' and in an uncorrelated disorder (${\xi=0}$), the distributions $\bar{\mathcal{P}} \argc{\bar{F}}$ and $\bar{\mathcal{P}} \argc{\eta}$ are Gaussian and their two-points correlators are exactly known
\begin{equation}
 \bar{C}_{\xi=0}(\infty,y)
 = \widetilde{D}_{\infty} \cdot \valabs{y}
 \, , \;
 \bar{R}_{\xi=0}(\infty,y)
 = \widetilde{D}_{\infty} \cdot R_{\xi=0}(y)
 \label{eq-infty-time-FPsolution}
\end{equation}
with~$\widetilde{D}_{\infty} = cD/T$ and~${R_{\xi=0}(y)=\delta(y)}$.
The Dirac $\delta$-function of $\bar{R}$ encodes the infinite-`time' amnesia of the DP with respect to the remoteness of its initial condition ${t=0}$,
and the absolute value of $\bar{C}$ encodes the scale invariance of this steady state characterized by the scaling in distribution~$\bar{F}(y) \sim y^{1/2}$.
%
This steady-state solution of the KPZ equations \eqref{eq-FeynmanKac-FV}-\eqref{eq-FeynmanKac-FbarV} for a $\delta$-correlated ${V(t,y)}$ actually yields the prediction $\zeta_{\text{KPZ}}=2/3$ for the asymptotic roughness exponent \cite{huse_henley_fisher_1985_PhysRevLett55_2924} (as discussed later in Sec.~\ref{section-DPtoymodel-D}).

Actually at ${\xi=0}$ the distribution of the total free-energy $F_V(t,y)$ itself, given by the KPZ equation with `sharp-wedge' initial condition, is exactly known at all `times' in terms of a Fredholm determinant with an Airy kernel \cite{calabrese_2010_EPL90_20002,dotsenko_2010_JStatMech_P03022,
amir_arXiv:1003.0443,
sasamoto_2010_NuclPhysB_834_523,
dotsenko_2010_EPL90_20003}.
It is non-Gaussian and at asymptotically large `times' it tends to the Gaussian-Unitary-Ensemble (GUE) Tracy-Widom distribution \cite{craig_widom_1994_CommMathPhys159_151,amir_arXiv:1003.0443}, but at strictly infinite `time' it eventually yields back a Gaussian distribution.
Its second cumulant corresponds to our correlator $\bar{C}(t,y)$ (for $\bar{F}_V(t,y)$) and is exactly known asymptotically as the correlator of an Airy$_2$ process \cite{praehofer-spohn_2002_JStatPhys108_1071}.
Schematically the asymptotic ${\bar{C}_{\xi=0}(t,y)}$ displays additional saturation `wings' compared to the absolute value \eqref{eq-infty-time-FPsolution} as pictured in~Fig.~\ref{fig:graphDPtoymodel}. These `wings' appear when ${y^2 \sim \overline{\moy{y(t)^2}}}$, \textit{i.e.} where the transverse displacement is defined by the roughness $B(t)$ at a given `time' $t$~\cite{agoritsas-2012-FHHpenta}.

At finite `time' and/or in a correlated disorder \eqref{eq-def-moydis} with ${\xi>0}$, the distributions $\bar{\mathcal{P}} \argc{\bar{F},t}$ and $\bar{\mathcal{P}} \argc{\eta,t}$ are thus \textit{a priori} not Gaussian but we can still focus on the two-point correlator $\bar{R}(t,y)$ properties.
We know in particular that its integral must satisfy \cite{agoritsas-2012-FHHpenta}:
\begin{equation}
 \int_{\mathbb{R}}dy \, \bar{R}(t,y) = 0
 \label{eq-new-integral-Rbar-1}
\end{equation}
with the exception of strictly infinite `time':
\begin{equation}
 \int_{\mathbb{R}}dy \, \bar{R}(\infty,y) \equiv \widetilde{D}_{\infty} >0
 \label{eq-new-integral-Rbar-2}
\end{equation}
These exact properties of its curvature~${\bar{R}(t,y)=\frac12 \partial_y^2 \bar{C}(t,y)}$ actually \textit{require} the existence of saturation `wings' of the asymptotic $\bar{C}_{\xi=0}(t,y)$, which are pushed to ${y \to \pm \infty}$ as ${t\to\infty}$.

The infinite-`time' solution~\eqref{eq-infty-time-FPsolution} was obtained in~Ref.~\cite{huse_henley_fisher_1985_PhysRevLett55_2924} as defining the steady-state solution of the Fokker-Planck (FP) equation.
However, as detailed in Appendix~\ref{A-FokkerPlanck-equation-xi0}, the steady-state solutions of the FP equation for ${\partial_t \bar{\mathcal{P}} \argc{\bar{F},t} }$  and ${\partial_t \bar{\mathcal{P}} \argc{\eta,t} }$ at ${\xi=0}$ are Gaussian distributions with the correlators~\eqref{eq-infty-time-FPsolution} \textit{only at strictly infinite `time'}, and by imposing ${\widetilde{D}_{\infty} = cD/T}$ and the boundary conditions ${\eta_V(t,y) \vert_{y = \pm \infty} =0}$.
The correlator ${\bar{R}(\infty,y)}$ of the random phase \eqref{eq-def-corr-etaeta} then coincides with the transverse correlator ${R_{\xi=0}(y)}$ of the microscopic disorder \eqref{eq-def-moydis}, up to the overall amplitude ${\widetilde{D}_{\infty} }$. 
Note that the KPZ non-linear term ${-\frac{1}{2c} \argc{\partial_y \bar{F}_V(t,y)}^2}$ in \eqref{eq-FeynmanKac-FbarV} plays no role in the determination of this asymptotic amplitude, since its contribution disappears completely with the chosen boundary conditions.

We now transpose this FP scheme from the uncorrelated case (${\xi=0}$) to the correlated case (${\xi>0}$):
we assume a Gaussian ${\bar{\mathcal{P}}_{\text{G}} \argc{\bar{F},t}}$ of correlator $\bar{R}^{\text{lin}}(t,y)$, and we impose ${\widetilde{D}_{\infty} = cD/T}$ and the boundary condition ${\eta_V(t,y) \vert_{y = \pm \infty} =0}$.
Using this set of assumptions, we show in Appendix~\ref{A-FokkerPlanck-equation-xinonzero} that at infinite `time' the correlator
\begin{equation}
 \bar{R}^{\text{lin}}(\infty,y)
 = \widetilde{D}_{\infty} \cdot R_{\xi}(y)
 \, , \;
 \widetilde{D}_{\infty}  = \frac{cD}{T}
 \label{eq-infty-time-Rxi-lin}
\end{equation}
defines a steady-state solution but for the \textit{linearized} FP equation, where the KPZ non-linear term ${-\frac{1}{2c} \argc{\partial_y \bar{F}_V(t,y)}^2}$ has been neglected.
This result is compatible with~\eqref{eq-new-integral-Rbar-2} and coincides remarkably with the exact solution for the uncorrelated case~\eqref{eq-infty-time-FPsolution}.
It emphasizes that any non-Gaussianity in the steady-state can only stem from the KPZ nonlinearity in~${\partial_t \bar{F}_V(t,y)}$~\eqref{eq-FeynmanKac-FbarV}.

\subsection{Solution of the linearized tilted KPZ equation for a generic~${R_{\xi=0}(y)}$} \label{section-DPtoymodel-C}

The steady-state of the FP equation, that we have discussed in the previous section, characterizes the infinite-`time' properties of the DP (hence the macroscopic lengthscales for the static 1D interface).
We now consider its finite-`time' properties by computing exactly the full solution of the linearized correlator~${\bar{R}^{\text{lin}}(t,y)}$ --~first introduced in~\eqref{eq-infty-time-Rxi-lin}~-- for a generic RB disorder correlator~${R_{\xi}(y)}$ and its complete derivation can be found in~Appendix~\ref{A-short-time-dynamics-Fbar-generic}.

As discussed after the Feynman-Kac evolution equation~${\partial_t \bar{F}_V(t,y)}$~\eqref{eq-FeynmanKac-FbarV}, linearizing this \textit{tilted} KPZ equation is not equivalent to the EW equation~\cite{edwards_wilkinson_1982_ProcRSocLondA381_17}, because it still contains a contribution stemming from the KPZ nonlinearity~${\frac{1}{2c} \argc{\partial_y F_V (t,y)}^2}$ via the (linear) tilt ${-\frac{y}{t} \partial_y \bar{F}_V (t,y)}$.
This approximation is physically correct at least for sufficiently short `times', for which the nonlinearity~${\frac{1}{2c} \argc{\partial_y \bar{F}_V (t,y)}^2}$ can be neglected compared to the tilt.
At larger `times' however this approximation eventually breaks down. The linearized correlator  will then bear a trace of the short-`time' diffusive behavior, as an artifact of the linearization.

The disorder free-energy distribution ${\bar{\mathcal{P}}_{\text{lin}} \argc{\bar{F},t}}$ is predicted to be Gaussian --~consistently with the assumption needed for the derivation of~\eqref{eq-infty-time-Rxi-lin}~-- and thus fully characterized by~${\bar{R}^{\text{lin}}(t,y)}$.
The full solution decomposes as follows:
\begin{equation}
\argp{\frac{cD}{T}}^{-1} \bar{R}^{\text{lin}}(t,y)
 = R_{\xi}(y) - b^{\text{lin}}(t,y)
 \label{eq-decomposition-Rbarlin-1}
\end{equation}
with
\begin{equation}
\begin{split}
 b^{\text{lin}}& (t,y) \sqrt{B_{\text{th}}(t)} \\
 = & - \frac{y}{\sqrt{B_{\text{th}}(t)}} R_{\xi/\sqrt{B_{\text{th}}(t)}}^{(-1)}{\argp{y/\sqrt{B_{\text{th}}(t)}}} \\
 & + \int_0^{\infty} dw \, w^2 e^{-w \argc{w+y/\sqrt{B_{\text{th}}(t)}}} R_{\xi/\sqrt{B_{\text{th}}(t)}}^{(-1)}(w) \\
 & + \int_{y/\sqrt{B_{\text{th}}(t)}}^{\infty} \!  \!  \!  \!  \!  \! dw \, w^2 e^{-w \argc{w-y/\sqrt{B_{\text{th}}(t)}}} R_{\xi/\sqrt{B_{\text{th}}(t)}}^{(-1)} (w)
\end{split}
\label{eq-decomposition-Rbarlin-2}
\end{equation}
where $R^{(-1)}_{\xi}(y)$ denotes the primitive of the disorder correlator,
all the rescaling is purely diffusive with as usual ${B_{\text{th}}(t)=\frac{Tt}{c}}$
and ${\lim_{t \to \infty}  b^{\text{lin}}(t,y) = 0}$ so that the asymptotic correlator \eqref{eq-infty-time-Rxi-lin} is indeed recovered.
A remarkable property of the linearized solution is that all the `time'-dependence in ${b^{\text{lin}} (t,y)}$ is described by an overall factor $\sqrt{B_{\text{th}}(t)}$ and the rescaling of the transverse lengthscales $y$ ad $\xi$ by the same factor, as shown explicitly in \eqref{eq-decomposition-Rbarlin-2}.
As an example, the graphs of ${\bar{R}^{\text{lin}}(t,y)}$ , ${\frac12 \partial_y \bar{C}^{\text{lin}}(t,y)}$ and ${\bar{C}^{\text{lin}}(t,y)}$ for ${R_{\xi}(y)}$ taken as a Gaussian function of variance~${2\xi^2}$ are given in~Fig.~\ref{fig:finitetimescaling-Cbarlin-Rbarlin}.

Let us emphasize that the infinite-`time' contribution ${R_{\xi}(y)}$ in~\eqref{eq-decomposition-Rbarlin-1} arises from the non-analyticity of the kernel relating this correlator to ${\bar{R}^{\text{lin}}(t,y)}$.
It thus requires a careful treatment of the boundary terms in the corresponding convolution formula~\eqref{eq-res-Cbarlin}.

The exact relations~\eqref{eq-new-integral-Rbar-1}-\eqref{eq-new-integral-Rbar-2} are satisfied by~${\bar{R}^{\text{lin}}(t,y)}$, however two artefacts of the linearization can be identified:
on one hand the distribution ${\bar{\mathcal{P}}_{\text{lin}} \argc{\bar{F},t}}$ is Gaussian, whereas the exact ${\bar{\mathcal{P}} \argc{\bar{F},t}}$ is known to display non-Gaussian features;
on the other hand, ${\bar{R}^{\text{lin}}(t,y)}$ rescales with respect to the diffusive roughness~${B_{\text{th}}(t)}$ at all `times', whereas it should rescale with respect to the asymptotic roughness~${\sim t^{4/3}}$ at large `times'~\cite{agoritsas-2012-FHHpenta}.
These two artefacts point out the crucial role played of by the nonlinearity~$\argc{\partial_y \bar{F}_V}^2$ in the non-Gaussianity and the `time'-dependence of the free-energy fluctuations.
The linearized solution~\eqref{eq-decomposition-Rbarlin-1}-\eqref{eq-decomposition-Rbarlin-2} can nevertheless be considered as a qualitative benchmark for the correlated disorder case~${\xi>0}$.

\begin{figure}
 \includegraphics[width=0.95\columnwidth]{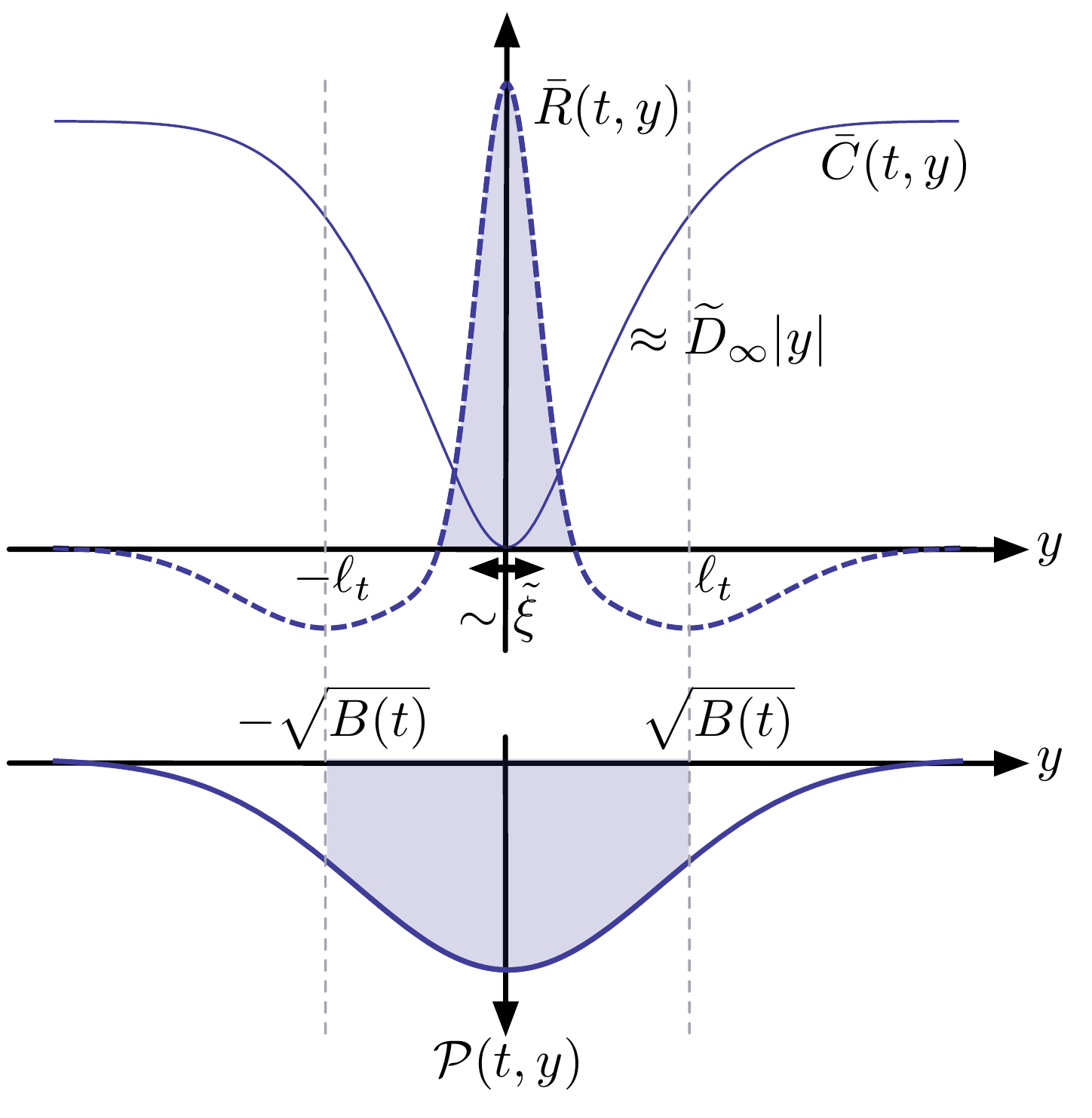}
 \caption{
 \textit{Top}:~Schematic graphs of the two-point correlators ${\bar{C}(t,y)}$ and ${\bar{R}(t,y)}$ (respectively in full and dashed curves) at fixed `time' ${t>t_{\text{sat}}}$, which suggests the generic decomposition~\eqref{eq-infty-time-Rxi-bumps-1}-\eqref{eq-infty-time-Rxi-bumps-2};
 they display the two characteristic lengthscales $\tilde{\xi}$ and $\ell_t$ in the {$y$-direction};
 the dashed area below the central peak of $\bar{R}$ corresponds roughly to the saturation amplitude $\widetilde{D}_{\infty}$ and translates into the slope of the intermediate linear behavior of $\bar{C}$ (since ${\partial_y^2 \bar{C}(t,y) = 2 \bar{R}(t,y)}$ by \eqref{eq-def-corr-CbarRbar-integral}).
 \textit{Bottom}:~Corresponding PDF ${\mathcal{P}(t,y)}$ whose variance is the roughness ${B(t)}$; the dashed area emphasizes the most probable positions of the DP endpoint, which exclude the large $y$ and thus the saturation `wings' of $\bar{C}$ or  the negative bumps of ${\bar{R}}$. 
 }
 \label{fig:graphDPtoymodel}
\end{figure}

The form of the linearized solution suggests indeed the following generic decomposition at finite `time' and for a generic RB disorder correlator~${R_{\xi}(y)}$:
\begin{eqnarray}
 &\bar{R}(t,y)
 = \widetilde{D}_{\infty} \cdot \argc{ \mathcal{R}_{\xi}(y) - \frac{b_{+}(t,y)+b_{-} (t,y)}{2}}
 & \label{eq-infty-time-Rxi-bumps-1} \\
 & \lim_{t \to \infty} b_{\pm}(t,y) = 0
 \, \Rightarrow \,
 \bar{R}(\infty,y) \equiv \widetilde{D}_{\infty} \, \mathcal{R}_{\xi}(y) 
 & \label{eq-infty-time-Rxi-bumps-3} \\
 &  \int_{\mathbb{R}} dy \, \mathcal{R}_{\xi}(y) \equiv 1
 \, \Rightarrow \,
 \int_{\mathbb{R}} dy \, b_{\pm}(t,y) = 1 \, \forall t
 & \label{eq-infty-time-Rxi-bumps-2}
\end{eqnarray}
suited for the asymptotically large `times' where we conjecture the function ${\mathcal{R}_{\xi}(y)}$ to tend towards the microscopic-disorder transverse correlator ${R_{\xi}(y)}$ at high temperature. The distinct corrections $b_{\pm}(t,y)$ correspond to the `wings' in $\bar{C}(t,y)$ and move to large $y$ with increasing scale ${\ell_t \sim \sqrt{B(t)}}$~\cite{agoritsas-2012-FHHpenta}.
Fig.~\ref{fig:graphDPtoymodel} summarizes schematically this phenomenology (to be compared to~Fig.~\ref{fig:finitetimescaling-Cbarlin-Rbarlin}), with~${\tilde{\xi} \sim \xi}$ being the rounding of the correlators due to the microscopic disorder correlation ${\xi>0}$.

Finally, the asymptotic amplitude $\widetilde{D}_{\infty}$ in \eqref{eq-infty-time-Rxi-bumps-1}  is predicted to be $cD/T$ in the limit $\xi \to 0$; however this prediction is non-physical in the limit ${T\to 0}$ so it must break down for temperatures below ${T_c(\xi)=(\xi c D)^{1/3}}$ as discussed in Ref.~\cite{agoritsas_2012_ECRYS2011}.
We will examine from now on the assumption that for ${T<T_c}$ the decomposition \eqref{eq-infty-time-Rxi-bumps-1} remains valid but with ${\widetilde{D}_{\infty}=cD/T_c}$, justifying it with scaling and saddle-point arguments.
Since the full linearized solution is known explicitly,
we know that such a saturation of the asymptotic amplitude can only arise from the KPZ non-linear contribution at `times' below the Larkin length, before the corrections $b_{\pm}(t,y)$ separate from the microscopic disorder correlator $R_{\xi}(y)$.

\subsection{DP Toymodel} \label{section-DPtoymodel-B}

We do not know exactly the distributions $\bar{\mathcal{P}}\argc{\bar{F},t}$ and $\bar{\mathcal{P}}\argc{\eta,t}$, or even their correlators $\bar{C}(t,y)$ and $\bar{R}(t,y)$, for a generic disorder transverse correlator \eqref{eq-def-moydis}.
As we have just seen,
neglecting the KPZ non-linearity in the Feynman-Kac equations, it is however possible to go beyond \eqref{eq-infty-time-Rxi-lin} and actually compute at all `times' the correlators $\bar{C}^{\text{lin}}(t,y)$ and $\bar{R}^{\text{lin}}(t,y)$, starting from a generic RB correlator $R_{\xi}(y)$ (cf. \eqref{eq-decomposition-Rbarlin-2});
they reconnect of course with the infinite lengthscale limit \eqref{eq-infty-time-Rxi-lin} but their corresponding corrections $b_{\pm}(t,y)$ encode a pure thermal scaling of the roughness, inherited from the small lengthscales and kept at all lengthscales \cite{agoritsas-2012-FHHpenta}.

Taking an opposite point of view, we have considered a DP toymodel constructed from the asymptotically large `times' properties of the random-phase $\eta_V$.
This construction is based on the main assumption that there exists a characteristic `time' $t_{\text{sat}}$ above which the fluctuations of~${\eta_V(t,y)}$ have reached a saturation regime. This regime can be minimally characterized via its two-point correlator behavior around~$y=0$, \textit{i.e.} ${\bar{R}(t,y) \sim \mathcal{R}(y)}$ as defined in~\eqref{eq-infty-time-Rxi-bumps-1}-\eqref{eq-infty-time-Rxi-bumps-2} and depicted in~Fig.~\ref{fig:graphDPtoymodel}.
For the DP geometrical fluctuations, the characteristic scale traditionally invoked is the Larkin length~$L_c$, defined below \eqref{eq-def-roughness-zeta} as the `time' marking the beginning of the asymptotic powerlaw regime for the roughness.
The scale invariance thus displayed for the \textit{geometrical fluctuations} can only be achieved for scales where the \textit{free-energy fluctuations} have saturated, hence for `times' larger than~$t_{\text{sat}}$.
This argument yields consistently the upper bound~${t_{\text{sat}}\leq L_c}$.

First we assume that the effective disorder at fixed `time', $\bar{F}_V(t,y)$ and $\eta_V(t,y)$, have \textit{Gaussian distributions} accordingly to their linearized FP equation.
So they are fully described by their two-points correlators \eqref{eq-def-corr-FbarFbar2} and \eqref{eq-def-corr-etaeta} (translational-invariant by the STS \eqref{eq-STS-PFbarV}) and their mean values ${\overline{\eta_V(t,y)}=0}$ and ${\overline{\bar{F}_V(t,y)}=-\frac{t}{2c}\bar{R}(t,0)}$ by \eqref{eq-mean-etaV} and \eqref{eq-mean-FbarV} (which play however no role in the computation of statistical averages).

Secondly we assume that \textit{the random-phase correlator has a stable normalized function $\mathcal{R}$}, all the possible `time'-dependence being generically hidden in two effective parameters $\widetilde{D}_t$ and $\tilde{\xi}_t$:
\begin{equation}
 \bar{R}(t,y)
 \approx \widetilde{D}_t \cdot \mathcal{R}_{\tilde{\xi}_t}(y)
 \, , \;
 \int_{\mathbb{R}} dy \cdot \mathcal{R}_{\tilde{\xi}_t}(y) \equiv 1
 \label{eq-toymodel-def-Rbar-functional}
\end{equation}
This form generalizes the decomposition \eqref{eq-infty-time-Rxi-bumps-1} but neglecting the corrections $b_{\pm}(t,y)$.
This is a self-consistent approximation since those corrections and the corresponding `wings' of $\bar{C}(t,y)$ appear at ${y^2 \sim B(t)}$, and by definition of the roughness it corresponds to an improbable position of the DP end-point of decreasing weight $\mathcal{P}(t,y)$ as illustrated in~Fig.~\ref{fig:graphDPtoymodel}.

Finally we assume that \textit{the function $\mathcal{R}_{\tilde{\xi}}(y)$ coincides with the transverse correlator $R_{\xi}(y)$} of the microscopic disorder, as for the linearized FP equation at infinite `time' \eqref{eq-infty-time-Rxi-lin}.
The effective width $\tilde{\xi}_t$ and amplitude $\widetilde{D}_t$ are kept generic though but under the asymptotic constraint
\begin{equation}
 \widetilde{D}_{\infty}(T,\xi) \equiv f(T,\xi) \cdot \frac{cD}{T}
 \label{eq-Dtilde-infty-finterp}
\end{equation}
where ${f(T,\xi)}$ is an interpolating parameter such that we recover the correct ${\xi=0}$ limit \eqref{eq-infty-time-FPsolution} with ${f(T,0) \equiv 1}$.
A weaker assumption would be to assume the rescaling  ${\mathcal{R}_{\tilde{\xi}}(y)=\frac{1}{\tilde{\xi}} \mathcal{R}_1 (y/\tilde{\xi})}$ with~$\mathcal{R}_1$ decaying as fast as a RB disorder correlator.

In Ref.~\cite{agoritsas_2010_PhysRevB_82_184207},
we have obtained for this DP toymodel a set of GVM predictions for the roughness and the Larkin length,
with $R_{\xi}(y)$ taken specifically as a Gaussian function of variance $2\xi^2$ (cf. Appendix~\ref{A-GVM-PRB2010}).
Those predictions are constructed centered on the full-RSB cutoff ${u_c(T,\xi)}$ of equation \eqref{equa-DPtoy-roughness-4}.
Assuming ${\tilde{\xi}_t \approx \xi}$ and ${\widetilde{D}_t \approx \widetilde{D}_{\infty}}$, the form \eqref{eq-Dtilde-infty-finterp} and the definition ${f(T,\xi) \equiv \frac43 u_c(T,\xi) }$ yields a a self-consistent equation for the interpolating parameter:
\begin{equation}
 f^6 = 4 \pi \argc{\frac{T}{T_c(\xi)}}^6 (1-f)
 \, , \;
 T_c (\xi) \equiv (\xi c D)^{1/3}
 \label{eq-equa-interpf-uc}
\end{equation}
that connects monotonously the low- and high-temperature scaling of $\widetilde{D}_{\infty}$ at ${f(T_c,\xi)\approx 0.94}$:
\begin{eqnarray}
 T \ll T_c \, : \; & f \approx (4 \pi)^{1/6} \, \frac{T}{T_c} & \; \Rightarrow \widetilde{D}_{\infty} (0,\xi) \sim cD/T_c \label{eq-toy-Dtilde-lowT-uc} \\
 T \gg T_c \, : \; & f \lesssim 1  & \; \Rightarrow \widetilde{D}_{\infty} (T,0) \sim cD/T  \label{eq-toy-Dtilde-highT-uc}
\end{eqnarray}
and hence for the Larkin length \eqref{equa-DPtoy-roughness-3} and the asymptotic roughness \eqref{equa-DPtoy-roughness-1} beyond $L_c$:
\begin{eqnarray}
 L_c (T,\xi)
 &=& 4 \pi \cdot \frac{T^5}{cD^2} \cdot f(T,\xi)^{-5}
 \label{eq-def-Larkin-GVM-DPtoymodel} \\
 B_{\text{asympt}}(t)
 & \approx &
 \frac{3}{2^{2/3}\pi^{1/3}} \argc{\frac{{\widetilde{D}_{\infty}(T,\xi)}}{c^2}}^{2/3} \! \! \! t^{4/3}
 \label{eq-Basympt-DPtoymodel}
\end{eqnarray}
%
According to these GVM predictions, the amplitude of the geometrical fluctuations ${y(t)^2 \sim A(c,D,T,\xi) \, t^{2\zeta}}$ at large lengthscales has a temperature dependence which is damped as sufficiently low $T$ below ${T_c(\xi)>0}$, whereas the superdiffusing scaling ${\zeta=\frac23}$ remains unchanged.

For the DP toymodel, there is thus a physically deep connection between the full-RSB cutoff ${u_c}$ in the GVM computation, the asymptotic amplitude of the random-phase correlator at small $\valabs{y}$ (${\bar{R}(t,0) \stackrel{(t \to \infty)}{\sim} \widetilde{D}_{\infty}/\xi}$),
the Larkin length $L_c$ and the amplitude of the roughness at large lengthscales, this last quantity being typically accessible in experiments.
Let us emphasize the physical meaning of the Larkin length: as defined below~\eqref{eq-def-roughness-zeta}, $L_c$ is the lengthscale or `time' which marks the beginning of the asymptotic `random-manifold' regime for the roughness, \textit{i.e.} for `times' larger than $L_c$ the roughness follows the powerlaw~${B(t)\sim t^{4/3}}$. This promotes $L_c$ to a characteristic scale for the \textit{asymptotic} fluctuations and properties of the DP and 1D interface, a crucial point that will be fully exploited in the scaling analysis of next section.

Note finally that the DP toymodel we propose is an improved version of a previous toymodel, where the equivalent of ${\bar{F}(t,y)}$ is a double-sided Brownian motion in $y$ (see \textit{e.g.}~Refs~\cite{schulz_1988_JStatPhys51_1, ledoussal_2003_PhysicaA317_140}).
In other words it matches the infinite-`time' and ${\xi=0}$ limit~\eqref{eq-infty-time-FPsolution}.
Our first and main new ingredient is to implement the finite disorder correlation length in a rounding of ${\bar{C}(t,y)}$ at small~$y$ and to attribute it to a similarity between the correlator curvature (${\propto \bar{R}(t,y)}$) and the microscopic disorder correlator.
Our second new ingredient is to introduce generically a `time'-dependence of the effective parameters~${\tilde{\xi}_t}$ and~${\widetilde{D}_t}$, that will be discussed in~Sec.~\ref{section-Dtildeinfty-T-xi}.


\section{Scaling analysis of the temperature-dependence of the asymptotic roughness} \label{section-scaling-saddle}

Now that we have an efficient effective model, it is important to relate its parameters with those of the original 1D interface. We perform such an identification in this section using scaling arguments, by making explicit 
the relations between the DP toymodel effective parameters and the 1D interface parameters $\arga{c,D,T,\xi}$ in the two limits of low- versus high-temperature, and extrapolate a continuous crossover between those two regimes via the temperature dependence of the GVM Larkin length \citep{agoritsas_2010_PhysRevB_82_184207}.

We conclude this construction by sketching two saddle-point arguments for the roughness, which use either the large lengthscale $t$ or the zero-temperature limit of $\frac{1}{T}$ as a control parameter in order to accredit our asymptotic assumptions for a short-range correlated disorder ($\xi>0$).

\subsection{Scaling arguments} \label{section-DPtoymodel-C}

Coming back to a full path-integral representation for the roughness and more generally for any average of observables depending exclusively on the  DP end-point, we present thereafter scaling arguments such as sketched in Ref.~\cite{agoritsas_2012_ECRYS2011} for the 1D interface model defined in Sec.~\ref{section-def-DES}. We also refer to Ref.~\cite{nattermann_renz_1988_PhysRevB38_5184} for a previous approach. Note that we systematically disregard the numerical prefactors in the whole section.
Assuming that the random potential $V$ scales in distribution consistently with its two-point transverse correlator ${R_{\xi}(y)}$ and that ${R_{\xi}(ay) = a^{-1} R_{\xi/a}(y)}$,
the rescaling of the spatial coordinates and of the energy yields \textit{exactly} for the roughness:
\begin{equation}
 B(r;c,D,T,\xi) = a^2 \, \bar{B}(r/b;1,1,T/\widetilde{E},\xi/a)
 \label{eq-scaling-roughness}
\end{equation}
where $\bar{B}$ is the roughness function with \textit{adimensional} parameters, provided the scalings factors
satisfy the two relations involving the Flory exponent ${\zeta_{\text{F}}^{\text{1D}}=3/5}$:
\begin{eqnarray}
 & a = (D^{1/3} c^{-2/3} b )^{3/5} \Leftrightarrow b = (D^{-1/5} c^{2/5} a )^{5/3}  & \label{eq-scaling-ab-Flory} \\
 & \widetilde{E} \equiv ca^2/b = (cD^2 b)^{1/5} = (acD)^{1/3} \label{eq-scaling-Etilde-ab} &
\end{eqnarray}

Fixing one of the scaling factors to a characteristic scale of the model gives three possible choices, each suited for the description of a particular temperature regime (high-$T$, low-$T$ and their connection), with the \textit{ad hoc} assumptions on the scaling function ${\bar{B}(\bar{r};1,1,\bar{T},\bar{\xi})}$.
Firstly with respect to the temperature $T$:
\begin{eqnarray}
 & \widetilde{E} = T  \, , \;  b = r_*(T) \equiv \frac{T^5}{cD^2}  \, , \;  a= \xi_{\text{th}}(T)  \equiv \frac{T^3}{cD} &
 \label{eq-rescalingsB-for-highT} \\
 & B(r;c,D,T,\xi) \stackrel{(T \gg T_c)}{\approx} \xi_{\text{th}}(T)^2 \bar{B} \argp{\frac{r}{r_*(T)};1,1,1,0}
 \label{eq-rescaledB-for-highT} &
\end{eqnarray}
catches the high-$T$ scalings if the function ${\bar{B}(\bar{r};1,1,1,0)}$ is properly defined (cf. Sec.~\ref{section-DPtoymodel-D}).
Secondly with respect to the finite width or disorder correlation length $\xi$:
\begin{eqnarray}
 & a= \xi \, , \; \widetilde{E} = T_c(\xi) \equiv (\xi c D)^{1/3} \, , \; b = r_*(T_c) 
 & \label{eq-rescalingsB-for-lowT} \\
 & B(r;c,D,T,\xi) \stackrel{(T \ll T_c)}{\approx} \xi^2 \bar{B} \argp{\frac{r}{r_*(T_c)};1,1,0,1}
 &  \label{eq-rescaledB-for-lowT}
\end{eqnarray}
catches the low-$T$ scalings if the function ${\bar{B}(\bar{r};1,1,0,1)}$ is properly defined (cf. Sec.~\ref{section-DPtoymodel-D}),
with ${r_*(T_c)=\xi^{5/3} c^{2/3} D^{-1/3}}$.
Thirdly with respect to the Larkin length ${L_c(T,\xi)}$, defined as the beginning of the asymptotic `random-manifold' regime \cite{Larkin_model_1970-SovPhysJETP31_784} (as discussed first after~\eqref{eq-def-roughness-zeta} and then in~Sec.~\ref{section-DPtoymodel-B}):
\begin{eqnarray}
 && b = L_c (T,\xi) \equiv r_* (T) / f(T,\xi)^5 \label{eq-rescaling-Lc-1} \\
 && a = \xi_{\text{eff}} (T,\xi) = \xi_{\text{th}} (T) / f(T,\xi)^3 \label{eq-rescaling-Lc-2} \\
 && \widetilde{E} = T / f(T,\xi) \label{eq-rescaling-Lc-3} \\
 && B(r;c,D,T,\xi) = \xi_{\text{eff}}^2 \bar B ({r}/{L_c};1,1,f,{\xi}/{\xi_{\text{eff}}})  \label{eq-rescaling-Lc-4}
\end{eqnarray}
with $f(T,\xi)$ an interpolating function between the high-$T$ and low-$T$ regimes for both the Larkin length and its corresponding effective width:
\begin{eqnarray}
 f(T,0) = 1 \; &,& \quad f(0,\xi) = T/T_c \label{eq-interp-param-scaling} \\
 L_c(T,0) = r_*(T)  \; &,& \quad L_c(0,\xi) = r_*(T_c) \equiv r_0(\xi) \label{eq-interp-param-scaling-2} \\
 \xi_{\text{eff}}(T,0) = \xi_{\text{th}} (T)   \; &,& \quad \xi_{\text{eff}} (0,\xi)=\xi \label{eq-interp-param-scaling-3} 
\end{eqnarray}

We can now focus on the roughness itself, and discuss the consequences of a powerlaw behavior at larges lengthscales, which is known to be governed by the roughness exponent $\zeta^{\text{exact}}_{\text{RM}}=\frac23$. A behavior such as
\begin{equation}
 \bar{B}_{\text{asympt}}(\bar{r};1,1,\bar{T},\bar{\xi})
 \stackrel{(\bar{r} \gg 1)}{\approx} \bar{r}^{2 \zeta_{\text{RM}}}
 \label{eq-hyp-powerlaw-RM}
\end{equation}
\textit{without} any other parameter dependence
(this constraint can actually be taken as another definition of the Larkin length \eqref{eq-rescaling-Lc-1} for ${\bar{r}=r/L_c}$)
implies for the rescaling \eqref{eq-rescaling-Lc-1}-\eqref{eq-rescaling-Lc-3}:
\begin{equation}
B_{\text{asympt}}(r;c,D,T,\xi)
 \approx \underbrace{\argc{\frac{\xi_{\text{eff}} (T,\xi) }{L_c (T,\xi)^{\zeta_{\text{RM}}}}}^2}_{\equiv A(c,D,T,\xi)} r^{2 \zeta_{\text{RM}}}
 \label{eq-Basympt-GVM}
\end{equation}
An artefact of the GVM framework is that it predicts the Flory exponent of the model for the asymptotic roughness exponent;
for the 1D interface ${\zeta_{\text{F}}^{\text{1D}}=\frac35}$ whereas for the DP toymodel ${\zeta_{\text{F}}^{\text{toy}}=\frac23=\zeta_{\text{RM}}^{\text{exact}}}$.
So either the asymptotic GVM exponent coincides with the Flory exponent of \eqref{eq-scaling-ab-Flory} and all the temperature dependence is cancelled in $A(c,D,T,\xi)$, 
or they do not and the scaling prediction
\begin{equation}
 A(c,D,T,\xi)
 = \argp{\frac{D^{3/10}}{c^{3/5} L_c^{1/10}}}^{4/3}
 = \argc{\frac{D}{cT} f(T,\xi) }^{2/3}
 \label{eq-Basympt-GVM-amplitude}
\end{equation}
matches the GVM result for the DP toymodel \eqref{eq-Basympt-DPtoymodel} with \eqref{eq-Dtilde-infty-finterp}.
It is important to emphasize that the Flory exponent $\frac35$ is imposed by the rescaling procedure of the full model of a 1D interface, whereas the exact RM exponent $\frac23$ is the true physical roughness exponent at large lengthscales and is predicted by assuming only that the scaling of the disorder free-energy  is dominated by ${\bar{F}(t,y)^2\sim \widetilde{D}_{\infty} \valabs{y}}$ as in \eqref{eq-infty-time-FPsolution} hence ${\zeta_{\text{F}}^{\text{toy}}=\zeta_{\text{RM}}^{\text{exact}}}$ (cf. Sec.~\ref{section-DPtoymodel-D}).

If we try boldly the rescaling ${b=r}$ in order to catch the large lengthscales behavior, we obtain:
\begin{eqnarray}
 & b= r \, , \; a=(D^{1/3} c^{-2/3} r )^{3/5} \, , \; \widetilde{E} =(cD^2 r)^{1/5}
 & \label{eq-rescalingsB-for-Flory} \\
 & B(r;c,D,T,\xi) \stackrel{(r \to \infty)}{\approx} \argp{\frac{D r^{3}}{c^2}}^{2/5}  \bar{B} (1;1,1,0,0)
 &  \label{eq-rescaledB-for-Flory}
\end{eqnarray}
that would predict the asymptotic roughness exponent $\zeta_{\text{F}}^{\text{1D}}=\frac35$ if the function ${\bar{B}(\bar{r};1,1,0,0)}$ was properly defined, but this is not the case since the two limits ${T \to 0}$ and ${\xi \to 0}$ cannot be exchanged or taken simultaneously.

The quantity ${f(T,\xi)}$ has been introduced here in order to interpolate between the two limits \eqref{eq-rescalingsB-for-highT} and \eqref{eq-rescalingsB-for-lowT}, in the only way compatible with the rescaling procedure \eqref{eq-scaling-ab-Flory}-\eqref{eq-scaling-Etilde-ab}.
We argue however that ${f(T,\xi)}$ is the same parameter defined in \eqref{eq-Dtilde-infty-finterp} for $\widetilde{D}_{\infty}$ in our DP toymodel.
Actually all the scalings \eqref{eq-rescalingsB-for-highT}-\eqref{eq-rescaledB-for-lowT} are properly recovered in a GVM approximation of the Hamiltonian \citep{agoritsas_2010_PhysRevB_82_184207}, cf. \eqref{equa-1D-roughness-1}-\eqref{equa-1D-roughness-4}, with the identification ${f(T,\xi) \equiv \frac65 v_c(T,\xi) }$ that transforms the equation \eqref{equa-1D-roughness-3} for the full-RSB cutoff ${v_c(T,\xi)}$ into
\begin{equation}
 f^6 = \frac{16 \pi}{9} \argc{\frac{T}{T_c(\xi)}}^6 (1-f)
 \label{eq-equa-interpf-vc}
\end{equation}
So ${f(T,\xi)}$ turns out to be the key quantity for the connection of our scaling arguments and the two sets of GVM predictions, centered either on the Hamiltonian or on the pseudo free-energy at a fixed lengthscale, both recalled in Appendix~\ref{A-GVM-PRB2010}.
The numerical discrepancy between the equations \eqref{eq-equa-interpf-vc} and \eqref{eq-equa-interpf-uc} for $f(T,\xi)$ can be either reabsorbed in the definition ${\tilde{\xi}_t \approx \frac23 \xi}$ for the latest, or more safely attributed to the GVM approximation.

\subsection{Saddle-point arguments} \label{section-DPtoymodel-D}

The previous scaling arguments are based on the presumed existence of specific limits, which can be precised in a path-integral reformulation of the roughness functions \eqref{eq-scaling-roughness}.
We present thereafter two saddle-point arguments which provide a controlled validation of our different assumptions at ${T>0}$ and ${\xi>0}$.

Firstly we use $\frac{1}{T}$ as a large parameter at low temperature in order to argue the existence of a proper limit for $\bar{B}(\bar{r};1,1,0,1)$ in \eqref{eq-rescaledB-for-lowT} (the high-temperature case \eqref{eq-rescaledB-for-lowT} is already well controlled);
this is not obvious in the usual conventions of mathematicians regarding the DP (${c=T}$), see Appendix~\ref{A-saddle_scalings-maths}.

Secondly we revisit the original derivation of the exponent ${\zeta_{\text{RM}}^{\text{exact}}=\frac23}$ by Huse, Henley and Fisher \cite{huse_henley_fisher_1985_PhysRevLett55_2924} from the point of view of our DP toymodel and using the lengthscale $t$ as large parameter for the saddle point.

\subsubsection{Zero-temperature roughness of the 1D interface}

The low-temperature limit in \eqref{eq-rescaledB-for-lowT} can be made explicit coming back to the path-integral definition of the roughness and performing the rescaling \eqref{eq-rescalingsB-for-lowT} with ${t_*(T) \equiv \frac{T^5}{cD^2}}$ as in \eqref{eq-rescalingsB-for-highT}:
\begin{align}
\!\!  B&(t_1;c,D,T,\xi) = \xi^2 \bar{B} \argp{\frac{t_1}{t_*(T_c)};1,1,\frac{T}{T_c},1 }
\label{eq-rescalingB_forsaddle_lowT}
\\
& \!= \xi^2\, \overline{
\frac
{\displaystyle 
\int_{y(0)=0}%
\!\!\!\!\!\!\!\!\!\!\!\!\!\!\mathcal Dy\  y(\tfrac{t_1}{t_*(T_c)})^2 \,
  e^{\mbox{\scriptsize$\displaystyle-\frac{T_c}{T}\!\int_{0}^{\frac{t_1}{t_*(T_c)}}\! dt\:\Big[ \tfrac 12 {(\partial_ty)^2} +V_1\big(t,y(t)\big)\Big]$}}
}
{\displaystyle 
\int_{y(0)=0}
\!\!\!\!\!\!\!\!\!\!\!\!\mathcal Dy \
  e^{\mbox{\scriptsize$\displaystyle-\frac{T_c}{T}\!\int_{0}^{\frac{t_1}{t_*(T_c)}}\! dt\:\Big[ \tfrac 12 {(\partial_ty)^2} +V_1\big(t,y(t)\big)\Big]$}}
} }
\label{eq-explicitBsaddle_lowT}
\end{align}
where $V_1(t,y(t)) \equiv V(t,y(t))|_{D=1,\xi=1}$.
In the path integrals, the trajectories $y(t)$ have a fixed starting point ${y(0)=0}$ but a free endpoint ${y(t_1)}$.
Since all temperature-dependence has been gathered in a single and large prefactor ${\frac{T_c}{T}}$, the path integrals are dominated by a
\textit{common} optimal trajectory $y^\star(t)$, which, assuming that it exists, does not depend on temperature since it minimizes
$\int_{0}^{t_1/t_*(T_c)}\! dt\:\big[ \tfrac 12 {(\partial_ty)^2}+V_1\big(t,y(t)\big)\big]$.
The saddle trajectory endpoint is then reached at some optimal endpoint $y_1^\star=y_1^\star\big(\frac{t_1}{t_*(T_c)},V_1\big)$, common to the numerator and denominator and independent of $T$.
Finally, one obtains from \eqref{eq-explicitBsaddle_lowT} that in \eqref{eq-rescalingB_forsaddle_lowT}
$\lim_{T\to 0} \bar{B} ( \frac{t_1}{t_*(T_c)};1,1,\frac{T}{T_c},1 )$ is finite, being equal to $\overline{ y_1^\star (\frac{t_1}{t_*(T_c)},V_1)^2 }$.
\textit{So if the optimal path $y^{\star}$ does exist and if its variance at fixed lengthscale $t_1$ is finite, the zero-temperature limit is well-defined.}
See Appendix~\ref{A-saddle_scalings-maths} for a discussion on this last point.

\subsubsection{DP toymodel scaling argument, asymptotic roughness and Flory exponent} \label{section-DPtoymodel-D-DPtoymodel}

The scaling arguments of the previous section, established on the full model of a 1D interface, have of course their counterpart for our DP toymodel.
The main assumption is that the large `time' scaling of ${\bar{F}(t,y)}$ is governed by its infinite-`time' correlator ${\bar{C} (t,y)=\widetilde{D}_{\infty} \valabs{y}}$ \eqref{eq-infty-time-FPsolution} with the amplitude being essentially a constant $\widetilde{D}_t \approx \widetilde{D}_{\infty}$ (denoted thereafter simply by $\widetilde{D}$) and similarly ${\tilde{\xi}_t \approx \tilde{\xi}}$.
This ensures that upon the change of variable ${y=a \bar{y}}$ and ${t=b\bar{t}}$, the following free-energy is equal in distribution to
\begin{align}
 F_{\text{th}}(t,y) + \bar{F}(t,y) 
 & \stackrel{d}{=} \frac{a^{2}}{b}  \frac{c \bar{y}^2}{2\bar{t}} +  a^{\frac 12} \widetilde D^{\frac 12} \bar{F}_1(\bar{t}, \bar{y})
\label{eq-rescalingFFthF1dis}
\end{align}
where ${\bar{F}_1(t,y) \equiv \bar{F}(t,y) \vert_{\widetilde{D}=1,\tilde{\xi}/a}}$.
The argument of Ref.~\cite{huse_henley_fisher_1985_PhysRevLett55_2924} can then be summarized as follows: the free-energy and roughness fluctuation exponents $\chi$ and $\zeta$ are respectively defined as ${\bar{F}(y)\sim b^{\chi} \bar{F}(\bar{y})}$ and ${y\sim b^{\zeta}}$ (which amounts to take $a\sim b^{\zeta}$).
The fact that in distribution ${\bar{F}(t,y)\stackrel{d}{\sim} a^{\frac 12}\bar F(\bar y)}$ implies ${\chi =\frac 12 \zeta}$ while equating the thermal and disorder contributions in \eqref{eq-rescalingFFthF1dis} yields ${\chi = 2\zeta -1}$.
These two equations fully determine the values of the exponents: $\chi=\frac13$ and $\zeta=\frac23$.
Taking care of the prefactors of those powerlaws, we define the following rescaling procedure:
\begin{eqnarray}
 & B(t;c,\widetilde{D},T,\tilde{\xi}) = a^2 \, \bar{B}(t/b;1,1,T/\widetilde{E},\tilde{\xi}/a) &
  \label{eq-rescaledB-for-DPtoymodel} \\
 & a = (\tilde{D}/c^2)^{1/3} b^{2/3} \Leftrightarrow b = c \widetilde{D}^{-1/2} a^{3/2}  & \label{eq-scaling-ab-Flory-DPtoymodel} \\
 & \widetilde{E} \equiv ca^2/b = (\widetilde{D}^2 b/c)^{1/3} = \tilde{D}^{1/2} a^{1/2} & \label{eq-scaling-Etilde-ab-DPtoymodel}
\end{eqnarray}
where $\bar{B}$ is the roughness function with \textit{adimensional} parameters, if the scalings factors
satisfy the two relations involving the Flory exponent ${\zeta_{\text{F}}^{\text{toy}}=2/3}$.
To understand how this power counting can describe correctly the large `time' asymptotics, we chose the rescaling equivalent to \eqref{eq-rescalingsB-for-Flory}:
\begin{eqnarray}
 & b=t \, , \; a= (\widetilde{D}/c^2)^{\frac13}\, t^{\frac 23} \, , \; \widetilde{E} = (\widetilde{D}^2 t/c)^{1/3}
 & \label{eq-rescalingsB-for-lowT-toymodel}
\end{eqnarray} 
 which implies from the definition of the roughness $B(t;c,\widetilde D,T,\tilde{\xi})$ in \eqref{eq-def-moments-pdf-replicas-1}:
\begin{align}
 B(&t;c,\widetilde D,T,\tilde{\xi}) = \Big[\frac{\widetilde D }{c^2}\Big]^{\frac 23}t^{\frac 43} \nonumber \\
 & \times \overline{
 \frac{\int d\bar y\:\bar y^2 \exp\Big\{\!- \frac 1T \big[\frac {\widetilde{D}^2 }{c}t\big]^{\frac 13}\Big[ \frac{\bar y^2}{2} + 
     \bar F_1(1,\bar y)\Big]\Big\}}
  {\int d\bar y  \exp\Big\{\!- \frac 1T \big[\frac {\widetilde{D}^2 }{c}t\big]^{\frac 13}\Big[ \frac{\bar y^2}{2} + 
     \bar F_1(1,\bar y)\Big]\Big\}}
 }
 \label{eq-rescaling_B_zeta23}
\end{align}
where the overline denotes the average over the random $\bar{F}_1$.
The advantage of our specific choice of the rescaling parameters $a$ and $b$ is that the `time'-dependence of the exponentials in \eqref{eq-rescaling_B_zeta23} is then gathered in a single prefactor $t^{\frac 13}$.
For each fixed $\bar{F}_1$, one may thus evaluate the integrals in $\bar y$ through the saddle point method in the large $t$ limit. 
The integrals at the numerator and denominator of \eqref{eq-rescaling_B_zeta23} are dominated by the same $y^\star[\bar F_1]$ which minimizes 
${ \frac{\bar y^2}{2} + \bar F_1(1,\bar y) }$, ensuring that $y^\star[\bar F_1]$ \textit{is independent of $t$}.
We read from \eqref{eq-rescaling_B_zeta23} that
\begin{equation}
 B_{\text{asympt}}(t;c,\widetilde D,T,\tilde{\xi}) =\overline{(y^\star[\bar F_1])^2} \cdot (\widetilde D/c^2 )^{\frac 23}\,t^{\frac43}
 \label{eq-Basympt-scaling-prediction}
\end{equation}
\textit{i.e.} the roughness exponent is ${\zeta^{\text{exact}}_\text{RM}=\frac 23}$.
However all this construction breaks down at the very last when the scaling $\bar{F}(t,y)^2 \sim \widetilde{D} \valabs{y}$ ceases to be valid, at small ${\valabs{y} \leq \tilde{\xi}}$, \textit{\textit{i.e.}} when the scaling factor $a(t)$ matches with the effective width $\tilde{\xi}$.
This yields an alternative definition of the Larkin `time' $t_0$ as ${ a(t_0) \equiv \tilde{\xi}}$ or ${t_0 = ({c^2 \tilde{\xi}^3}/{\widetilde{D}})^{1/2}}$.
Coming from the large lengthscales, this asymptotic scaling breaks earlier due to thermal fluctuations, at the Larkin `time' ${t_c \geq t_0}$.
Identifying ${t_c}$ and ${L_c(\xi,T)}$,
generalizing ${\tilde{\xi} \approx \xi}$ to ${\xi_{\text{eff}}(T,\xi)}$ of \eqref{eq-interp-param-scaling-2}
and using finally ${\widetilde{D}=f(T,\xi) \frac{cD}{T}}$ of \eqref{eq-Dtilde-infty-finterp},
we recover consistently with \eqref{eq-def-Larkin-GVM-DPtoymodel} and \eqref{eq-interp-param-scaling} for the Larkin `time':
\begin{equation}
 a(t_c) \equiv \xi_{\text{eff}}
 \Leftrightarrow t_c = \argp{\frac{c^2 \xi_{\text{eff}}^3}{\widetilde{D}}}^{1/2} = \frac{T^5}{cD^2} f(T,\xi)^{-5}
\end{equation}
which has as a lower bound its low-temperature limit 
\begin{equation}
 t_0=\xi^{5/3} c^{2/3} D^{-1/3}=r_*(T_c(\xi))
\end{equation}

The large-`time' limit makes the scaling assumption ${\bar{F}_1(1,\bar{y})^2 \sim \valabs{y}}$ even more reliable, and the saddle point can be properly taken in this limit, yielding the Flory exponent of the DP toymodel $\zeta_{\text{F}}^{\text{toy}}=\frac23$.
This was not the case for the 1D interface in \eqref{eq-rescaledB-for-Flory}. Indeed, upon the rescalings \eqref{eq-rescalingsB-for-Flory} we obtain in a path-integral representation:
\begin{align}
\!\!  B&(t_1;c,D,T,\xi) \nonumber \\
&= 
\Big[\frac{D t_1^{3}}{c^2}\Big]^{ \frac 25}  \bar{B} \argp{1;1,1,\frac{T}{(cD^2 t_1)^{\frac 15}},\frac{\xi}{(D^{\frac 13} c^{-\frac 23} t_1 )^{\frac 35} }}
\\
& \!=\Big[\frac{D t_1^{3}}{c^2}\Big]^{ \frac 25}\, \overline{
\frac
{\displaystyle 
\int_{y(0)=0}
\!\!\!\!\!\!\!\!\!\!\!\!\!\!\mathcal Dy\  y(1)^2 \,
  e^{\mbox{\scriptsize$\displaystyle-\frac{\widetilde{E}}{T}\!\int_{0}^{1}\! dt\:\Big[ \tfrac 12 {(\partial_ty)^2} +V \big(\tfrac{t}{t_1},y(t)\big) \big\vert_{D=1,{\frac \xi a}} \Big]$}}
}
{\displaystyle 
\int_{y(0)=0}
\!\!\!\!\!\!\!\!\!\!\!\!\mathcal Dy \
  e^{\mbox{\scriptsize$\displaystyle-\frac{\widetilde{E}}{T}\!\int_{0}^{1}\! dt\:\Big[ \tfrac 12 {(\partial_ty)^2} +V\big(\tfrac{t}{t_1},y(t)\big) \big\vert_{D=1,{\frac \xi a}} \Big]$}}
}  }
\end{align}
 with
$ \; a=(D^{1/3} c^{-2/3} t_1 )^{3/5} \, , \; \widetilde{E} =(cD^2 t_1)^{1/5}$.
The large $t_1$ asymptotics cannot be taken directly from this expression since it is not in a saddle form and all scales are intertwined, contrarily to the study of the free-energy itself which corresponds to scales integrated up to `time' $t_1$.


\section{Synthetic outlook}
\label{section-synthetic-outlook}

We address analytically throughout this paper the consequences of a \textit{finite} correlation length ${\xi>0}$ of the microscopic disorder ${V(t,y)}$ explored by a 1+1 DP, or alternatively of a 1D-interface finite width which is always present in experimental systems.
On one hand, several analytical arguments yielding exact results at ${\xi=0}$ break down as such, questioning their generalization to ${\xi>0}$.
On the other hand, 
despite a lack of exact analytical expressions the finiteness of this quantity allows to control the scalings and the low-temperature limit of the model (see~Sec.~\ref{section-scaling-saddle}), avoiding the pathological and unphysical divergences that appear at ${\xi=0}$, in particular conjointly to the limit ${T\to 0}$.

In order to tackle the case at ${\xi>0}$,
the 1+1 DP formulation allows to follow effective quantities at fixed lengthscale or growing `time' as defined in~Sec.~\ref{section-def-fluctuations},
in an approach thus conceptually similar to the FRG which focuses on the flow and fixed points of the disorder correlator (denoted $\Delta$ or $R$ \cite{fisher_1986_PhysRevLett56_1964,balents-fisher_1993_PhysRevB48_5949,chauve_2000_ThesePC_PhysRevB62_6241}).
Considering the free-energy at fixed disorder (averaging over the thermal fluctuations but one step before the disorder average), it is thus possible to disconnect theoretically the two statistical averages, and even to focus on the pure disorder contributions thanks to the STS and the Feynman-Kac equations for ${\bar{F}_V}$ and its derivative ${\eta_V}$ (see Sec.~\ref{section-FeynmanKac}), paving the way to the numerical computation frame presented in~Ref.~\cite{agoritsas_2012_FHHtri-numerics}.
Those two quantities are not directly accessible experimentally (except for liquid crystals, as discussed later in~Sec.~\ref{section-discussion-exp}) and are \textit{a priori} more complex to handle since they encode more information than direct observables such as the geometrical fluctuations and the roughness.
However they actually display a simpler phenomenology by disconnecting the thermal and disorder effects and the different lengthscales, whereas the roughness $B(t)$ intricates all of them as illustrated by the combination of ${F_{\text{th}}(t,y)}$ and ${\bar{F}_V (t,y)}$ in~\eqref{eq-def-moments-pdf-replicas-1}.

Although ${\bar{\mathcal{P}}\argc{\bar{F}}}$ and ${\bar{\mathcal{P}}\argc{\eta}}$ are not Gaussian at ${\xi>0}$ not even in the infinite-`time' limit,
their main features are encoded in the two-point correlators ${\bar{C}(t,y)}$ and ${\bar{R}(t,y)}$,
\textit{i.e.} the scalings of ${\bar{C}}$ and ${\bar{R}}$ dominate the higher moments of the PDFs,
similarly to the case ${\xi=0}$ (see~Sec.~\ref{section-DPtoymodel-A}).
This supports consequently the construction of the toymodel of~Sec.~\ref{section-DPtoymodel-B}, which relies on the assumption that the PDFs can be approximated as Gaussian ones described by a given set of two-point correlators;
the GVM predictions derived from this DP toymodel~\cite{agoritsas_2010_PhysRevB_82_184207} (see~Appendix~\ref{A-GVM-PRB2010}) are actually found to be qualitatively in agreement with the numerical results presented in~Ref.~\cite{agoritsas_2012_FHHtri-numerics}.
The study of the two-point correlator ${\bar{R}}$ provides thus a vantage point on the DP properties,
first at asymptotically large `times'
(keeping in mind that the infinite-`time' limit simplifies the analytical treatment of the RM regime, \textit{i.e.} via a Fokker-Planck approach as in~Appendix~\ref{A-FokkerPlanck-equation})
and secondly at finite `time' with the connection to the short-`times' regime.

In order to characterize the asymptotic large-`times' behavior, which can potentially display universality, a central quantity is the saturation amplitude of the effective disorder \textit{i.e.} ${\widetilde{D}_{\infty}(T,\xi)}$.
At fixed ${\xi>0}$ it is equivalent to the maximum of the asymptotic correlator ${\bar{R}(\infty,y=0)}$
which is equal to ${\widetilde{D}_{\infty}/\tilde{\xi}_{\infty} \cdot \mathcal{R}_{\tilde{\xi}=1} (y=0)}$ and is measured numerically as the maximum of the saturation correlator ${\bar{R}_{\text{sat}}(y=0)}$ in~Ref.~\cite{agoritsas_2012_FHHtri-numerics}.
However for a $\delta$-correlated microscopic disorder ${R_{\xi=0}(y)=\delta(y)}$, ${\bar{R}(\infty,y=0)}$
diverges whereas the amplitude ${\widetilde{D}_{\infty}/\tilde{\xi}_{\infty}}$ remains well-defined.
It is remarkable to notice that from the whole saturation correlator, the quantity $\widetilde{D}_{\infty}$  is the only feature that eventually plays a role in the asymptotic roughness in GVM or scaling arguments \textit{e.g.}  in~\eqref{eq-Basympt-DPtoymodel}, the specificity of the (normalized) RB disorder correlator ${R_{\xi}(y)}$~\eqref{eq-def-moydis} and thus of the function ${\mathcal{R}(y)}$ being then gathered into a numerical constant.
\textit{${\widetilde{D}_{\infty}(T,\xi)}$ appears to be the relevant quantity for a universal description of the crossover between low- and high-$T$ asymptotic DP fluctuations}, from an analytical point of view
and in a remarkable agreement with the numerical results of~Ref.~\cite{agoritsas_2012_FHHtri-numerics}.
Going one step further, the crossover from its ${\xi=0}$ (or high-$T$) limit is better described by the interpolating parameter ${f(T,\xi)=\widetilde{D}_{\infty}(T,\xi)/\argp{\frac{cD}{T}}}$  first introduced in \eqref{eq-Dtilde-infty-finterp} and expected to rescale also the characteristic scales such as ${L_c(T,\xi)}$ according to the relations \eqref{eq-rescaling-Lc-1}-\eqref{eq-rescaling-Lc-4} obtained by pure scaling arguments.
The GVM framework yields the two predictions \eqref{eq-equa-interpf-uc} and \eqref{eq-equa-interpf-vc} for ${f(T,\xi)}$ derived from the value of the full-RSB cutoff (cf. Appendix~\ref{A-GVM-PRB2010}), and a third analytical prediction will be presented in the next subsection~\ref{section-Dtildeinfty-T-xi};
all of them predict a monotonous crossover connecting the limits ${T=0}$ and ${\xi=0}$ \eqref{eq-interp-param-scaling}-\eqref{eq-interp-param-scaling-3} with a polynomial equation on~${f(T,\xi)}$.
This prediction can be checked to be qualitatively consistent with the numerical results in~Ref.~\cite{agoritsas_2012_FHHtri-numerics} but the comparison will anyway suffer quantitatively from the variational approximation and from several corrective factors due to the numerical procedure.

Note that although the two communities of physicists and mathematicians work with two different conventions, respectively at fixed elastic constant $c$ (the choice ${c=1}$ essentially fixing the units of energy) \textit{versus} at ${c=T}$ (as discussed in Appendix~\ref{A-saddle_scalings-maths}),
all the above discussion remains valid in both conventions, although the choice ${c=T}$ leads to other limits in temperature.
The two opposite limits at high-$T$ (or ${\xi \approx 0}$) ${\widetilde{D}_{\infty} \approx \frac{cD}{T}}$, and at low-$T$ (or ${\xi>0}$ and below ${T_c(\xi)=(\xi c D)^{1/3}}$)  ${\widetilde{D}_{\infty} \approx \frac{cD}{T_c} = (c^{2} D^{2} \xi^{-1})^{1/3}}$
translate with the convention ${c=T}$ into ${\widetilde{D}_{\infty} \approx D}$ and ${\widetilde{D}_{\infty} \approx T^{2/3} D^{2/3} \xi^{-1/3}}$ respectively above and below ${T_c^{\mathfrak{m}}=\sqrt{\xi D}}$ (deduced self-consistently from ${T_c^{\mathfrak{m}}=(\xi c D)^{1/3} = (\xi T_c^{\mathfrak{m}} D)^{1/3}}$).
In the course of the study of the rescaling of the correlator ${\bar{C}(t,y)}$ with respect to the roughness ${B(t)}$ in~Ref.~\cite{agoritsas-2012-FHHpenta}, it has been noticed that in the regime ${\valabs{y} \lesssim \sqrt{B(t)}}$ we have numerically as expected a linear behavior ${\bar{C}(t,y) \propto \valabs{y}}$ but with a by-product prefactor that corresponds precisely to our ${\widetilde{D}_{\infty}}$.
Taking as a criterion the collapse of the curves ${\bar{C}(t,y)}$ at different temperatures on an arbitrary chosen curve, the temperature dependence of this prefactor is consistent with all our analysis on the origin and interpretation of ${\widetilde{D}_{\infty}(T,\xi)}$
(see the insets in Fig.~10 and Fig.~14 of Ref.~\cite{agoritsas-2012-FHHpenta}, which illustrate respectively the conventions ${c=T}$ \textit{versus} independently fixed ${c=1}$ and~$T$).

As for the finite-`time' behavior,
especially at short-`times' it is \textit{a priori} crucially sensitive to the specific microscopic disorder correlator, thus compromising a possible universality.
We speculate that the `time'-evolution of the free-energy fluctuations displays essentially two regimes on the fluctuations of the disorder free-energy, separated by the saturation `time' ${t_{\text{sat}}}$ first introduced in the course of our DP toymodel definition in~Sec.~\ref{section-DPtoymodel-B}:
starting from the initial condition ${\bar{R}(0,y)\equiv 0}$ imposed by~\eqref{eq-initial-FbarV},
the central peak of the correlator ${\bar{R}(t,y)}$ develops itself keeping the integral ${\int_{\mathbb{R}} dy \, \bar{R}(t,y) = 0}$ constant,
until it reaches the saturation shape ${\widetilde{D}_{\infty} \, \mathcal{R}_{\tilde{\xi}}(y)}$ compensated by negative bumps according to the generic decomposition \eqref{eq-infty-time-Rxi-bumps-1}-\eqref{eq-infty-time-Rxi-bumps-2} as illustrated in~Fig.~\ref{fig:graphDPtoymodel}.
Note that an independent criterion to determine ${t_{\text{sat}}}$ is provided by \eqref{eq-mean-FbarV}: ${\partial_t \overline{\bar{F}_V (t,y)}=-\frac{1}{2c} \bar{R}(t,y=0)}$ should self-consistently be a constant above ${t_{\text{sat}}}$, as it has been observed numerically in~Ref.~\cite{agoritsas_2012_FHHtri-numerics}.
Since ${\partial_y^2 \bar{C}(t,y) = 2 \bar{R}(t,y)}$,
after the double integration \eqref{eq-def-corr-CbarRbar-integral} the correlator ${\bar{C}(t,y)}$ starts from the initial condition ${\bar{C}(0,y)\equiv 0}$ (also imposed by~\eqref{eq-initial-FbarV}), and at fixed `time' above ${t_{\text{sat}}}$ it is rounded at ${\valabs{y} \lesssim \xi}$, increases then linearly ${\bar{C}(t,y) \approx \widetilde{D}_{\infty} \valabs{y}}$ at ${\xi \lesssim \valabs{y} \sqrt{B(t)}}$ and is constant for any larger $\valabs{y}$ (see again~Fig.~\ref{fig:graphDPtoymodel}).
The position $\ell_t$ of these `wings' of ${\bar{C}(t,y)}$ or equivalently of the negative bumps of ${\bar{R}(t,y)}$ is  discussed at length in~Ref.~\cite{agoritsas-2012-FHHpenta} and identified to correspond physically to the typical position of the DP endpoint, ${\ell_t \approx \sqrt{B(t)}}$, in the different roughness regimes and even below the Larkin length $L_c$.

What happens below ${t_{\text{sat}}}$ cannot be understood without taking into account the whole microscopic disorder correlator ${R_{\xi}(y)}$, whose feedback via the KPZ nonlinearity at small $\valabs{y}$ modifies the amplitude ${\widetilde{D}_{\infty}(T,\xi)}$ and the shape ${\mathcal{R}_{\tilde{\xi}}(y)}$ before the saturation is achieved, especially in the low-$T$ regime and in any case with ${t_{\text{sat}} \leq L_c}$.
Neglecting the KPZ nonlinearity yields the prediction \eqref{eq-decomposition-Rbarlin-1}-\eqref{eq-decomposition-Rbarlin-2} which mixes different limits:
the ${\xi=0}$ (high-$T$) amplitude ${\widetilde{D}_{\infty}=\frac{cD}{T}}$, the same correlator as the microscopic disorder ${\mathcal{R}_{\tilde{\xi}}(y)=R_{\xi}(y)}$, and the `wings' rescaled with respect to the pure thermal roughness ${B_{\text{th}}(t)}$ at all lengthscales (this diffusive behavior being for sure an artefact of the linearization).
In the low-$T$ regime, we believe that by generating relevant non-Gaussian correlations such as $\bar{R}_3$ and $\bar{C}_3$ (defined in~\eqref{eq-R3-def}-\eqref{eq-C3-def}) below ${t_{\text{sat}}}$,
the KPZ nonlinearity introduces an effective kernel for ${\bar{R}}$ that modifies simultaneously ${\widetilde{D}_{\infty}}$ and ${\mathcal{R}_{\tilde{\xi}}(y)}$, with in particular the saturation below $T_c$ of the amplitude ${\widetilde{D}_{\infty} \approx \frac{cD}{T_c}}$ as predicted by scaling arguments in~Sec.~\ref{section-DPtoymodel-C}.

\textit{The phenomenology of the DP fluctuations is simpler from the point of view of the disorder free-energy $\bar{F}_V$, via its two-point correlators $\bar{R}$ and $\bar{C}$} because they display these two `time'-regimes separated by $t_{\text{sat}}$, as we have speculated here and then checked numerically in~Ref.~\cite{agoritsas_2012_FHHtri-numerics}.
From the competition between the typical $\bar{F}_V$ and the thermal $F_{\text{th}}$ the resulting roughness ${B_{\text{dis}}(t)}$ should also display two regimes as observed numerically in~Ref.~\cite{agoritsas_2012_FHHtri-numerics}.
However, when recombined with the pure thermal effect the roughness $B(t)$ displays two or three `time'-regimes respectively at high-$T$ (thermal and RM regimes) and low-$T$ (with an additional intermediate `Larkin-modified' regime), with $L_c$ at the beginning of the RM regime, as predicted by GVM and again observed numerically.

The disorder free-energy is an effective quantity which encodes the microscopic disorder explored by the polymer, there is consequently a feedback between the geometrical fluctuations ${\mathcal{P}(t,y)}$ and the free-energy correlations ${\bar{C}(t,y)}$:
firstly the existence of `wings' in ${\bar{C}(t,y)}$  are imposed physically by the finite variance of the PDF ${\mathcal{P}(t,y)}$  (the polymer does not explore often regions ${\valabs{y}>\sqrt{B(t)}}$ so these regions do not contribute much to the correlator ${\bar{C}(t,y)}$);
secondly the PDF ${\mathcal{P}(t,y)}$ is deduced from the competition between ${F_{\text{th}}}$ and ${\bar{F}_V \sim \bar{C}^{1/2}}$, the maximum of the typical ${\bar{F}_V}$ being precisely fixed by the `wings' of ${\bar{C}}$;
thirdly the `wings' of $\bar{C}$ or the bumps in $\bar{R}$ can be skipped for a GVM computation of the roughness, providing a self-consistent justification of our DP toymodel.
Beyond the scaling in `time' of those fluctuations, which we plainly understand physically now, their temperature dependence at all `times' is finally determined by the integrated disorder up to ${t_{\text{sat}} \leq L_c}$, where the KPZ non-linearity plays a crucial role below ${T_c(\xi)}$.


\section{Effective evolution in `time' and temperature of the amplitude~$\tilde{D}_t$}
\label{section-Dtildeinfty-T-xi}

Having this global picture in mind, we can now gather all the physical intuition we have obtained and construct the following analytical argument in order to obtain an evolution equation for the an effective `time'-dependent amplitude $\widetilde{D}_t$ as a refinement of our DP toymodel.

The evolution of the correlator $\bar{R}(t,y)$, given by the `flow' equation~\eqref{eq-noclose-Rbar}, cannot be solved directly since it brings into play the three-point correlation function $\bar{R}_3(t,y)$, a hallmark of the KPZ non-linearity.
To extract an exact information from this flow one should in principle solve the full hierarchy of equations connecting the whole set of $n$-point correlation functions, a task which seems however out of reach.
%
As we will detail thereafter, the restriction of the flow to the vicinity of ${y=0}$ leads in fact to an (approximate) closed equation on the height of the two-point correlator ${\bar{R}(t,0)=\widetilde{D}_t \cdot \mathcal{R}_{\tilde{\xi}_t}(y)}$ according to our DP toymodel~\eqref{eq-toymodel-def-Rbar-functional}.
It will allow to pinpoint the role of the non-linearity in the temperature-dependence of the asymptotic ${\widetilde{D}_{\infty}(T,\xi)}$ and its interpolating parameter ${f(\xi,T)}$ defined by \eqref{eq-Dtilde-infty-finterp},
and give more insight into the short-`time' behavior of $\widetilde{D}_t$ and $\overline{\bar{F}_V(t,y)}$ (with respect to \eqref{eq-mean-FbarV}).

\subsection{Rescalings of $R$, $\bar{R}$, $\bar{R}_3$ and ${\eta_V}$}
\label{subsection-Dtildeinfty-T-xi-rescalings}

From~\eqref{eq-noclose-Rbar}, the flow of $\bar{R}(t,y)$ in $y=0$ reads
\begin{equation}
 \partial_t \bar{R}(t,0) =
 \frac Tc \bar{R}''(t,0)
 -\frac 1c \bar{R}_3'(t,0)
 -\frac 1 t \bar{R}(t,0)
 -DR''_\xi(0)
 \label{eq:evoltRbart0bis}
\end{equation}
(throughout this section we denote for short the derivative with respect to $y$ by a prime).
Although this equation is exact, it cannot be solved directly since the three-point correlator $\bar{R}_3$ is not known.
To go further and try to find out what relations between the physical parameters it might nevertheless imply, one has to surmise a (minimal) scaling form of the different correlators and their first derivatives.

Let us first consider the known scaling of the microscopic disorder correlator ${R_{\xi}(y)}$:
\begin{equation}
  D R_\xi(y)
 \stackrel{(y\to 0)}{\approx} 
  D \Big[ R_\xi(0)+ R''_\xi(0) \frac{y^2}{2} \Big] 
 =
  c_0\frac D\xi \Big[ 1 - c_1 \frac{y^2}{2\xi^2} \Big]
  \label{eq:expansion_Rxi_small-y_bis}
\end{equation}
where $c_0= R_\xi(0)|_{\xi=1}$ and $c_1 = -\frac{R''_\xi(0)}{R_\xi(0)}\Big|_{\xi=1}$ are numerical constants, independent of $\xi$ and reflecting the specific geometry of the correlator around the origin. For instance when the correlator is a Gaussian function ${R_{\xi}(y) = e^{-y^2/(4\xi^2)}/\sqrt{4 \pi \xi^2}}$
(\textit{i.e.} used to generate the Fig.~\ref{fig:finitetimescaling-Cbarlin-Rbarlin}),
one has $c_0=\frac1{\sqrt{4\pi}}$ and $c_1=\frac 12$.

By analogy with~\eqref{eq:expansion_Rxi_small-y_bis}
and supported by the numerical test of our DP toymodel in~Ref.~\cite{agoritsas_2012_FHHtri-numerics},
we now assume that the correlator $\bar{R}(t,y)$ scales around $y\approx 0$ as
\begin{equation}
  \bar{R}(t,y) \stackrel{(y\to 0)}{\approx} 
  c_2\frac{\widetilde D_t}\xi\Big(1-c_3\frac{y^2}{2\xi^2}\Big)
 \label{eq:expansion_Rbar_small-y_bis}
\end{equation}
Here, $c_2$ and $c_3$ are numerical constants independent of the parameters $(c,D,T,\xi,t)$:
\begin{equation}
  c_2= \bar{R}(t,0)\Big|_{\footnotesize\parbox[b]{0pt}{$\substack{\xi=1\\ \widetilde D_t =1}$}}
\qquad , \qquad
  c_2c_3= -\bar{R}''(t,0)\Big|_{\footnotesize\parbox[b]{0pt}{$\substack{\xi=1\\ \widetilde D_t =1}$}}
 \label{eq:expansion_Rbar_small-y_bis-bis}
\end{equation}
while $c_2\frac{\widetilde D_t}\xi$ is the height of the central peak $\bar{R}(t,0)$ assumed to capture all the  dependence in the parameters.
$c_2$ is actually defined so that the $\xi\to 0$ limit \eqref{eq-infty-time-FPsolution} is recovered:
\begin{eqnarray}
 && \widetilde D_\infty(T,\xi)
 \equiv \lim_{t\to\infty} \widetilde D_t
 = \int_{\mathbb{R}} dy \cdot \bar{R}(\infty,y) >0
  \label{eq:condition-for-c2-FHH-bis} \\ 
 && \lim_{\xi\to 0} \widetilde D_\infty(T,\xi)
 = \frac{cD}T
 \label{eq:condition-for-c2-FHH}
\end{eqnarray}
%
which is known to hold exactly, without any additional numerical constant.
The constant $c_2$ depends on global properties of the infinite-`time' limit of the correlator, in the sense that it is constrained by \eqref{eq:condition-for-c2-FHH-bis}-\eqref{eq:condition-for-c2-FHH}. 
The main assumption in the scaling form~\eqref{eq:expansion_Rbar_small-y_bis} is actually that the curvature of the correlator $\bar{R}(t,y)$ at the top of its central peak happens on a scale $\xi/\sqrt{c_3}$ which corresponds to $\widetilde\xi_\infty$ and is independent of `time' and $\arga{c,D,T}$.
This assumption is not exact at all `times', but we expect that it
captures anyway the main features of the geometry of the correlator ${\bar{R}(t,y)}$ close to its central peak.

Finally the three-point correlation function is assumed to scale in $\widetilde D_t$ and $\xi$ in the same way as it naively does merely by counting the number of occurrences of $\eta$ in the definition~\eqref{eq:defR3} of~$\bar{R}_3(t,y)$ (\textit{i.e.} inferred from ${\bar{R}\sim \overline{\eta \eta}}$, ${\bar{R}_3 \sim \overline{\eta \eta \eta}}$ and rescaling also the derivative $\partial_y$):
\begin{equation}
  \bar{R}_3'(t,0)= c_4 \frac{\widetilde D_t^{3/2}}{\xi^{5/2}}
\quad\text{with}\quad 
c_4=\bar{R}_3'(t,0)\Big|_{\footnotesize\parbox[b]{0pt}{$\substack{\xi=1\\ \widetilde D_t =1}$}}
 \label{eq:scalingR3prime_c4}
\end{equation}
Here $c_4$ is also assumed to be a numerical constant. 
If this form is again not expected to be exact at all `times', it can still be thought as a reasonable approximation provided the three-point correlator $\bar{R}_3'(t,y)$ is analytic around ${y=0}$. This last assumption is justified for instance in view of the zero-temperature and infinite-`time' limit of
the `flow' equation~\eqref{eq-noclose-Rbar} (under the stationarity condition $\partial_t \bar{R}=0$), which yields that the three-point correlator $\lim_{t\to\infty}\bar{R}_3'(t,y)$ is merely proportional to ${R''_\xi(y)}$, which is analytic around $y=0$.

Note finally that the scalings~\eqref{eq:expansion_Rxi_small-y_bis}, \eqref{eq:expansion_Rbar_small-y_bis}, \eqref{eq:scalingR3prime_c4} are all compatible with the following rescaling in distribution (also inferred from~${\overline{\eta\eta}\sim\bar R}$)
\begin{equation}
  \eta(t,y) \stackrel{(d)}= \bigg(\frac{\widetilde D_t}{a}\bigg)^\frac12 \eta(t, \tfrac ya)\big|_{\xi/a}
\end{equation}
for all values of the rescaling parameter $a$.
A way to reformulate the definitions~\eqref{eq:expansion_Rbar_small-y_bis-bis} and \eqref{eq:scalingR3prime_c4} of the numerical constants $\arga{c_2,c_3,c_4}$ is thus to identify those constants with the corresponding derivatives of the correlator $\bar{R}(t,y)$ taken at $a=\xi$ and $\widetilde D_t=1$, which ensures their independence with respect to the other parameters.

\subsection{Evolution of $\widetilde{D}_t$ and prediction for $\widetilde{D}_{\infty}$ and ${f(T,\xi)}$}
\label{subsection-Dtildeinfty-T-xi-Dtildetime}

\begin{figure}[t]
  \centering
  \includegraphics[height=.292\columnwidth]{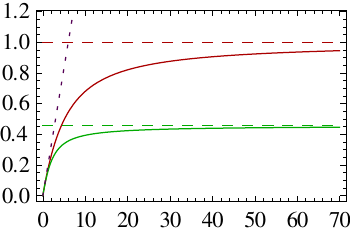}
  \includegraphics[height=.28\columnwidth]{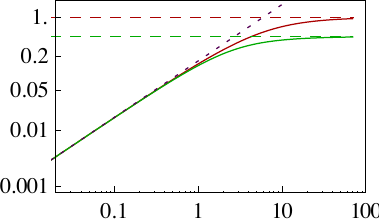}
  \caption{(Color online) $\widetilde D_t$ as a function of $t$ (\textit{left}: linear scale, \textit{right}: logarithmic scale).
  In green (bottom curve), the solution of the full differential equation~\eqref{eq:eq-evol-Dtilde-t};
  in red (top curve), the solution of the equation without the non-linear KPZ term;
  the dashed horizontal lines are the corresponding large-`time' asymptotics;
  in purple dotted, the short-`time' asymptotics~\eqref{eq:Dtilde_shorttime}.
  Chosen parameters: $c=D=T=\xi=1$, $c_0=c_2=\frac 1{\sqrt{4\pi}}$, $c_1=c_3=\frac 12$, $c_4=\frac 14$.
\label{fig:Dtilde-of-t_KPZornot}
}
\end{figure}

Substituting the rescalings~\eqref{eq:expansion_Rxi_small-y_bis}, \eqref{eq:expansion_Rbar_small-y_bis}, \eqref{eq:scalingR3prime_c4} into~\eqref{eq:evoltRbart0bis}
transforms the equation for ${\bar{R}(t,0)}$ into an effective closed evolution equation for the amplitude:%
\begin{equation}
  \partial_t \widetilde D_t + \frac 2 t \widetilde D_t 
  = -c_3 \frac{T}{c\xi^2}\widetilde D_t
  - \frac{c_4}{c_2} \frac{1}{c\xi^{3/2}} \widetilde D_t^{3/2}
  +\frac{c_0c_1}{c_2}\frac{D}{\xi^2}
\label{eq:eq-evol-Dtilde-t}
\end{equation}
The non-linear KPZ term of the equation of evolution for $F(t,y)$ corresponds to the term $\propto \widetilde D_t^{3/2}$.
The numerical solution of this equation is plotted in~Fig.~\ref{fig:Dtilde-of-t_KPZornot}.

%
To tackle the infinite-`time' case, we have introduced in~Sec.~\ref{section-DPtoymodel-B}-\ref{section-DPtoymodel-C} the interpolating parameter $f$ and identified it as the full-RSB cutoff in the GVM predictions (cf.~Appendix~\ref{A-GVM-PRB2010}).
By cancelling the right-hand side of~\eqref{eq:eq-evol-Dtilde-t} and introducing the interpolating
parameter ${f(c,D,T,\xi)=\widetilde D_\infty\big/\tfrac{cD}{T}}$ as in~\eqref{eq-Dtilde-infty-finterp}, one obtains a new equation for $f$:
\begin{equation}
  f^{3/2}= \frac{c_2 c_3}{c_4} \argc{\frac{T}{T_c(\xi)}}^{3/2}\Big(\frac{c_0c_1}{c_2c_3}-f\Big)
 \label{eq:equation0-for-fudging}
\end{equation}
where $T_c(\xi)=(\xi cD)^{1/3}$ is the same characteristic temperature as the one obtained in~Sec.~\ref{section-DPtoymodel-C} by scaling.

Noting that $c_2$ is defined in~\eqref{eq:expansion_Rbar_small-y_bis-bis} precisely so that it absorbs all the quantitative contribution of the geometry of $\bar{R}$, the condition \eqref{eq:condition-for-c2-FHH} guarantees that $\lim_{\xi\to 0} f = 1$.
The $\xi\to 0$ solution of~\eqref{eq:equation0-for-fudging} is then $f=\frac{c_0c_1}{c_2c_3}$ so one obtains eventually that $\frac{c_0c_1}{c_2c_3} =1$.
This is valid for all values of the parameters if the $\arga{c_i}$ are indeed parameter-independent.
However those numerical constants are constrained only by the geometry of ${\bar{R}(t,y)}$, and we know from the numerical study in~Ref.~\cite{agoritsas_2012_FHHtri-numerics}
that at large `times' the correlator saturates to a ${\bar{R}_{\text{sat}}(y) \approx \widetilde{D}_{\infty}\mathcal{R}(y) }$ with only a slight $T$-dependence of the function $\mathcal{R}$.

Disregarding this possible but small modification of the numerical constant depending on $T$,
the equation for the interpolating parameter finally reads
\begin{equation}
 f^{\gamma}= \frac{c_0 c_1}{c_4} \bigg(\frac{T}{T_c}\bigg)^{\gamma}\big(1-f\big)
 \quad\text{with}\quad \gamma = \frac 32
 \label{eq:equation-for-fudging}
\end{equation}
Strikingly, it takes a form very similar to the equations~\eqref{eq-equa-interpf-uc} and~\eqref{eq-equa-interpf-vc} obtained in Sec.~\ref{section-DPtoymodel} from the GVM approach, with an exponent $\gamma=6$ instead of $\gamma=\frac 32$.
In fact the value of this exponent only influences the specific monotonous crossover from the high-$T$ regime, where the KPZ term has little influence ($f \lesssim 1$) to the low-$T$ asymptotics where $f$ is linear in $T$ as ${f\stackrel{(T\to 0)}{\sim} \big(\frac{c_0 c_1}{c_4}\big)^\frac1\gamma\frac{T}{T_c}}$.

The value of $\gamma$ modifies the numerical constants in this last regime but does not influence the power-law dependence in the physical parameters, gathered in $T_c(\xi)$.
The equation \eqref{eq:equation-for-fudging} is a consistency check with respect to the two GVM predictions \eqref{eq-equa-interpf-uc} and \eqref{eq-equa-interpf-vc},
and the strictly monotonous behavior of ${\bar{R}(t,0)}$ and ${\widetilde{D}_{\infty}}$ observed numerically in~Ref.~\cite{agoritsas_2012_FHHtri-numerics}.

\subsection{Short-`time' evolution of $\widetilde{D}_t$ and saturation at $t_{\text{sat}}$}
\label{subsection-Dtildeinfty-T-xi-shorttime-lin}

Leaving the infinite-`time' case, we consider now the opposite regime of short `times' where the evolution~\eqref{eq:eq-evol-Dtilde-t} of $\widetilde D_t$ can be solved first in the absence of the non-linear KPZ term, predicting that $\widetilde D_t$ is linear in $t$ at short `times':
\begin{equation}
 \widetilde D_t \stackrel{(t\to 0)} \simeq
 \frac{c_0 c_1}{3c_2}\frac{D }{ \xi^2} t
 \label{eq:Dtilde_shorttime}
\end{equation}
This behavior can also be checked by the naked eye directly on~\eqref{eq:eq-evol-Dtilde-t} searching for a solution $\widetilde D_t\propto t$.
Note the factor 3 in the denominator, due to the two terms in the left-hand side of~\eqref{eq:eq-evol-Dtilde-t}.
Assuming that the solution $\widetilde{D}_t$ is also linear at short `times' while keeping the KPZ term actually yields the same result (since then ${\widetilde D_t^{3/2} \ll \widetilde D_t \ll t^0}$).

One checks that this self-consistent hypothesis is correct by solving~\eqref{eq:eq-evol-Dtilde-t} numerically
(cf.~Fig.~\ref{fig:Dtilde-of-t_KPZornot}).
Note that \eqref{eq:Dtilde_shorttime} predicts the short-`time' regime $\widetilde{D}_t$ to be temperature-independent.
The behavior~\eqref{eq:Dtilde_shorttime} should thus hold in generality, and allow us to define a saturation scale $t_\text{sat}$ at which $\widetilde D_t$ reaches its asymptotic value
\begin{equation}
  \widetilde D_\infty =  \frac{c_0 c_1}{3c_2}\frac{D }{ \xi^2} t_\text{sat}
  \quad \text{\textit{i.e.}} \quad t_\text{sat} =  \frac{3c_2}{c_0 c_1}\frac{ c\xi^2}{T} f
\end{equation}
\textit{Accordingly to this equation, the saturation occurs earlier at higher temperatures, the thermal fluctuations actually  smoothing the evolution of the disorder correlator.
}

%
Another consequence of the short-`time' behavior~\eqref{eq:Dtilde_shorttime} deals with the short-`time' dynamics of the mean-value ${-2c\, \overline{\bar{F}_V(t,y)}}$.
Inserting the short-`time' ${\widetilde D_t}$ of \eqref{eq:Dtilde_shorttime} into the assumed scaling ${\bar{R}(t,0)=c_2\frac{\widetilde D_t}{\xi}}$ of~\eqref{eq:expansion_Rbar_small-y_bis} and into the exact relation \eqref{eq-mean-FbarV}, one obtains from the initial condition ${\bar{F}_V(t,y)\equiv 0}$ that
${- 2c\,\partial_t \overline{\bar{F}_V(t,y)}  \stackrel{(t\to 0)} = \frac{c_0c_1}3 \frac{D}{\xi^3}t}$
hence the prediction
\begin{equation}
 - 2c\,\overline{\bar{F}_V(t,y)}  \stackrel{(t\to 0)} = 
 \frac{2c_0c_1}3 \frac{D}{\xi^3}t^2
\end{equation}
This quadratic behavior in $t$ thus predicts a superlinear short-`time' regime for ${\overline{\bar{F}_V(t,y)}}$.

This saturation `time' is different from the characteristic `time' ${t^*=\frac{t^5}{cD^2}}$ recently discussed in~Ref.~\cite{gueudre_2012_PhysRevE86_041151} for the evolution of the free-energy fluctuations in the high-temperature regime and which corresponds to the Larkin length appearing by scaling at high-$T$ or ${\xi=0}$ as discussed in~\eqref{eq-rescalingsB-for-highT}. This scale ${t^*}$ allows us to draw apart a short-`time' diffusive and the large-`time' KPZ regime in the evolution of the fluctuations of ${F(t,y)}$ (see also~Ref.~\cite{agoritsas-2012-FHHpenta} for a related study), while the scale ${t_{\text{sat}}}$ we examine in this section, singular in the limit ${\xi\to 0}$, captures the short-`time' effects inherently due to the finiteness of~$\xi$ via the KPZ non-linearity.

To summarize, at finite $\xi$ the non-linear KPZ term does not modify the short-`time' regime but induces a saturation of $\widetilde D_t$ at shorter `times' with increasing $T$ and to an asymptotic value $\widetilde D_\infty< \frac{cD}T$.
The high- and low-temperature asymptotic regimes $\widetilde D_\infty\!\!\! \stackrel{(T\gg T_c)}= \frac{cD}T$ and $\widetilde D_\infty \!\!\! \stackrel{(T\ll
  T_c)}= \frac{cD}{T_c}$ are both independently well-controlled, and the evolution equation~\eqref{eq:eq-evol-Dtilde-t} we presented thus allows us to tackle the crossover from one regime to the other, but only in an effective way.
An interesting open question is to provide a proper analytical derivation of the full temperature crossover and to fix the value of its exponent~$\gamma$.

%

\section{Link to experiments} \label{section-discussion-exp}

We discuss in this last section the consequences of the finite ${\xi>0}$ and its associated low-$T$ regime for two specific experiments.

On one hand domain walls in ultrathin magnetic films \cite{lemerle_1998_PhysRevLett80_849,repain_2004_EurPhysLett68_460,metaxas_2007_PhysRevLett99_217208} are well described by the DES model of a 1D interface defined in~Sec.~\ref{section-def-DES},
and both their static geometrical properties and their quasistatic dynamical properties (in the so-called `creep' regime) are thus captured by the DP endpoint ${y(t)}$ fluctuations.

On the other hand, a second instance of experiments encompassed by the KPZ theory is provided by interfaces in liquid crystals~\cite{takeuchi_2010_PhysRevLett104_230601,takeuchi_2011_scientificReports1_34,takeuchi_2012_JStatPhys147_853}, whose geometrical fluctuations are directly described by the DP free-energy ${F(t,y)}$ properly recentered.

\subsection{Temperature-dependence of the asymptotic roughness} \label{section-discussion-exp-A}
As emphasized in Sec.~\ref{section-DPtoymodel-C} a prominent feature of the DP in a correlated disorder is that the amplitude of the roughness~${B(t)}$ is modified by the microscopic length~$\xi$ even at very large lengthscales (in the RM regime), provided that the temperature is lower than the characteristic temperature ${T_c=(\xi c D)^{1/3}}$. 
Interfaces in ferromagnetic thin films are a prototype system~\cite{agoritsas_2012_ECRYS2011} for the experimental study of 1D interfaces, and in particular their roughness exponent has been measured~\cite{lemerle_1998_PhysRevLett80_849,metaxas_2007_PhysRevLett99_217208} to be ${\zeta_{\text{RM}}\approx 0.66}$ in agreement with the KPZ exponent ${\zeta_{\text{RM}}=\frac 23}$.
As estimated in Ref.~\cite{agoritsas_2010_PhysRevB_82_184207} (Sec.~VII B), the order of magnitude of~$T_c$ could be of room-temperature for these systems, which makes it even more relevant to determine whether they lie in the low-$T$ or in the high-$T$ regime.

One could in principle distinguish between those two temperature regimes through the prefactor ${A_{\text{RM}}}$ of the asymptotic roughness at large lengthscales:
\begin{equation}
 A_{\text{RM}}
\stackrel{(T \gg T_c)}{\sim}
 \Big(\frac{D}{cT}\Big)^{\frac 23}
\qquad
 A_{\text{RM}}
\stackrel{(T \ll T_c)}{\sim}
 \Big(\frac{D^2}{c^4\xi}\Big)^{\frac 29}
\end{equation}
as given by scaling arguments \eqref{eq-Basympt-GVM}-\eqref{eq-Basympt-GVM-amplitude}, predicted by GVM in~\eqref{eqGVM-fullRSBcutoff-twolimits}-\eqref{eqGVM-amplitude-twolimits}, and consistent with the numerical study of the roughness in~Ref.~\cite{agoritsas_2012_FHHtri-numerics}.

The main problem regarding such a study is that the elastic constant $c$ and possibly the disorder strength $D$ may depend themselves on temperature, making it difficult to characterize the low-$T$ regime by a temperature-independent ${A_{\text{RM}}}$, or to interpret a measure of the scaling in temperature ${A_{\text{RM}}\sim T^{2 \text{\thorn}}}$ with {\thorn} the \textit{thorn} exponent.
In any case, a change of regime in the temperature-dependence of ${A_{\text{RM}}}$ would provide a strong evidence for a low-$T$ to high-$T$ crossover.
Promising experiments have actually been performed regarding the temperature dependence in ultrathin Pt/Co/Pt films, as analyzed in Ref.~\cite{bustingorry_kolton_2012_PhysRevB85_214416} where a fine study  is devoted to the thermal rounding at the depinning transition.

In case of an elastic constant depending linearly on the temperature ${c = \kappa T}$, we would expect a roughness amplitude similar to $A^{\mathfrak{m}}_{\text{RM}}$ obtained under the usual mathematicians convention ${c=T}$, as exposed in Appendix~\ref{A-saddle_scalings-maths} and with an additional dependence in $\kappa$.

\subsection{Quasistatic creep regime} \label{section-discussion-exp-B}

The creep motion of 1D interfaces, describing the quasistatic but non-linear response of the interface to an external driving field, could also be an interesting benchmark for the $\xi>0$ DP model predictions.

Even though its description~\cite{balents-fisher_1993_PhysRevB48_5949,chauve_2000_ThesePC_PhysRevB62_6241} is not directly covered by the equilibrium statistical properties of the interface we have presented here, it  happens that the characteristic lengthscales governing its scaling are actually believed to be the static ones. As detailed in Ref.~\cite{agoritsas_2010_PhysRevB_82_184207} (Sec.~VII B), those lengthscales are modified in the low-$T$ regime, implying that the characteristic free-energy barriers scale differently with the temperature above or below $T_c$.
This behavior could provide an experimental criterion to distinguish between the low-$T$ and high-$T$ regime, especially since the exponent of the creep law has been successfully tested on domain walls in ultrathin magnetic films on several order of magnitude in the velocity \cite{lemerle_1998_PhysRevLett80_849,repain_2004_EurPhysLett68_460,metaxas_2007_PhysRevLett99_217208}.

A challenging situation would be that of an interface moving in a gradient of temperature, as studied numerically in~Ref.~\cite{candia_albano_2011_PhysRevE84_050601}, with a gradient spanning $T_c$ itself and with an elastic constant $c$ (or the disorder strength $D$) which could again depend or not on the temperature.

\subsection{High-velocity regime in liquid crystals} \label{section-discussion-exp-C}

Another experimental system where our approach might prove instructive is that of growing interfaces in liquid crystal turbulence~\cite{takeuchi_2010_PhysRevLett104_230601,takeuchi_2011_scientificReports1_34,takeuchi_2012_JStatPhys147_853}.
The corresponding setup consists in a thin layer of liquid crystal subjected to a constant voltage~$U$ and to an alternating electric field which can generate two distinct turbulent modes (called `dynamical scattering modes') DSM1 and DSM2, the later being more stable than the former. Starting from a dot (respectively a line) of DSM2 in a DSM1 background, one thus observes the growth of circular (respectively flat) interface.
We refer the reader to Ref.~\cite{takeuchi_2012_JStatPhys147_853} for a complete account of the phenomena at hand.

The fluctuations of this interface are actually remarkably well described by the KPZ theory, providing a benchmark for its predictions not only about scaling exponents but also about scaling functions. Contrarily to the case of magnetic interfaces, the fluctuations of the interface \textit{position} are described by the random variable ${\bar{F} (t,y)}$ that plays the role of the disorder free-energy in the context of the DP, as defined by~\eqref{eq-def-free-energy-1}-\eqref{eq-def-free-energy-2}. Besides, for the liquid crystal interface,  $t$ is the physical time and $y$ the longitudinal direction of the interface.

For the circular interface, it has been shown that the experimentally measured ${\bar{C} (t,y)}$ defined by \eqref{eq-def-corr-FbarFbar2}, once properly rescaled, matches very well the prediction of the KPZ theory at $\xi=0$: ${\bar{C} (t,y)}$ can be fitted by the corresponding {Airy$_2$} correlator \cite{praehofer-spohn_2002_JStatPhys108_1071}.
Although the lengthscale $\xi$ of the disorder correlations is below the optical resolution of the experiment --~no rounding of the cusp of ${\bar{C} (t,y)}$ is observed~-- it is still finite and should be relevant by inducing a crossover between two `temperature' regimes to identify.
The (non-centered) displacement $h(t,y)$ of the interface evolves according to
\begin{equation}
  \partial_t h(t,y) = v_\infty + \nu \partial_y^2h(t,y) +\frac \lambda 2 \big[\partial_y h(t,y)\big]^2 +V(t,y)
  \label{eq-KPZ-experimental-stuff}
\end{equation}
where $v_\infty$ represents the mean displacement velocity. 
Comparing this equation to the flow of ${F_V(t,y)}$~\eqref{eq-FeynmanKac-FV}, one thus reads the correspondence ${\lambda=\frac 1c}$, ${\nu=\frac{T}{2c}}$, or equivalently ${c=\frac 1\lambda}$, ${T=\frac{2\nu}\lambda}$.
It can be argued~\cite{takeuchi_2011_scientificReports1_34,takeuchi_2012_JStatPhys147_853} that in the experimental configuration of the growing dot, the parameter $\lambda$ is directly given by the mean velocity $\lambda=v_\infty$. This velocity itself is well described by a growing affine function of the applied constant voltage $U$, in the probed voltage range $26\,\text{V} < U < 30\,\text{V}$ (see Fig.~20 in Ref.~\cite{takeuchi_2012_JStatPhys147_853}).
One may surmise that the parameter $\nu$, which describes the diffusive fluctuations of the interface, is independent of the constant voltage~$U$ (although this point is not discussed in Ref.~\cite{takeuchi_2012_JStatPhys147_853}), and similarly for the amplitude $D$ of the disorder fluctuations.

If those assumptions are true, one can deduce from the correspondence ${c=\frac 1\lambda}$, ${T=\frac{2\nu}\lambda}$ that the \textit{low-temperature} regime of our description (where $\xi$ matters in the scaling of the fluctuations amplitude) corresponds to a \textit{high-velocity} $\lambda=v_\infty$ regime for the liquid crystal interface --~and inversely for the low-velocity range being described by our high-temperature regime.
A natural question is thus to determine whether the experiments are done in the high- or low-velocity regime.
A test observable is provided by the amplitude of the fluctuations of ${h(t,y)}$, denoted $\Gamma$ in  Ref.~\cite{takeuchi_2012_JStatPhys147_853}, and defined in the large-time regime from the scaling 
\begin{equation}
 h(t,y) \stackrel{(d)}= v_\infty t + (\Gamma t)^{\frac 13} \chi_2 \argp{y/(\Gamma^{1/6} \lambda^{1/2} t^{2/3} )}
\end{equation}
where $\chi_2$ is the {Airy$_2$} process and its argument is rescaled with respect to the asymptotic roughness~${\sim t^{2/3}}$. In our notations, one has
\begin{equation}
  \Gamma = \frac{\widetilde D^2}{c}
  \label{eq:linkGammaDtilde}
\end{equation}
This is read for instance from the factor $a^{\frac 12} \widetilde D^{\frac 12}$ which rescales the disorder free-energy in~\eqref{eq-rescalingFFthF1dis} with the choice $a= (\widetilde{D}/c^2)^{\frac13}\, t^{\frac 23}$ of~\eqref{eq-rescalingsB-for-lowT-toymodel}. Equivalently, one has $(\Gamma t)^{\frac 13} =\widetilde E$ where $\widetilde E$ is the factor in~\eqref{eq-rescalingsB-for-lowT-toymodel} which rescales the free-energy in~\eqref{eq-rescaling_B_zeta23}. 
From our intuition based on scaling arguments, the relation~\eqref{eq:linkGammaDtilde} should thus remain valid on the whole temperature range.
The high-$T$ and low-$T$ results
${\widetilde D \stackrel{(T\gg T_c)}= \frac{cD}T}$ and ${\widetilde D \stackrel{(T\ll  T_c)}= \frac{cD}{T_c}}$ thus imply for $\Gamma$
\begin{equation}
\left\{\begin{aligned}
 \lambda\ll \lambda_c: &\quad \Gamma^{\text{low\,$\lambda$}} = \frac{D^2 \lambda}{\nu^2}
\\
 \lambda\gg \lambda_c: &\quad \Gamma^{\text{high\,$\lambda$}} = \frac{D^{\frac 43}}{ \lambda^{\frac 13}\xi^{\frac 23}}
\end{aligned}\right.
\text{with }
\lambda_c=\Big(\frac{\nu^3}{\xi D}\Big)^{\frac12}
\label{eq-link-exp-lambda}
\end{equation}
The crossover between those two regimes occurs at a characteristic $\lambda_c$ which is similar to the inverse of the characteristic temperature~${T_c^{\mathfrak m}=\sqrt{\xi D}}$ in the conventions of mathematicians (see Appendix~\ref{A-saddle_scalings-maths}), as expected since ${T \propto \frac{1}{\lambda}}$ and ${c=\frac{1}{\lambda}}$.
Strikingly, the crossover occurs between a low-$\lambda$ regime (our high-$T$) where $\Gamma$ is an increasing function of $\lambda$ and a high-$\lambda$ regime (our low-$T$) where $\Gamma$ is a decreasing function of $\lambda$.
Note that the limit ${\lambda \to 0}$ in \eqref{eq-KPZ-experimental-stuff} is exactly the `linearized' KPZ problem whose evolution has been solved at all `times' in~\eqref{eq-decomposition-Rbarlin-1}-\eqref{eq-decomposition-Rbarlin-2} and should thus be described exactly the microscopic disorder correlator ${R_{\xi}(y)}$ at asymptotically large `times' \eqref{eq-infty-time-Rxi-lin}.

In Ref.~\cite{takeuchi_2012_JStatPhys147_853}, the authors have measured $\Gamma$ as a function of the voltage $U$ and found a decreasing dependence, attributing it to a dependence in $U$ of the parameters $D$ or $\nu$ in the expression $ \Gamma^{\text{low } \lambda}  = \frac{D^2 \lambda}{\nu^2}$ (tackled with ${\xi=0}$ in our language, when ${\xi \gtrsim 0}$ can be neglected).
In that spirit, \eqref{eq-link-exp-lambda} was first announced at the very conclusion of~Ref.~\cite{agoritsas-2012-FHHpenta} but with respect to the parameter $\nu$ and with the two regimes being separated by the characteristic value ${\nu_c = (\xi \lambda^2 D)^{1/3}}$.
We propose here another scenario where $D$ and $\nu$ are independent of $U$, and $\Gamma$ follows its high-$\lambda$ expression ${\Gamma^{\text{high\,$\lambda$}}=\frac{D^{\frac 43}}{ \lambda^{\frac 13}\xi^{\frac 23}}}$.
If confirmed (\textit{e.g.} by an independent measure of the parameter $\nu$), this would provide a clear evidence that the system lies deep in the
high-$\lambda$ regime, and would to our knowledge constitute the first example of a phenomenon depicted by the `low-temperature' KPZ regime.
%



\section{Conclusion} \label{section-conclusion}

In this paper, we have studied analytically the consequences of a finite disorder correlation length~$\xi$ on the static properties of a 1D interface depending on the lengthscale --~or equivalently a growing 1+1 directed polymer at a fixed `time'~-- in a random-bond quenched disorder accounting for a weak collective pinning.
The two-point correlator ${\bar R(t,y)}$ of  the \textit{derivative} of the disorder free-energy (the `random phase' $\eta_V(t,y)$) at fixed `time'~$t$ emerged as a central quantity in the determination of the dependence on the temperature $T$ and the other parameters of the model (the disorder correlation length $\xi$, the elastic constant $c$ and the disorder strength $D$), as summarized in~Sec.~\ref{section-synthetic-outlook}.

The characteristic temperature ${T_c=(\xi c D)^{1/3}}$ separates two temperature-regimes in the `time'-evolution of ${\bar R(t,y)}$,
which is characterized by its amplitude ${\widetilde{D}_t}$ and the shape of its central peak $\mathcal{R}(y)$, from the point of view of our DP toymodel which focuses on the small transverse displacements $\valabs{y}$ (which are the most probable to be visited by the polymer and thus the more relevant ones).
Although \textit{a priori} the full shape of the correlator ${\bar R(t,y)}$ should matter, we showed that the asymptotic height of its central peak, parametrized by ${ \widetilde D_\infty(T,\xi)/\tilde{\xi}}$ with ${\tilde{\xi} \approx \xi}$, captures the main features of the crossover between low-$T$ and high-$T$ in the large-`time' random-manifold regime;
the amplitude of the geometrical fluctuations in this asymptotic regime are actually not affected by $\xi$ at high-$T$, whereas on the contrary at low-$T$ it plays a crucial role and keeps this amplitude bounded.
We showed that the ratio ${f(T,\xi)}$ between the actual value of ${\widetilde D_\infty(T,\xi)}$ at finite $\xi$ and its value ${\frac{cD}T}$ at $\xi=0$ is directly related to a replica-symmetry-breaking cutoff parameter $u_c$ appearing in the GVM replica approach, endowing this cutoff with an unsuspected physical meaning (see Sec.~\ref{section-DPtoymodel-B}).
Moreover, ${\widetilde D_t}$ evolves at finite $t$  according to two `time' regimes, an initial regime and a saturation regime separated by a single `time'-scale $t_\text{sat}$ which allows us to understand how, depending on the temperature, the geometrical fluctuations depicted by the roughness $B(t)$ display in turn two (at high $T$) or three (at low $T$) `time' regimes.
Finally we believe that the function ${\mathcal{R}(y)}$ for `times' above ${t_{\text{sat}}}$ is closely connected to the microscopic disorder correlator ${R_{\xi}(y)}$, but such a connection still needs to be characterized with some additional kernel to be determined analytically, beyond the linearized case ${\bar{R}^{\text{lin}}(t,y)}$ that we have solved exactly.

The picture we have derived in this paper is in full agreement with extensive numerical results on the ${\xi>0}$ KPZ equation that will be presented in a companion paper~Ref.~\cite{agoritsas_2012_FHHtri-numerics}.
In agreement with an effective equation that we put forward for ${\widetilde{D}_{\infty}(T,\xi)}$, it appears that this quantity presents a crossover from low- to high-$T$, and not a phase transition that would be characterized by a non-analytic dependence of ${\widetilde{D}_{\infty}}$ in~$T$. 
We cannot strictly exclude nonetheless the scenario of a phase transition because of inherent numerical imperfections in the simulations and because the analytical equation on ${\widetilde{D}_{\infty}(T,\xi)}$ is either effective~\eqref{eq:equation-for-fudging} or obtained in the GVM approximation \eqref{eq-equa-interpf-uc} and \eqref{eq-equa-interpf-vc}. A rigorous procedure is needed to elucidate this alternative.
Possible approaches encompass from an extension to the ${\xi>0}$ case of recent results~\cite{calabrese_2010_EPL90_20002,amir_arXiv:1003.0443,sasamoto_2010_NuclPhysB_834_523,dotsenko_2010_EPL90_20003} in the ${\xi=0}$ KPZ class, to an adaptation of FRG arguments~\cite{balents-fisher_1993_PhysRevB48_5949,chauve_2000_ThesePC_PhysRevB62_6241,bustingorry_2010_PhysRevB82_140201} to the study of the correlator ${\bar R(t,y)}$.

More broadly, it would be interesting to identify possible connections between the `time' equation of evolution~\eqref{eq-noclose-Rbar} of the correlator  ${\bar R(t,y)}$, and the FRG flow equation of evolution with respect to scale $\ell$ of the renormalized disorder correlator in a FRG approach. Although the FRG framework holds perturbatively in $\epsilon=4-d$ (hence $\epsilon=3$ in our settings) the two varieties of flow equations bear striking resemblance. In particular, the high-temperature regime has been studied in Ref.~\cite{bustingorry_2010_PhysRevB82_140201} by neglecting the non-linear contributions to the FRG flow, allowing us to recover the same scalings as ours for $T\gg T_c$. On the other hand, neglecting the non-linearity of the flow in our settings (Appendix~\ref{A-short-time-dynamics-Fbar-generic}) indeed yields the correct high-temperature scaling $\widetilde D_\infty=\frac{cD}T$ but does not provide for instance the correct roughness exponent ${\zeta_{\text{RM}}=\frac 23}$ (when we consider the scaling of the negative bumps at large $y$, as in~\eqref{eq-decomposition-Rbarlin-2}). One would thus need a controlled expansion in order to draw a meaningful unified picture.
Another related question deals with the $T=0$ singularity of the FRG renormalized correlator which appears at a finite scale $\ell_c$ and plays an important role in the physical description of metastability in random manifolds. Our result should help to understand in which order the limits ${T\to 0}$
and ${\xi\to 0}$ are to be taken.
Last, a FRG approach, through the possible existence of a zero and a finite temperature fixed point, may allow us to distinguish
between the crossover and the phase-transition scenari.

The richness of the KPZ universality class also allows us to translate the occurrence of low- \textit{vs} high-temperature regimes into different languages.
We have illustrated this fact in the analysis of a liquid crystal experiment~\cite{takeuchi_2010_PhysRevLett104_230601,takeuchi_2011_scientificReports1_34,takeuchi_2012_JStatPhys147_853} (Sec.~\ref{section-discussion-exp}) where the low-temperature regime corresponds to a high-velocity regime --~yet to be ascertained experimentally.
On the mathematical side, keeping $\xi$ finite amounts to generalize the {Airy$_2$} process in a non-trivial way (see also Ref.~\cite{agoritsas-2012-FHHpenta}).
In the language of replicae~\cite{kardar_1987_NuclPhysB290_582}, it corresponds to solving the problems of bosons with attractive but non-$\delta$ interactions in one dimension, for which the $\xi=0$ Bethe Ansatz solution is not known to generalize.
The same search for an extension also applies to the {Airy$_1$} process, which describes a point-to-line DP problem (with `flat' initial conditions), which finds an experimental incarnation for instance in flat interfaces in liquid crystals.

One may finally wonder how the existence of a low-temperature regime extends to phenomena which do not fall \textit{a priori} into the KPZ universality class. 
A basic assumption made at the very start is that the disorder is Dirac $\delta$-correlated along the longitudinal `time' direction. The disorder is nevertheless always correlated in both directions in physical systems; one generalization of our results would be to take this property into account. Such correlations may also account for overhangs
present in interfaces, once smoothed out by a change of scale.
The generalization is non-trivial in the sense that the equations of evolution become non-local in `time'.
More broadly, the cases of non-RB disorder and/or higher dimensions, where a mapping to a directed-path problem is not always possible, are still open.


\begin{acknowledgments}
  We would like to thank Sebastian Bustingorry, Gregory Schehr, Francis Comets and Jeremy Quastel for fruitful discussions. This work was supported in part by the Swiss NSF under MaNEP and Division II, and by the ANR~2010 BLAN~0108.
\end{acknowledgments}


\appendix


\section*{Appendices}


\section{Reminder of previous GVM roughness predictions} \label{A-GVM-PRB2010}

For completeness, here we recall and adapt the roughness predictions for the static 1D interface obtained in Ref.~\cite{agoritsas_2010_PhysRevB_82_184207},
first assuming that the disorder correlator $R_{\xi}(x)$ in \eqref{eq-def-moydis} was a normalized Gaussian function of variance $2\xi^2$,
then using the replica trick in order to average over disorder
and finally performing a  Gaussian Variational Method (GVM) with a full replica-symmetry-breaking (RSB) variational Ansatz as introduced by M\'ezard and Parisi in Ref.~\cite{mezard_parisi_1991_replica_JournPhysI1_809,mezard_parisi_1992_JPhysI02_2231},
further investigated also by Goldschmidt and Blum in Ref.~\cite{goldschmidt-blum_1993_PhysRevE48_161}.
We emphasize in particular the role of the interpolating parameter ${f(T,\xi)}$ between the high- and low-temperature regimes.

For the full DES model of the 1D interface, we had obtained for the variance of the relative displacements
${B(r) \equiv \overline{\moy{\Delta u(r)^2}}}$ at the lengthscale $r$:
\begin{eqnarray}
 B(r)
 &=& \frac{Tr_0}{c} \argp{\frac{r}{r_0} + \bar{B}_{\text{dis}}\argp{\frac{r}{r_0}}} \label{equa-1D-roughness-1} \\
 \bar{B}_{\text{dis}}(\bar{r})
 &=& \frac{1}{v_c} \sum_{k=2}^{\infty} \frac{(-\bar{r})^k}{k!} \argc{\frac{1}{5k-6}+(1-v_c)} \label{equa-1D-roughness-2} \\
 r_0 &=& \frac{5^5 \pi}{3^7} \frac{1}{cD^2} \argp{\frac{T}{v_c}}^5 \label{equa-1D-roughness-3} \\
 v_c^6 &=& \widetilde{A}_1 (5/6 - v_c) \, , \, \widetilde{A}_1=\frac{5^5 \pi}{2 \times 3^7} \argp{\frac{T}{T_c}}^6 \label{equa-1D-roughness-4} \\
 T_c & \equiv & (\xi c D)^{1/3} \label{equa-1D-roughness-5}
\end{eqnarray}
where the four DES parameters $\arga{c,D,T,\xi}$ are respectively the elastic constant $c$ (elastic energy per unit of length along the interface), the disorder strength $D$ (the typical amplitude of the random potential), the temperature $T$ and the disorder correlation length $\xi$ (or width of the interface).
$r_0$ is the Larkin length introduced in \eqref{eq-rescaling-Lc-1} and marking the beginning of the asymptotic `random-manifold' regime, $v_c(T,\xi)$ the full-RSB cutoff and $T_c$ the characteristic temperature separating the low- and high-temperature regimes.

As for our DP `rounded' toymodel, assuming that the effective disorder correlator of $\eta_V(t,y)$ \eqref{eq-def-corr-etaeta} is of the form ${\bar{R}(t,y)=\widetilde{D} \cdot \mathcal{R}_{\tilde{\xi}}(y)}$ we have performed a GVM procedure on the following statistical average obtained from \eqref{eq-def-moments-pdf-replicas-1} using replic\ae \, to average over disorder:
\begin{equation}
\begin{split}
 \overline{\moy{y(t)^k}}
 = \lim_{n \to 0} & \int dy_1 (\cdots) dy_n \cdot y_1^k
 \cdot e^{-\sum_{a=1}^{n} F_{\text{th}}(t,y_a)/T} \\
 & \cdot \exp \argc{-\frac{\widetilde{D}}{2} \sum_{a,b=1}^{n} \mathcal{R}_{\tilde{\xi}}(y_a-y_b)/T^2}
\end{split}
\label{eq-def-moments-pdf-replicas-2}
\end{equation}
With the function $\mathcal{R}_{\tilde{\xi}}(y)$ the same normalized Gaussian function of variance $2\tilde{\xi}^2$, we have obtained for the variance of the DP's endpoint fluctuations
${B_{\text{DP}}(t) \equiv \overline{\moy{y(t)^2}}}$ after a growing `time' $t$:
\begin{eqnarray}
B_{\text{DP}}(t \geq t_c)
 &=& \frac{3}{2} \argp{\frac{2 \widetilde{D}^2}{\pi c^4}}^{1/3} t^{4/3} - \tilde{\xi}^2 \label{equa-DPtoy-roughness-1} \\
B_{\text{DP}}(t \leq t_c)
 &=& \frac{Tt}{c} + \frac{\widetilde{D}}{c^2 \sqrt{\pi}} \cdot t^2 \argp{\tilde{\xi}^2 + \frac{Tt}{c}}^{-1/2} \label{equa-DPtoy-roughness-2} \\
 t_c &=& \frac{3^3 \pi}{2^4} \frac{c}{\widetilde{D}^2} \argp{\frac{T}{u_c}}^3 \label{equa-DPtoy-roughness-3} \\
 u_c^4 &=& \widetilde{A}_2 (3/4 - u_c) \, , \, \widetilde{A}_2=\frac{3^3 \pi}{2^4} \frac{T^4}{(\tilde{\xi} \widetilde{D})^2} \label{equa-DPtoy-roughness-4}
\end{eqnarray}
where, as before, the four parameters $\arga{c,T,\widetilde{D},\tilde{\xi}}$ are respectively the elastic constant $c$, the temperature $T$, the effective strength of disorder $\widetilde{D}$, the effective interface width $\tilde{\xi}$ and ${u_c(T,\xi)}$ the full-RSB cutoff.

However, the second GVM computation was performed \textit{at a fixed `time' $t$}, so the effective parameters $\widetilde{D}_t$ and $\tilde{\xi}_t$ have \textit{a priori} also a `time' dependence.
Assuming that they both saturate quite quickly, we can safely compare the two sets of predictions
\eqref{equa-1D-roughness-1}-\eqref{equa-1D-roughness-5}
and \eqref{equa-DPtoy-roughness-1}-\eqref{equa-DPtoy-roughness-4}
with the translation of `time' $t$ into the lengthscale $r$,
the approximation ${\tilde{\xi} \approx \xi}$,
and the identification of ${t_c}$ with ${r_0}$ for the Larkin length.
As for the full-RSB cutoffs, the definitions ${f^{\text{1D}}(T,\xi) \equiv \frac65 v_c(T,\xi)}$ and ${f^{\text{toy}}(T,\xi) \equiv \frac43 u_c(T,\xi)}$  yield then two similar equations for the interpolating parameter $f(T,\xi)$, respectively \eqref{eq-equa-interpf-vc} and \eqref{eq-equa-interpf-uc},
if we impose by hand ${\widetilde{D}=\frac{cD}{T} f^{\text{toy}}(T,\xi)}$ in order to match the Larkin lengths $t_c$ and $r_0$.
Their numerical discrepancy can be safely attributed to the GVM approximation.
The structure ${f^6 \propto (T/T_c)^6 (1-f)}$ stems in both cases from the comparison of ${\xi_{\text{th}}(T)^2=(\frac{T^3}{cD})^2}$ and ${\xi^2=\xi_{\text{th}}(T_c)^2}$, the equation \eqref{equa-1D-roughness-4} being initially of the form:
\begin{equation}
 \xi^2 + \frac{16\pi}{9} \xi_{\text{th}}(T)^2 \argp{\frac65 v_c}^{-6} \argp{\frac65 v_c -1} = 0
\end{equation}

As discussed in Ref.~\cite{agoritsas_2012_ECRYS2011}, the Larkin length is a physical benchmark for the roughness,
firstly as the beginning of its asymptotic `random-manifold' regime,
secondly with its relation to the maximum value of the GVM self-energy $\argc{\sigma}(v_c)$ and thus to the full-RSB cutoff itself ${v_c(T,\xi)}$,
and thirdly consistent with its original definition by Larkin \citep{Larkin_model_1970-SovPhysJETP31_784} as the lengthscale at which the typical relative displacement of the 1D interface corresponds to its effective width ${ B(L_c(T,\xi)) \approx \xi_{\text{eff}}(T,\xi)^2 }$ which fixes the amplitude of the RM roughness $B(r>L_c)\approx A(c,D,T,\xi) r^{4/3}$:
\begin{equation}
 A_{(c,D,T,\xi)}^{\text{GVM}}
 \leftrightarrow L_c^{\text{GVM}}(T,\xi)
 \leftrightarrow \argc{\sigma} (v_c)  ^{\text{GVM}}
 \leftrightarrow v_c 
\end{equation}
Using \eqref{equa-1D-roughness-4}, \eqref{equa-1D-roughness-3} and \eqref{eq-Basympt-GVM}, we give the GVM predictions for the quantities
\begin{small}
\begin{eqnarray}
 L_c^{\text{GVM}} (T,\xi)
 & \stackrel{\eqref{equa-1D-roughness-1}}{=} & \frac{32 \pi}{9} \argp{\frac65 v_c}^{-5} r_*(T)
 \, , \; r_*(T) \equiv \frac{T^5}{cD^2} \\
 A_{(c,D,T,\xi)}^{\text{GVM}}
 & = & \argc{D^{3/10} c^{-3/5} L_c^{\text{GVM}}(T,\xi)^{-1/10} }^{4/3} \\
 & = & \argc{\argp{\frac{D}{cT}}^{1/3} \cdot \argp{\frac{9}{32 \pi}}^{1/15} \cdot \argp{\frac65 v_c}^{1/3} }^2 \nonumber
\end{eqnarray}
\end{small}
at low- \textit{versus} high-temperatures:
\begin{small}
\begin{equation}
 v_c \stackrel{(\xi \to 0)}{\approx} \frac56
 \; , \:
 v_c \stackrel{(T \to 0)}{\approx} \frac56 \argp{\frac{16 \pi}{9}}^{1/6} \frac{T}{T_c} \approx 1.11 \frac{T}{T_c}
 \label{eqGVM-fullRSBcutoff-twolimits}
\end{equation} 
\begin{eqnarray}
 L_c^{\text{GVM}} (T,0) & \approx & \frac{32 \pi}{9} \frac{T^5}{cD^2} \approx 11.17 \cdot r_*(T) 
 \label{eqGVM-Larkinlength-twolimits} \\
 L_c^{\text{GVM}} (0,\xi) & \approx & \frac{32 \pi}{9} \! \argp{\frac{16 \pi}{9}}^{-5/6} \! \! \frac{T_c^5}{cD^2} \approx 2.66 \cdot r_*(T_c)
 \nonumber
\end{eqnarray}
\begin{eqnarray}
 A_{(c,D,T,0)}^{\text{GVM}} & \approx & \argp{\frac{9}{32 \pi}}^{2/15} \argp{\frac{D}{cT}}^{2/3} \approx 0.72 \, \argp{\frac{D}{cT}}^{2/3} 
 \label{eqGVM-amplitude-twolimits} \\
 A_{(c,D,0,\xi)}^{\text{GVM}} & \approx & \argp{\frac{9}{32 \pi}}^{2/15} \! \! \!  \argp{\frac{16 \pi}{9}}^{1/9} \! \! \! \argp{\frac{D}{cT_c}}^{2/3} \! \! \!  \approx 0.88 \, \argp{\frac{D}{cT_c}}^{2/3}
 \nonumber
\end{eqnarray}
\end{small}


\section{Statistical Tilt Symmetry (STS)} \label{A-STS-bythebook}

We justify in this appendix the decomposition of the free-energy ${F_V(t,y)}$ defined by \eqref{eq-def-free-energy-1} into the sum of a disorder-independent term ${F_{V\equiv 0}(t,y)}$ and a translationally invariant term ${\bar
F_V(t,y)}$ (\ref{eq-def-free-energy-2}-\ref{eq-def-free-energy-5}), from the complementary viewpoints of path-integrals and of stochastic differential equations.
This symmetry arises from three ingredients: the precise form of the elastic energy density ${\frac c2 (\partial_ty)^2}$, the invariance in distribution of the disorder $V(t,y)$ by translation along $y$ and the continuum nature of the transverse direction $y$.
We refer the reader to Ref.~\cite{fisher_huse_1991_PhysRevB43_10728,hwa_1994_PhysRevB49_3136,ledoussal_2003_PhysicaA317_140,amir_arXiv:1003.0443} for previous discussions of the STS.

The weight of trajectories $\arga{y(t)}$ starting in $(0,0)$ and arriving in ${(t_1,y_1)}$ can be compared to the weight of those arriving in $(t_1,0)$ from the change of coordinates:
\begin{equation}
  \bar{y}(t) \equiv y(t)-\frac{y_1}{t_1}t
  \label{eq-tiltedtraj}
\end{equation}
with now the initial and final conditions $\bar y(0)=\bar y(t_1)=0$.
Introducing the tilted disorder 
\begin{equation}
 \mathcal{T}_{y_1}^{t_1} V(t,y)\equiv V\big(t,y+\frac{y_1}{t_1}t\big) 
\end{equation}
and performing the change~\eqref{eq-tiltedtraj} in the path-integral of the unnormalized weight~\eqref{eq-def-unnorm-Boltzmann-Wv} one obtains:
\begin{equation}
 W_V(t_1,y_1)= e^{-y_1^2/(2B_{\text{th}}(t_1))} W_{\mathcal{T}_{y_1}^{t_1} V}(0,t_1)
 \label{eq-pathintegral-unnorm-Boltzmann}
\end{equation}
thanks to the form $\frac c2 (\partial_ty)^2$ of the elastic energy density which allows to identify a translated path integral over $\arga{\bar{y}(t)}$ and to single out the thermal contribution with ${B_{\text{th}}(t)=\frac{Tt}{c}}$.
The measure of the path-integral remains unchanged (${\mathcal D y(t)=\mathcal D \bar y(t)}$) since the change of
variable~\eqref{eq-tiltedtraj} is a translation. 
Using the definitions~\eqref{eq-def-free-energy-2}, this equality writes for the disorder free-energy
\begin{equation}
  \bar F_V(t_1,y_1)=  \bar F_{\mathcal{T}_{y_1}^{t_1} V}(t_1,0)
  \label{eq-transl-invariance-FbarV}
\end{equation}
This means that the whole dependence in the arrival point $y_1$ of the disorder free-energy can be absorbed into
a tilt $\mathcal{T}_{y_1}^{t_1} $ of the disorder potential.
Using that at fixed final time $t_1$ the disorder is translation-invariant in $y$ \textit{i.e.} $\bar{\mathcal{P}}\argc{V}=\bar{\mathcal{P}}\argc{\mathcal{T}_{y_1}^{t_1}V}$
one obtains the symmetry
${\bar{\mathcal{P}} \argc{\bar{F}_V (t,y+Y)} = \bar{\mathcal{P}} \argc{\bar{F}_V (t,y)}}$
as announced in~\eqref{eq-STS-PFbarV}.
This translation invariance in the $y$-direction is valid for any functional of ${\bar{F}_V(t,x)}$, and yields in particular for the $k$-point correlators:
\begin{align}
  &\overline{\bar F_V(t_1,y_1+Y)\ldots \bar F_V(t_1,y_k+Y)} 
\nonumber \\
   &\qquad \qquad \qquad=  \overline{\bar F_V(t_1,y_1)\ldots \bar F_V(t_1,y_k)}
\label{eq-barF_invariance-translation}
\end{align}

The same result~\eqref{eq-STS-PFbarV} can also be deduced from the strict point of view of stochastic differential equations.
Note first that the initial condition $\frac{W_V (0,y)}{\Wbar_{V\equiv 0}(0)}=\delta(y)$ translates for $\bar F_V$ as ${\bar F_V(0,y) \equiv 0}$ which is trivially translation-invariant along the $y$-direction. Showing the STS~\eqref{eq-STS-PFbarV} thus amounts to checking that this property is preserved in time through the evolution equation evolution~\eqref{eq-FeynmanKac-FbarV} for $\bar F_V$.
Consider for this purpose a `Galilean transformation' of ${\bar F_V(t,y)}$, which consists in defining ${\bar F^v_V(t,y)}$ through
\begin{equation}
  \bar F_V(t,y)
 \equiv \bar F^v_V(t,y-vt) \label{eq-defFbarV}
\end{equation}
where $v$ represents the `velocity' of the tilt $y\mapsto y+vt$. 
One sees directly from~\eqref{eq-FeynmanKac-FbarV} that the terms in $v$ in the equation of evolution for $\bar F^v_V(t,y)$ compensate between left and right hand side: $\bar F^v_V(t,y)$ verifies the \textit{same} equation of evolution as $\bar F_V(t,y)$, but in a tilted disorder $V^v(t,y)=V(t,y+vt)$.
Since the initial condition is also the same, one obtains:
\begin{equation}
  \bar F^v_V(t_1,y_1) =  \bar F_{V^v}(t_1,y_1) \label{eq-FbarvVvV}
\end{equation}
Choosing $v=y_1/t_1$ in~\eqref{eq-defFbarV} and~\eqref{eq-FbarvVvV} yields again~\eqref{eq-transl-invariance-FbarV}.

Another but less general incarnation of the STS arises when considering the geometrical fluctuations instead of the free-energy.
Defining the `generating function' of the moments of $y(t_1)$
\begin{equation}
  W_V^\lambda(t_1)   
 = \int_{y(0)=0} \mathcal{D}y(t) \, e^{-\frac 1T \mathcal{H} \argc{y,V;t_1}+\lambda y(t_1) }
\end{equation}
where the final-time condition is free (or equivalently, $y(t_1)$ is integrated over),
the disorder average of the variance of the endpoint $y(t_1)$ is given by:
\begin{equation}
 \overline{\moy{y(t_1)^2}_c}
 \equiv \overline{\langle y(t_1)^2\rangle-\langle y(t_1)\rangle^2}
 = \partial^2_\lambda{\vphantom{\int}}\Big|_{\lambda=0}\!\!\!\! \overline{\log W_V^\lambda(t_1)}
 \label{eq-fromWlambdatovariance}
\end{equation}
We note that performing an appropriate tilt encoding the thermal roughness,
namely ${y(t)=\bar y(t)+\lambda B_{\text{th}}(t)}$ one obtains for the generic function:
\begin{equation}
 W_V^{\lambda}(t_1)
 = e^{B_{\text{th}}(t_1) \lambda^2/2} \, W_{\tilde V}^{\lambda=0} (t_1)
\end{equation}
where
$\tilde V(t,y)\equiv V(t,y+\tfrac{\lambda T}{c}t)$ is a tilted disorder.
Taking the logarithm and averaging over disorder, one obtains
for~\eqref{eq-fromWlambdatovariance}:
\begin{equation}
 \overline{\moy{y(t_1)^2}_c}
 = \frac{T t_1}{ c} = B_\text{th}(t_1)
 \label{eq-STS-second-cumulant-disorder}
\end{equation}
and
${ \overline{\moy{y(t_1)^k}_c}= \partial^k_\lambda \big\vert_{\lambda=0} \overline{\log W_V^\lambda(t_1)} = 0 }$
for ${k>2}$.
It would be tempting to conclude from these averaged cumulants that the average of the whole distribution $\overline{P_V(t_1,y_1)}$ is a normal law ${\mathcal{N}(0,B_{\text{th}}(t_1))}$; this is only true in the trivial case without disorder where ${P_{V \equiv 0}(t,y)=P_{\text{th}}(t,y)=\mathcal{N}(0,B_{\text{th}}(t_1))}$.
Indeed the normalization at fixed disorder $\bar{W}_V(t_1)$ of definition \eqref{eq-def-PDF-yt} prevents a direct disorder average on \eqref{eq-pathintegral-unnorm-Boltzmann}.
As for the roughness $B(t)$, the property \eqref{eq-STS-second-cumulant-disorder} can be reformulated in an intrinsic way comparing the cumulants with respect to the thermal, disorder and the joint disorder-thermal distributions:
\begin{eqnarray}
 B(t)
 &\equiv & \overline{\moy{y(t)^2}}
 = \overline{\moy{y(t)^2}_c\vphantom{\moy{y(t)}^2}} + \overline{\moy{y(t)}^2}^c
 \label{eq-STS-roughness-1} \\
 B_{\text{dis}}(t)
 &\equiv & B(t)-B_\text{th}(t)=\overline{\langle y(t)\rangle^2}^c
 \label{eq-STS-roughness-2} 
\end{eqnarray}
Physically this decomposition allows to focus on the pure disorder contribution of the roughness ${B_{\text{dis}}(t)}$, which is equivalent to the full roughness at large `times'.
Such a simple relation is possible for the second cumulant thanks to the following generic identity valid in presence of two probability laws $p_1$ and $p_2$ (respectively thermal and disorder distribution in our case):
\begin{equation}
 \operatorname{Var}_{p_2 \circ \, p_1}
 = \mathbb{E}_{p_2} \, \circ \, \operatorname{Var}_{p_1} + \operatorname{Var}_{p_2}\,  \circ \, \mathbb{E}_{p_1}
\end{equation} 
%


\section{Derivation of the Feynman-Kac `time'-evolution equations} \label{A-FHHderivation}

We rederive in this appendix the `time'-evolution equation \eqref{eq-FeynmanKac-Wvnorm}
of the weight $Z_V(t,y)=\frac{W_V (t,y)}{\Wbar_{V\equiv 0}(t)}$, defined with respect to the pseudo free-energy ${F_V(t,y)}$ in \eqref{eq-def-free-energy-1}.
Given in Ref.~\cite{huse_henley_fisher_1985_PhysRevLett55_2924}, this evolution equation is the starting point of the Feynman-Kac equations of Sec.~\ref{section-FeynmanKac} which define univocally the pseudo free-energy quantities $F_V(t,y)$, $\bar{F}_V(t,y)$ and its random phase $\eta_V(t,y)$ with their \textit{ad hoc} initial conditions.
However, the normalization of the weight $Z_V(t,y)$ is not conserved in presence of disorder, so it requires a careful treatment in order to yield its continuous formulation \eqref{eq-FeynmanKac-Wvnorm}.

Using infinitesimal `propagators' in a path-integral formulation of the weight $Z_V$ \cite{feynman_1948_RevModPhys20_367},
we derive thereafter the evolution equation
first using its propagation equation in continuous time
and secondly constructing explicitly the weight with the propagators in discretized time.
We rederive this result for completeness, to pinpoint the required hypotheses and the possible issues in a generalization regarding the form of the elasticity.

\subsection{Propagation equation in continuous time}

We first define the following propagator (with ${\beta \equiv \frac{1}{T}}$):
\begin{align}
  Z_V(t_1,y_1|t_0,y_0)
  &= \int_{y(t_0)=y_0}^{y(t_1)=y_1} \widetilde {\mathcal D}y(t) \:
  e^{ -\beta \mathcal H[y,V;t_0,t_1]} \, \,
  \label{eq-def_ZVtwotimes}
\\
 \mathcal H[y,V;t_0,t_1]
 &= \int_{t_0}^{t_1} dt\:\Big[ \frac c2 (\partial_ty)^2+V(t,y(t))\Big]
 \label{eq-defH_ZVtwotimes}
\end{align}
which represents the weight of trajectories starting in~$y_0$ at time $t_0$ and ending in $y_1$ at time $t_1>t_0$.
Note that by definition $Z_V(t,y)=Z_V(t,y|0,0)$.
Paths $y(t)$ are weighted by a measure $\widetilde{\mathcal D}y(t)$ ensuring that for purely thermal paths the weight is normalized.
Explicitly, with the normalization $\Wbar_{V}(t)$  defined by \eqref{eq-WV-normalization}, the measure 
\begin{equation}
 \widetilde{\mathcal D}y(t)
 = \frac{{\mathcal D}y(t)}{\Wbar_{V\equiv 0}(t_1-t_0)}
 \label{eq-deftildeDyt}
 \end{equation}
in~\eqref{eq-def_ZVtwotimes} ensures that
 \begin{equation}
 \int dy_1\; Z_{V\equiv 0}(t_1,y_1|t_0,y_0)=1
 \label{eq-normZVt0t1}
\end{equation}
a property which is not true anymore for any $V$.  Here, $\int dy$ denotes $\int_{-\infty}^{+\infty} dy$ as in the rest of this appendix.
The advantage of using this choice of normalization is that $Z_V(t,y)$ obeys the so-called \textit{stochastic heat equation}~\eqref{eq-FeynmanKac-Wvnorm}, which can be shown using the Feynman-Kac formula.

From its definition, we see that the propagator~\eqref{eq-def_ZVtwotimes} presents several useful properties.
For $t_1\to t_0$ it goes to a Dirac delta:
\begin{equation}
  \lim_{t_1\to t_0} Z_V(t_1,y_1|t_0,y_0) = \delta(y_1-y_0)
  \label{eq-deltalimitZ}
\end{equation}
since the trajectory endpoint $y_1$ remains very close to its departure point at small times; to be more precise, we read from~\eqref{eq-def_ZVtwotimes} the expression of the infinitesimal propagator $\delta Z_V$, valid for $t_1$ close to $t_0$:
\begin{align}
  \delta Z_V(t_1&,y_1|t_0,y_0) \stackrel{(t_1\approx t_0)}{=} \nonumber \\
 &\frac 1{\sqrt{2\pi}}\sqrt{\frac{\beta c}{t_1-t_0}} 
e^{-\frac{\beta c}{2}\frac{(y_1-y_0)^2}{t_1-t_0} -\beta (t_1-t_0)V(t_1,y_1)}
  \label{eq-deltaZ}
\end{align}
The prefactor ensures the normalization condition~\eqref{eq-normZVt0t1} and also yields~\eqref{eq-deltalimitZ} in the limit $t_1\to t_0$.  It is rather important not to overlook this prefactor since it ensures that the infinitesimal propagator~\eqref{eq-deltaZ} evolves in time according to the forward and backward equations (already close to the final one on $Z_V$):
\begin{eqnarray}
 \partial_{t_1} \delta Z_V
 &=& + \argc{ \frac 1{2\beta c} \partial^2_{y_1} - \beta V(t_1,y_0)} \delta Z_V
  \label{eq-forward_evol_deltaZ}
\\
 \partial_{t_0} \delta Z_V
 &=& - \argc{\frac 1{2\beta c} \partial^2_{y_0} - \beta V(t_1,y_1)} \delta Z_V
  \label{eq-backward_evol_deltaZ}
\end{eqnarray}
Last, the non-infinitesimal time evolution is described by the propagation equation
\begin{equation}
 Z_V(t_1,y_1|t_0,y_0)
 = \int dy \, Z_V(t_1,y_1|t,y) Z_V(t,y|t_0,y_0)
 \label{eq-propgZeq}
\end{equation}
which expresses that the path integral~\eqref{eq-def_ZVtwotimes} can be cut at any time $t_0<t<t_1$ provided the intermediate values~$y$ of the path at time $t$ are integrated upon.
Using~\eqref{eq-deltalimitZ}, the limit $t\to t_1$ of the equation of propagation~\eqref{eq-propgZeq} yields a trivial identity.
To go further, we can differentiate~\eqref{eq-propgZeq} with respect to time $t$
and then use that for $t$ close to $t_1$ the infinitesimal propagator verifies the backwards evolution~\eqref{eq-backward_evol_deltaZ} in order to write
\begin{align}
   \int &d y\,   \Big[\frac 1{2\beta c} \partial^2_{y} - \beta V(t_1,y_1)\Big] \delta Z_V(t_1,y_1|t,y)\ Z_V(t,y|t_0,y_0)
\nonumber \\
 &=\int d y\ \delta Z_V(t_1,y_1|t,y) \partial_t Z_V(t,y|t_0,y_0)  
\end{align}
Integrating by parts and taking the limit $t\to t_1$ thanks to~\eqref{eq-deltalimitZ} finally yields the expected stochastic heat equation
\begin{equation}
 \partial_t Z_V(t,y|t_0,y_0) = \Big[ \frac 1{2\beta c} \partial^2_{y} - \beta V(t,y)\Big] Z_V(t,y|t_0,y_0) 
 \label{eq-SHE_propagator}
\end{equation}

Note that a propagator obeying such an equation of evolution cannot keep its normalization constant since in general
\begin{equation}
 \partial_{t}\!\int \! dy\, Z_V(t,y|t_0,y_0) = 
  -\beta \int \!dy\, V(t,y)  Z_V(t,y|t_0,y_0) 
\end{equation}
is non-zero.
We also remark that considering an elastic energy including higher powers of $\partial_ty$ than $(\partial_ty)^2$ in~\eqref{eq-defH_ZVtwotimes} would be problematic for finding the equation of evolution: the infinitesimal propagator~\eqref{eq-deltaZ}
would contain terms of the form $\frac{(y_1-y_0)^p}{(t_1-t_0)^{p-1}}$, making it not obvious to determine the equivalent of~(\ref{eq-forward_evol_deltaZ}-\ref{eq-backward_evol_deltaZ}) and deriving the evolution corresponding to~\eqref{eq-SHE_propagator}.

\subsection{Explicit propagator in discretized time}

A second approach to explicit the normalization properties of the path integral is to work in discretized time.
A path is going from ${y_\text{i}}$ to ${y_\text{f}} $ between time ${t_\text{i}} $ and ${t_\text{f}} $ in $N$ time steps $\delta t\equiv \frac{{t_\text{f}} -{t_\text{i}} }{N}$, so at times $t_k={t_\text{i}} +k\frac{{t_\text{f}} -{t_\text{i}} }{N}$.
We define the weight (or probability density) of a free path as
\begin{align}
 \mathcal P[y_0\ldots y_N] &= \left[\frac{\beta c}{2\pi\delta t}\right]^{\frac N2}
 \! \! \exp\bigg[\! -\beta\!\!\!\sum_{0\leq k<N} \! \! \! \delta t \frac c2 \Big(\frac{y_{k+1}-y_k}{\delta t}\Big)^2\bigg]
\nonumber
\\
 &= \prod_{0\leq k<N} g\big(\delta t,y_{k+1}-y_k\big)
 \label{eq-def_weight-trajectory}
\end{align}
where we have denoted the microscopic propagator by
\begin{equation}
g(t,y)=\sqrt{\frac{\beta c}{2\pi t}}\,e^{-\beta c \frac{y^2}{2t}}
\label{eq-thermal-micro-propagator}
\end{equation}
One defines the discrete equivalent in It\=o's discretization to the continuous propagator~\eqref{eq-def_ZVtwotimes} as
\begin{align}
  {\mathcal Z}_V^N({t_\text{f}},{y_\text{f}} |{t_\text{i}},{y_\text{i}} ) =
 \int & dy_0\ldots dy_N\ \mathcal P[y_0\ldots y_N] 
\nonumber
\\
  & \quad\times \delta(y_0-{y_\text{i}}) \delta(y_N-{y_\text{f}} ) 
  \label{eq-weighted_propagator_def}
\\
  &\quad\times \exp\bigg[-\beta\!\!\!\displaystyle\sum_{0\leq k<N} \delta t \:V(t_k,y_k)\bigg]
\nonumber
\end{align}
with the expectation that this result does not depend on $N$ in the large $N$ limit.  
The following decomposition makes the link with the continuum formulation~(\ref{eq-def_ZVtwotimes}-\ref{eq-deftildeDyt}):
\begin{small}
 \begin{align}
 & dy_0\ldots dy_N\  \mathcal P[y_0\ldots y_N] 
 \exp \argc{-\beta\!\!\!\!\sum\limits_{0\leq k<N} \!\!\! \delta t \frac c2 \big(\frac{y_{k+1}-y_k}{\delta t}\big)^2} \nonumber \\
 &= 
 \underbrace{
 \frac{dy_0\ldots dy_N}{\left[\frac{\beta c}{2\pi\delta t}\right]^{-\frac N2}}
 }_{\makebox{\footnotesize$\equiv \widetilde{\mathcal{D}}y(t)$}}
 \underbrace{
  \exp \arga{-\beta \delta t\!\!\!\!\sum\limits_{0\leq k<N} \!\!   \Big[ \frac c2 \big(\frac{y_{k+1}-y_k}{\delta t}\big)^2+V(t_k,y_k)\Big]}
 }_{\makebox{\footnotesize$\equiv e^{-\beta \mathcal H[y,V;t_0,t_1]}$}}
 \end{align}
\end{small}
At $V\equiv 0$ the normalization corresponding to~\eqref{eq-normZVt0t1} still holds: the writing~\eqref{eq-def_weight-trajectory} is a product of normalized probability densities.
The propagation equation \eqref{eq-propgZeq} is readily verified: for all intermediate times $t=t_\ell$ one has
\begin{small}
  \begin{align}
    & \int dy\, {\mathcal Z}_V^N({t_\text{f}},{y_\text{f}};t,y) {\mathcal Z}_V^N(t,y;{t_\text{i}},{y_\text{i}})
    \nonumber\\
    &= \int \!dy\int \!dy_0\ldots dy_{\ell}dy'_{\ell}\ldots dy_N\,
    \delta(y_0-{y_\text{i}})\delta(y_\ell-y)   \delta(y'_\ell-y)
    \nonumber \\ &\qquad\quad
    \times  \delta(y_N-{y_\text{f}})
    P[y_0\ldots y_N]\exp\bigg[-\beta\!\!\!\displaystyle\sum_{0\leq k<N}
    \delta t \:V(t_k,y_k)\bigg]
    \nonumber    \\
    &= \int dy_0\ldots dy_{\ell}\ldots dy_N\, \delta(y_0-{y_\text{i}})\delta(y_N-{y_\text{f}})
    \nonumber \\ &\qquad\quad
    \times P[y_0\ldots y_N]\exp\bigg[-\beta\!\!\!\displaystyle\sum_{0\leq k<N}
    \delta t \:V(t_k,y_k)\bigg]
    \nonumber \\
    &={\mathcal Z}_V^N({t_\text{f}},{y_\text{f}}|{t_\text{i}},{y_\text{i}} )
  \end{align}
\end{small}
\noindent The derivation of the stochastic heat equation can then be made explicit: fixing now ${y_\text{i}}$ and ${t_\text{i}} $ (and skipping them in the all following weights $Z_V(\dots|{t_\text{i}},{y_\text{i}})$), one compares two histories $(t_0,y_0; \ldots ; t_N,y_N)$ and $(t_0,y_0; \ldots ;t_N,y_N;t_{N+1},y_{N+1})$
\begin{align}
  {\mathcal Z}_V^{N+1}(t_{N+1},{y_\text{f}})=
  & \int dy_{N+1}\ldots dy_0 
  \delta(y_0-{y_\text{i}}) \delta(y_{N+1}-{y_\text{f}} ) 
  \nonumber \\ 
  & \times g(\delta t,y_{N+1}-y_N) e^{-\beta\delta t V(t_{N+1},y_{N+1})} 
  \nonumber \\ 
  &\times \mathcal P[y_0\ldots y_N]
  e^{-\beta \delta t\!\!\!\!\sum\limits_{0\leq k<N} \!\!\!   V(t_k,y_k)}
  \label{eq-evol_discreteZ_1timestep}
\end{align}
where we have isolated the contribution of the last time step.
Expanding the exponential $e^{-\beta\delta t V(t_{N+1},y_{N+1})}$ (which is valid at minimal order in $\delta t$), yields at order~$\delta t$:
\begin{align}
  {\mathcal Z}_V^{N+1}(&t_{N+1},{y_\text{f}})-{\mathcal Z}_V^N(t_{N},{y_\text{f}})\simeq
    -\delta t\beta V {\mathcal Z}_V^N(t_{N},{y_\text{f}})
\nonumber\\ &
 + 
  \int dy_{N+1}\ldots dy_0  \delta(y_0-{y_\text{i}}) \delta(y_{N+1}-{y_\text{f}} ) 
\nonumber\\ &
  \qquad\qquad \times\big[ g(\delta t,{y_\text{f}}-y_N)-\delta({y_\text{f}} -y_N)  \big] 
  \nonumber \\ &
  \qquad\qquad \times  \mathcal P[y_0\ldots y_N]
  e^{-\beta \delta t\!\!\!\!\sum\limits_{0\leq k<N} \!\!\!   V(t_k,y_k)}
\end{align}
where the third line reads
\begin{align}
 g({y_\text{f}}-y_N&,\delta t)-\delta({y_\text{f}} -y_N) \nonumber \\ &=
  g(\delta t,{y_\text{f}}-y_N)-g(0,{y_\text{f}}-y_N) \\ & 
  \approx  \delta t\, \partial_tg(\delta t,{y_\text{f}}-y_N)
\end{align}
Using now $ \partial_tg(\delta t,{y_\text{f}}-y_N)=\frac 1{2\beta  c}\partial^2_{{y_\text{f}}}g(\delta t,{y_\text{f}}-y_N)$ and, in the integral, $\partial^2_{{y_\text{f}}}=\partial^2_{y_{N+1}}$ and integrating by parts we get finally
\begin{align}
   &\frac{{\mathcal Z}_V^{N+1}(t_{N+1},{y_\text{f}})-{\mathcal Z}_V^N(t_{N},{y_\text{f}})}{\delta t}
   \approx
   \nonumber \\ & \qquad\qquad\qquad\qquad
   \Big[\frac 1{2\beta c} \partial^2_{{y_\text{f}} } - \beta V({t_\text{f}},{y_\text{f}})\Big] {\mathcal Z}_V^N(t_{N},{y_\text{f}})
\label{eq-discrete-SHE}
\end{align}
which corresponds to the continuum equation~\eqref{eq-SHE_propagator} in the limit $\delta t\to 0$.
Note moreover that from~\eqref{eq-evol_discreteZ_1timestep} the following exact recurrence equation for the propagator can be read:
\begin{align}
  {\mathcal Z}_V^{N+1}(&{t_\text{f}}+\delta t,{y_\text{f}})= \nonumber\\
  &\int \!dy\:
  g(\delta t,{y_\text{f}}-y) e^{-\beta\delta t V(t_{N},y)} 
  {\mathcal Z}_V^{N}({t_\text{f}},y)
  \label{eq-evol_discreteZ_recurrence}
\end{align}
and is the continuous analogue of the transfer matrix equation for a directed polymer constrained on a discretized lattice.

Note that similarly to the continuous case, the specific form of the short-range elasticity implies the Gaussian form of the microscopic propagator \eqref{eq-thermal-micro-propagator}, which thus satisfies the diffusion equation leading to the discrete stochastic heat equation \eqref{eq-discrete-SHE}.
However a different elasticity will in general not be Gaussian and consequently radically change its evolution equation.


\section{`Time'-evolution equations of averages using the functional It\=o formula} \label{A-flow_CbarRbar}

Aiming at the derivation the evolution equations for $\partial_t \bar{C}(t,y)$ and $\partial_t \bar{R}(t,y)$ given in Sec.~\ref{subsection-FK-Rbar-Cbar-Ito},
we first present in this appendix the functional It\=o formula \cite{vasiliev_1998_book-functional-QFT-StatPhys} applied to a field $X(t,y)$ obeying a generic Langevin equation, then we particularize it to the case of multi-point correlators, obtaining finally the flow equations \eqref{eq-noclose-Rbar}-\eqref{eq-noclose-Cbar}.

The Feynman-Kac evolution equations~\eqref{eq-FeynmanKac-FV}, \eqref{eq-FeynmanKac-FbarV} and \eqref{eq-FeynmanKac-etaV} of $F_V$, $\bar F_V$ and $\eta_V$ take the form of a generic Langevin equation for a field $X(t,y)$:
\begin{equation}
 \partial_{t} X(t,y) = {\mathcal G}[X(t,y);t,y]+ {\mathcal{V}}(t,y)
 \label{eq:Langevin_forFPfunct}
\end{equation}
where ${\mathcal{V}}(t,y)$ is a centered Gaussian noise with correlations 
\begin{equation}
\overline{{\mathcal{V}}(t,y) {\mathcal{V}}(t',y')}= D \delta(t'-t){\mathcal{R}}(y'-y)\label{eq:def_correl-mV}
\end{equation}
For $X(t,y)$ being $F_V(t,y)$, $\bar F_V(t,y)$ and $\eta_V(t,y)$ one reads $\mathcal G$ respectively as 
$\mathcal G[F]$ from~\eqref{eq-FP-KPZ-2}, 
$\bar {\mathcal G}[\bar F;t,y]$ from~\eqref{eq-FP-tiltedKPZ-2},
and $\mathcal G_\eta[\eta;t,y]$ from~\eqref{eq-FP-eta-2}.
For $F$ and $\bar F$ one has ${{\mathcal{R}}(y)=R_\xi(y)}$ while ${{\mathcal{R}}(y)=-R''_\xi(y)}$ for $X=\eta$ (note that this particular notation is specific only to this appendix).
In what follows one denotes ${\mathcal G}[X;t,y]$ for short instead of ${\mathcal G}[X(t,y);t,y]$.
Our aim is to deduce, from the Langevin equation~\eqref{eq:Langevin_forFPfunct}, evolution equations for the statistical average of functions of $\{X(t,y_i)\}$ at different points $\{y_i\}$, such as the two-point correlation function $\overline{X(t,y_1)X(t,y_2)}$.

This is fairly straightforward to obtain the time-derivative of $\overline{X(t,y)}$ by directly
averaging~\eqref{eq:Langevin_forFPfunct} which yields
$\partial_t \overline{X(t,y)} = \overline{{\mathcal G}[{X};t,y]}$ (\textit{e.g.} \eqref{eq-mean-etaV} and \eqref{eq-mean-FbarV}).
The same cannot be applied to $\overline{X(t,y)^2}$ since, when writing
\begin{align}
  \partial_t \overline{X(t,y)^2}
 & = 2 \overline{ X(t,y) \partial_t  X(t,y)}
\\ 
& =  2 \overline{X(t,y) {\mathcal G}[{X};t,y] } + 2 \overline{ X(t,y) {\mathcal{V}}(t,y)}
\label{eq:bad_ddt-meanX2}
\end{align}
one cannot easily eliminate ${\mathcal{V}}$ from the last term.
To tackle such correlation functions, one may use the functional It\=o formula which reads as follows, for the Langevin equation \eqref{eq:Langevin_forFPfunct} with continuous argument $y$ and non Dirac delta correlated random potential ${\mathcal{V}}$:
\begin{align}
  \partial_t \overline{g[X]}
  = \int& dy\: \overline{\mathcal G[X;t,y] \frac{\delta g[X]}{\delta X(y)}} 
\nonumber \\ &
  + \frac D2 \int dy dy'\: {\mathcal{R}}(y'-y) \overline{\frac{\delta^2 g[X]}{\delta X(y) \delta X(y')}}
 \label{eq:functionalItoformula}
\end{align}
where $g[X]$ is a functional of $X$.
For $g[X]={\mathcal{O}}(X(y_1))$ where $y_1$ is fixed (\textit{e.g.} an observable depending on the sole DP endpoint), one
thus finds
\begin{align}
  \partial_t \overline{ {\mathcal{O}}\big(X(t,y_1)\big) }
=\; &
 \overline{
 {\mathcal G}[{X};t,y_1]\partial_X {\mathcal{O}}(X(t,y_1)) 
 }
 \nonumber \\ &\qquad +
 \frac D2  {\mathcal{R}}(0)\, \overline{ \partial^2_X{\mathcal{O}}(X(t,y_1)) }
\label{eq:ddt_mean_O}
\end{align}
For ${\mathcal{O}}(X)=X^2$ one finds
\begin{equation}
 \partial_t \overline{ X(t,y_1)^2}
  =  2 \overline{{\mathcal G}[{X};t,y_1] X(t,y_1) }  + D {\mathcal{R}}(0)
  \label{eq:resdF2dt}
\end{equation}
which is the correct form of~\eqref{eq:bad_ddt-meanX2}. Note that the result is singular for $\xi\to 0$ in our cases of interest ${\mathcal{R}}=R_\xi$ and ${\mathcal{R}}=-R''_\xi$.

Another example is provided by the computation of the time evolution of the average of multiple-point correlators, for which~\eqref{eq:functionalItoformula} yields:
\begin{align}
   \partial_t \overline{ X(t,y_1)X(t,y_2)}
    = &\ \overline{
{\mathcal G}[{X};t,y_1] X(t,y_2)}
\label{eq:ddt-Xy1Xy2_langevin}
\\
&+
\overline{{\mathcal G}[{X};t,y_2] X(t,y_1)
}
+ D {\mathcal{R}}(y_2-y_1)
\nonumber 
\end{align}
which yields back~\eqref{eq:resdF2dt} for $y_1=y_2$.
More generally, one has, noting  $\partial_1{\mathcal{O}}$ (respectively $\partial_2{\mathcal{O}}$) the derivative of ${\mathcal{O}}$ with respect to its first (respectively second) argument:
\begin{align}
    \partial_t &\overline{ \mathcal{O} \big(X(t,y_1),X(t,y_2)\big) }
\nonumber\\    =&\,\overline{ 
      {\mathcal G}[{X};t,y_1] \partial_1 \mathcal
      O(\ldots)}
\nonumber + \overline{{\mathcal G}[{X};t,y_2] \partial_2 \mathcal
    O(\ldots) }
\nonumber\\
& + \frac D2 \Big[   {\mathcal{R}}(0)\partial_{11}+2  {\mathcal{R}}(y_2-y_1)\partial_{12}+  {\mathcal{R}}(0)\partial_{22}\Big] \overline{\mathcal{O} (\ldots)}
\label{eq:flowOX1X2}
\end{align}

We now derive the evolution equation for the correlator $\bar R(t,y)$ of $\eta_V(t,y)$. We first explicit some useful parity symmetry. The equation for $\eta_V(t,y)$ is the same as for $-\eta_{V^{\text{R}}}(t,-y)$ with a reflected disorder $V^R(t,y)=V(t,-y)$. This proves that
\begin{equation}
  \eta_V(t,y)=-\eta_{V^R}(t,-y)
\end{equation}
at all times. 
Since the distributions of $V$ and $V^R$ are the same, one can replace in averages \textsl{every} $\eta(t,-y)$ by $-\eta(t,y)$ without changing the result (so from now on we skip the index $V$). In other words, $\eta(t,y)$ is an odd
function of $y$ in distribution. We now define a three-point correlation function
\begin{equation}
  {\bar{R}}_3(t,y)=\overline{\eta(t,y)^2\eta(t,0)}
  \label{eq:defR3}
\end{equation}
To simplify the notations, and since one only considers one-time observables at time $t$, we now drop the dependence in $t$ and denote the derivation with respect to $y$ by a prime.
Using the noted parity (which also extend to the derivatives; \textit{e.g.}: $\eta'(y)$ is an even function of $y$ in
distribution
\footnote{This means that one can replace in averages {\textsl{every}} $\eta(-y)$ by $-\eta(y)$, simultaneously with
every $\eta'(-y)$ by $\eta'(y),\ldots$ every $\eta^{(k)}(-y)$ by $(-1)^{k+1} \eta^{(k)}(y),\ldots$ without changing the result.})
together with the statistical invariance by translation, one finds for instance, by translating all arguments by $-y$
\begin{align}
\!  {\bar{R}}''(y)&=\overline{\eta''(y)\eta(0)}
  \! \stackrel{\text{(tr.)}}{=} \!
  \overline{\eta''(0)\eta(-y)}
  \! \stackrel{\text{(par.)}}{=} \!
  \overline{\eta''(0)\eta(y)}\hspace{-1cm}
\label{eq:expressionsRbarprimeprime}
\\
  \frac 12 {\bar{R}}_3'(y)&=\overline{\eta'(y)\eta(y)\eta(0)}
  \stackrel{\text{(tr.)}}{=} 
  \overline{\eta'(0)\eta(0)\eta(-y)} ~~~~~~~~~~~~~~~~~~~~~
\nonumber
\\ & \!\!\!\!\phantom{.}\!\stackrel{\text{(par.)}}{=} 
  \overline{\eta'(0)\eta(0)\eta(y)}
\label{eq:expressionsRbar3prime}
\end{align}
We are now ready to determine the time-evolution of ${\bar{R}}(t,y)$ combining \eqref{eq:ddt-Xy1Xy2_langevin} and \eqref{eq-FP-eta-2}:
\begin{align}
  \partial_t {\bar{R}}(t,y)=\;&
  \partial_t \overline{\eta(0)\eta(y)}
\\
  \stackrel{\eqref{eq:ddt-Xy1Xy2_langevin}}{=}&
\frac{T}{2c}
\Big[\overline{\eta''(0)\eta(y)}+\overline{\eta''(y)\eta(0)}\Big]
\nonumber\\ &
-\frac 1c
\Big[\overline{\eta'(0)\eta(0)\eta(y)}+\overline{\eta'(y)\eta(y)\eta(0)}\Big]
\nonumber\\ &\qquad
-\frac 1t
\Big[2\overline{\eta(0)\eta(y)}+y\,\overline{\eta'(y)\eta(0)}\Big]
-D R''_\xi(y)
\nonumber
\end{align}
We eventually recognize thanks to~\eqref{eq:expressionsRbarprimeprime} and~\eqref{eq:expressionsRbar3prime} that
\begin{align}
\partial_t {\bar{R}}(t,y) =&\,
\frac Tc \partial_y^2 {\bar{R}}(t,y)
-\frac 1c \partial_y {\bar{R}}_3(t,y)
\nonumber \\ & 
-\!\frac 1 t \Big[{\bar{R}}(t,y)+\partial_y(y\,{\bar{R}}(t,y))\Big]
\!-\!DR''_\xi(y)
\label{eq:evoltRbarty}
\end{align}
This equation is valid at all times and would in principle allow to studying of the `flow' of ${\bar{R}}(t,y)$ starting from its initial condition ${\bar{R}}(0,y)\equiv 0$.  Due to the non-linear KPZ term, it is however non-closed on the two-point correlation function ${\bar{R}}(t,y)$ and brings into the game a three-point correlation function
${\bar{R}}_3(t,y)$, making it necessary to solve the full hierarchy of equations for the $n$-point functions to determine ${\bar{R}}(t,y)$.
Yet, using a scaling Ansatz in $y=0$, the equation~\eqref{eq:evoltRbarty} still enables us to determine the time evolution of the height of the ${\bar{R}}(t,y)$ in ${y=0}$ (see Sec.~\ref{section-Dtildeinfty-T-xi}).

Similarly, defining the three-point correlation function for $\bar F$
\begin{align}
 \bar{C}_3 (t,y) & \equiv -2 \overline{ \argc{\bar{F}(t,y)- \bar{F}(t,0)} \argc{\partial_y \bar{F} (t,0)}^2 }
\end{align}
one obtains the flow of $\bar C(t,y)$ using~\eqref{eq-FP-tiltedKPZ-2} in~\eqref{eq:flowOX1X2}
with ${\mathcal O(X_1,X_2)=(X_1-X_2)^2}$, ${X_1=\bar{F}(t,y)}$ and ${X_2=\bar{F}(t,0)}$:
\begin{equation}
\begin{split}
 \partial_t \bar{C}(t,y)
 =	& \frac{T}{c} \partial_y^2 \argc{\bar{C}(t,y) - \bar{C}(t,0)} - \frac{y}{t} \partial_y \bar{C}(t,y) \\
	&  - \frac{1}{c} \bar{C}_3 (t,y)  - 2D \argc{R_{\xi}(y) - R_{\xi}(0)}
\end{split}
\end{equation}


\section{Solution of the linearized dynamics of $\bar{F}$ for a generic disorder correlator ${R_{\xi}(y)}$} \label{A-short-time-dynamics-Fbar-generic}

In this appendix, we determine the explicit form of the correlator ${\bar{C}(t,y)}$ (resp. ${\bar{R}(t,y)}$) of the disorder free-energy ${\bar{F}}$ (resp. of the random phase ${\bar{\eta}}$), in the approximation where the `time'-evolution equation of those quantities is linearized. The crossover from finite to infinite `time' regime is discussed.

The evolution equations~\eqref{eq-noclose-Cbar} for ${\bar{C}(t,y)}$ and~\eqref{eq-noclose-Rbar} for
${\bar{R}(t,y)}$ 
are not closed because of the three-point correlation functions ${\bar{C}_3(t,y)}$ and ${\bar R_3(t,y)}$.
It is yet instructive to solve those equations in the approximation where those three-point functions are set to zero.
An equivalent alternative approach is to solve directly the equation~\eqref{eq-FeynmanKac-FbarV} for ${\bar{F}}$ or~\eqref{eq-FeynmanKac-etaV} for $\eta$ by neglecting again the non-linear terms in those equations (see also Appendix~C `Short-time dynamics (diffusive scaling)' of Ref.~\cite{agoritsas-2012-FHHpenta}, for the explicit case of Gaussian function for the microscopic disorder correlator ${R_{\xi}(y)}$).
We denote by ${\bar{C}^\text{lin}(t,y)}$ and ${\bar{R}^\text{lin}(t,y)}=\frac 12 \partial_y^2 {\bar{C}^\text{lin}(t,y)}$ their solutions, which are expected to be valid either at small `times' for all $y$ (because the initial condition ensures those functions vanish uniformly at `time' 0) or at all `times' but small $y$ in the high-temperature regime (see Sec.~\ref{section-Dtildeinfty-T-xi} for a discussion).

The equations at hand are linear and are thus solved \textit{e.g.} using Green functions,
and the solution takes the form
\begin{equation}
  \bar{C}^\text{lin}(t,y)=
  D\int dw\: K_t(y,w)R_\xi(w)
\end{equation}
where the kernel reads $ K_t(y,w)=\int_0^t ds K_s(y,w;t) $ with
\begin{align}
  K_s(y,w;t) &=
  \frac{\beta c t}{2\sqrt\pi}
  \frac 1{\sqrt{\beta c s t(t-s)}}
\\ & \times  \Big[
  2 e^{-\frac{\beta c t}{4}\frac{w^2}{s(t-s)}}
  - e^{-\frac{\beta c }{4}\frac{(tw-sy)^2}{ts(t-s)}}
  - e^{-\frac{\beta c }{4}\frac{(tw+sy)^2}{ts(t-s)}}
\Big]
\nonumber
\end{align}

The scaling analysis of this expression may be obtained by setting $s=t\tau$.
To this end, we assume the natural rescaling $R_\xi(a \bar y)=a^{-1}R_{\xi/a}(\bar y)$ of the disorder correlator.
One obtains that the free-energy correlator rescales \textit{purely diffusively} as
\begin{align}
  \bar{C}^\text{lin}(t,y)\  &= \ \frac{cD}T \: \sqrt{B_\text{th}(t)}\, \ \hat C_{\frac{\xi}{\sqrt{B_\text{th}(t)}}}\Big(\tfrac y{\sqrt{B_\text{th}(t)}}\Big)
  \label{eq:resClinscaling}
\end{align}
with as usual ${B_{\text{th}}(t)=\frac{Tt}{c}}$ and with the scaling function
\begin{align}
  \hat C_{\bar\xi}(\bar y)& = \int d\bar w\: \hat K(\bar y,\bar w) R_{\bar\xi}(\bar w)
\label{eq:generic-scaling-kernel-Chatxibarybar}
\\
  \hat K(\bar y,\bar w)&=\int_0^1 \frac{d\tau}{2\sqrt\pi}\: 
  \frac1{\sqrt{\tau(1-\tau)}}
\label{eq-defkernelKhat}
\\&\qquad\times \Big[
  2 e^{-\frac{\bar w^2}{4\tau(1-\tau)}}
  - e^{-\frac{(\bar w-\tau\bar y)^2}{4\tau(1-\tau)}}
  - e^{-\frac{(\bar w+\tau\bar y)^2}{4\tau(1-\tau)}}
\Big]
\nonumber
\end{align}
\begin{figure}
 \includegraphics[width=\columnwidth]{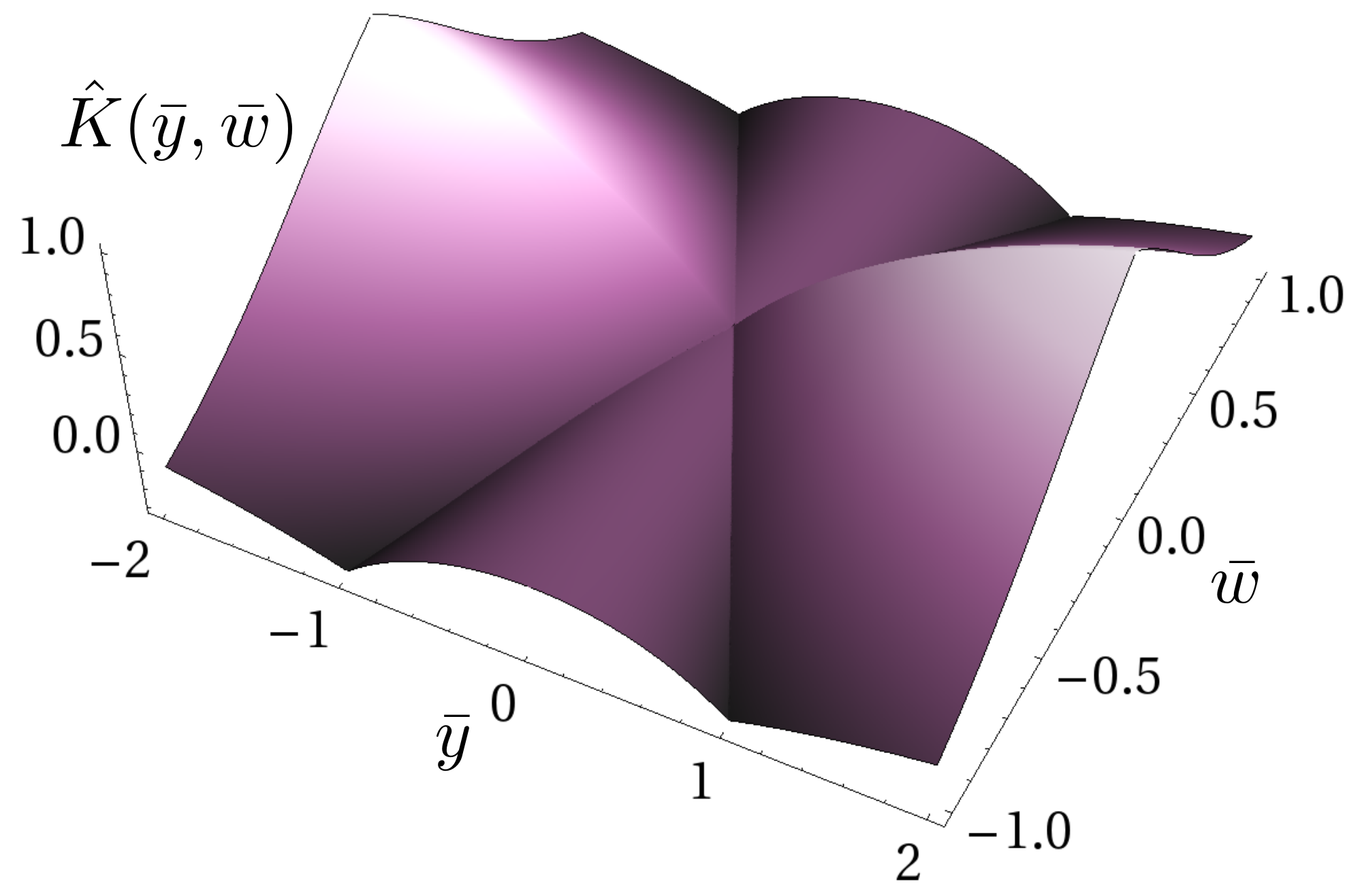}
 \caption{
 Kernel $\hat K(\bar{y},\bar{w})$ defined in~\eqref{eq-defkernelKhat} linking the linearized free-energy correlator ${\bar{C}^{\text{lin}}(t,y)}$ and the disorder correlator ${R_{\xi}(y)}$
 with the combination of \eqref{eq:resClinscaling}-\eqref{eq:generic-scaling-kernel-Chatxibarybar}.
 }
 \label{fig:kernelKhat}
\end{figure}%
%
Under this scaling, the whole $\{c,D,T\}$-dependence is absorbed in the prefactor $\frac{cD}T$.
The scaling kernel $\hat K(\bar y,\bar w)$, illustrated on Fig.~\ref{fig:kernelKhat} is continuous but non-analytical
on the lines $|\bar y|=|\bar w|$,
as can be seen from the direct computation of~\eqref{eq-defkernelKhat}, which, using the symmetry by even parity is
expressed for ${\bar y\geq 0}$ as:
\begin{small}
  \begin{align}
    \hat C_{\bar\xi}(\bar y)& \stackrel{(\bar{y}\geq 0)}= \int_0^{\bar{y}} d\bar w\:
    \hat K^<(\bar y,\bar w) R_{\bar\xi}(\bar w) 
    \label{eq-res-Cbarlin}
    \\& \qquad\qquad +
    \int_{\bar{y}}^\infty d\bar w\: \hat K^>(\bar y,\bar w) R_{\bar\xi}(\bar
    w) \nonumber
    \\
    \frac {\hat K^<(\bar y,\bar w)}{2\sqrt\pi}&= 1-\Erf \bar w-\frac
    {e^{\frac{\bar y^2}4}}2 \Big(2- \Erf\frac{\bar y}2-\Erf\frac{\bar
      y+2\bar w}2\Big) \nonumber
    \\
    \frac {\hat K^>(\bar y,\bar w)}{2\sqrt\pi}&= 1-\Erf \bar w-\frac
    {e^{\frac{\bar y^2}4}}2 \Big(2+ \Erf\frac{\bar y-2\bar
      w}2-\Erf\frac{\bar y+2\bar w}2\Big) \nonumber
  \end{align}%
\end{small}%
Those expressions describe through~\eqref{eq:resClinscaling} the complete transition from the initial regime where the correlator is close to zero to the infinite-time  asymptotic regime, where one should recover $\lim_{t\to\infty} \bar{R}^\text{lin}(t,y)=\frac {cD}T R_\xi(y)$ (see Sec.~\ref{A-FokkerPlanck-equation-xinonzero}).

As we now detail, this infinite-`time' limit is however not obvious to extract from~\eqref{eq-res-Cbarlin} and is in fact directly related to the non-analyticity of $\hat K(\bar y,\bar w)$.
Carefully integrating by part and differentiating with respect to ${\bar y}$ leads to the following expression, valid provided that $R_\xi(y)$ is bounded at infinity
\begin{align}
  \frac 12 \hat C'(\bar y)\stackrel{(\bar y\geq 0)}=&
  R_{\bar\xi}^{(-1)}(\bar y)
  +\int_0^\infty d\bar w\: \bar w e^{-\bar w(\bar w+\bar y)} R_{\bar\xi}^{(-1)}(\bar w)
 \nonumber \\ & -\int_{\bar y}^\infty d\bar w\: \bar w e^{-\bar w(\bar w-\bar y)} R_{\bar\xi}^{(-1)}(\bar w)
\label{eq:reshatCprime}
\end{align}
Here $R_{\bar\xi}^{(-1)}(\bar y)$ is the primitive of $R_{\bar\xi}(\bar y)$ which vanishes in~$0$.
It verifies the scaling relation ${R^{(-1)}_\xi(a\bar y)=R^{(-1)}_{\xi/a}(\bar y)}$ 
and its small $\bar\xi$ limit is half of the Heaviside step function ${\lim_{\bar\xi\to 0} R_{\bar\xi}^{(-1)}(\bar y) = \frac 12 \Theta(\bar y)}$, provided now that $R_{\bar\xi}(\bar y)$ describes a RB disorder.
\begin{figure}
 \includegraphics[width=\columnwidth]{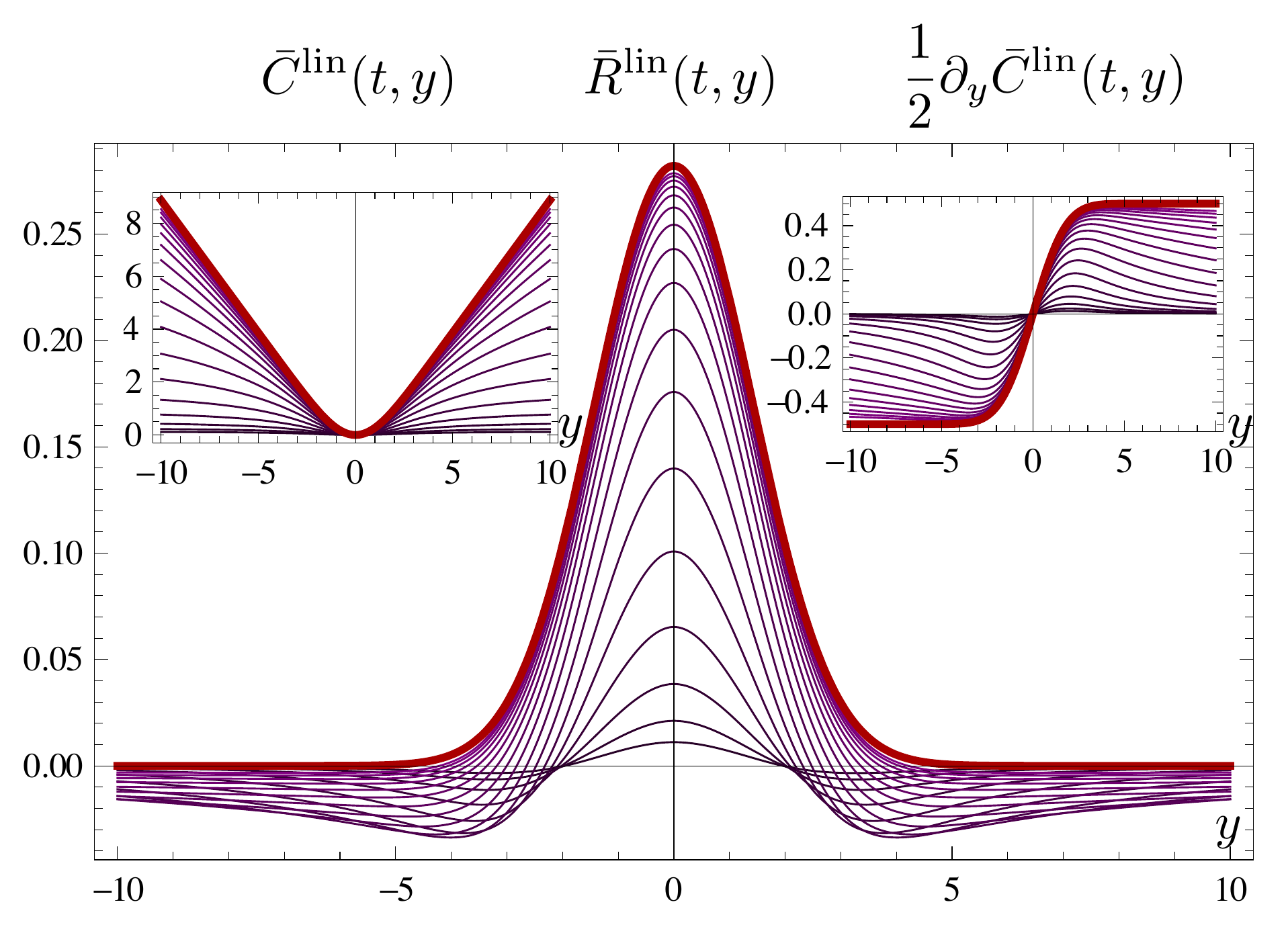}
 \caption{(Color online) The finite-`time' correlator $\bar{R}^\text{lin}(t,y)$ (thin purple lines) for a Gaussian disorder correlator ${R_{\xi}(y) = e^{-y^2/(4\xi^2)}/\sqrt{4 \pi \xi^2}}$, plotted as a function of $y$ for different `times', compared to its infinite-time limit $R_\xi(y)$ (thick red line), with ${\xi=1}$, ${c=1}$, ${D=1}$, and ${T=1}$.
 The central peak develops with increasing ‘times’ from the flat initial condition~${\bar{R}(0,y) \equiv 0}$.
   Larger `times' correspond to lighter colors.
   \textit{Left inset}: same behavior for $\bar{C}^\text{lin}(t,y)$.
   \textit{Right inset}: same behavior for 
$\frac 12 \partial_y \bar{C}^\text{lin}(t,y) = \int_0^y dy'\bar{R}^\text{lin}(t,y')$.
 }
 \label{fig:finitetimescaling-Cbarlin-Rbarlin}
\end{figure}%
Its occurrence as the first term of~\eqref{eq:reshatCprime} arises from the jump of the slope of $\hat K(\bar y,\bar w)$ in $\bar{y}=\bar{w}$, depicted in Fig.~\ref{fig:kernelKhat}.
Using those properties one obtains
  \begin{align}
\lim_{t\to\infty} \frac 12 &\hat C'_{\frac{\xi}{\sqrt{B_{\text{th}}(t)}}}\Big(\tfrac y{\sqrt{B_{\text{th}}(t)}}\Big)    
\nonumber
\\[-8mm] &=
R_{\xi}^{(-1)}( y)
+
\lim_{t\to\infty}
\bigg\{
 \overbrace{ 
\int_0^\infty \!\!\!\!\!d\bar w\: \bar w e^{-\bar w(\bar w+ y^t)} R_{\xi^t}^{(-1)}(\bar w)
 }^{\longrightarrow\:\frac 12 \int_0^\infty d\bar w\: \bar w e^{-\bar w^2 } \Theta(\bar w) }
\nonumber \\  & \qquad\qquad\qquad\qquad -
  \underbrace{ 
\int_{y^t}^\infty\!\!\!\!\! d\bar w\: \bar w e^{-\bar w(\bar w-y^t)} R_{\xi^t}^{(-1)}(\bar w)
 }_{\longrightarrow\:\frac 12 \int_0^\infty d\bar w\: \bar w e^{-\bar w^2} \Theta(\bar w) }
\bigg\}
\nonumber\\[-2mm]
&=
R_{\xi}^{(-1)}( y)
\end{align}
where we have denoted for short ${y^t=y/\sqrt{B_\text{th}(t)}}$ and
${\xi^t=\xi/\sqrt{B_\text{th}(t)}}$.  Differentiating with respect to~$y$, one
obtains the expected result $\lim_{t\to\infty}
\bar{R}^\text{lin}(t,y)=\frac {cD}T R_\xi(y)$, while keeping $t$
finite yields the
decomposition~(\ref{eq-decomposition-Rbarlin-1}-\ref{eq-decomposition-Rbarlin-2})
of ${\bar{R}(t,y)}$
announced in Sec.~\ref{section-DPtoymodel-A}.

The short to large-`time' behavior of the correlators is illustrated in Fig.~\ref{fig:finitetimescaling-Cbarlin-Rbarlin} for a Gaussian disorder correlator ${R_{\xi}(y) = e^{-y^2/(4\xi^2)}/\sqrt{4 \pi \xi^2}}$ as considered for the GVM computation in~Ref.~\cite{agoritsas_2010_PhysRevB_82_184207} whose predictions are recalled in~Appendix~\ref{A-GVM-PRB2010}.


\section{Fokker-Planck equations for the pseudo free-energy}
\label{A-FokkerPlanck-equation}

Starting from the path integral formulation of the pseudo free-energy, the average over thermal fluctuations yields the so-called `Feynman-Kac' equations for ${\partial_t W_V(t,y)}$, ${\partial_t \bar{F}_{\eta}(t,y)}$, ${\partial_t \eta_V (t,y)}$.
In this section, we reexamine the possible steady-state solutions of the Fokker-Planck equations, obtained after the disorder average over the random potential, and try to generalize them from the uncorrelated disorder (${\xi=0}$) to the case of a random-bond correlated disorder (${\xi>0}$ and short-range correlator).

\subsection{FP equations for $F_V$, $\bar{F}_V$ and $\eta_V$}
\label{A-FokkerPlanck-equation-def}

The pseudo free-energy ${F_V(t,y)}$ follows the KPZ equation \eqref{eq-FeynmanKac-FV}:
\begin{eqnarray}
 \partial_t F_V  (t,y)
 &=& \mathcal{G}\argc{F_V(t,y)}
	+ V(t,y)
 \label{eq-FP-KPZ-1} \\
 \mathcal{G}\argc{F}
 & \equiv & \frac{T}{2c} \partial_y^2 F (y)
	- \frac{1}{2c} \argc{\partial_y F(y)}^2
 \label{eq-FP-KPZ-2}
\end{eqnarray}
and similarly the disorder free-energy ${\bar{F}_V (t,y)}$ follows the \textit{tilted} KPZ equation \eqref{eq-FeynmanKac-FbarV} depending explicitly on $(t,y)$:
\begin{eqnarray}
 \partial_t \bar{F}_V  (t,y)
 &=& \bar{\mathcal{G}}\argc{\bar{F}_V(t,y);t,y}
	+ V(t,y)
 \label{eq-FP-tiltedKPZ-1} \\
 \bar{\mathcal{G}} \argc{\bar{F};t,y}
 & \equiv &  \mathcal{G}\argc{\bar{F}} - \frac{y}{t} \partial_y \bar{F}(y)
 \label{eq-FP-tiltedKPZ-2}
\end{eqnarray}
as its corresponding random phase ${\eta_V (t,y) = \partial_y \bar{F}_V (t,y)}$:
\begin{eqnarray}
 \partial_t \eta_V  (t,y)
 &=& \mathcal{G}_{\eta} \argc{\bar{\eta}_V(t,y);t,y}
	+ \partial_y V(t,y)
 \label{eq-FP-eta-1} \\
 \mathcal{G}_{\eta} \argc{\eta;t,y}
 & \equiv & \frac{T}{2c} \partial_y^2 \eta (y)
	- \frac{\partial_y \! \! \argc{ \eta (y)^2}}{2c}
	- \partial_y \! \! \argc{\frac{y}{t} \eta(y) }
 \label{eq-FP-eta-2}
\end{eqnarray}

Combining the Feynman-Kac equations \eqref{eq-FeynmanKac-FV}-\eqref{eq-FeynmanKac-etaV} and the random-potential disorder correlator:
\begin{equation}
 \overline{V(t,y) V(t',y')}
 = D \cdot \delta(t-t') \cdot R_{\xi} (y-y')
 \label{eq-FP-KPZ-3}
\end{equation}
the `time'-evolution of the free-energy distribution $\bar{\mathcal{P}}\argc{F,t}$ is then given by the functional FP equation (obtained \textit{e.g.} from It\={o}'s lemma):
\begin{equation}
\begin{split}
 \partial_t & \bar{\mathcal{P}}\argc{F,t}
 = \int dy \, \frac{\delta}{\delta F(y)} \arga{- \mathcal{G}\argc{F} \bar{\mathcal{P}}\argc{F,t}} \\ 
 & + \frac{D}{2} \int dy \, dy' \cdot R_{\xi} (y-y') \, \frac{\delta^2 \bar{\mathcal{P}}\argc{F,t}}{\delta F(y) \delta F(y')}
\end{split}
 \label{eq-FPequa-FV}
\end{equation}
with those two terms stemming respectively from the deterministic operator $\mathcal{G} $ and the remaining stochastic term in the Feynman-Kac equation.
The distribution $\bar{\mathcal{P}}\argc{\bar{F}_V,t}$ follows the same functional FP equation, with the tilt of the deterministic operator $\mathcal{G}\argc{F} \mapsto \bar{\mathcal{G}} \argc{\bar{F};t,y}$, whereas the random-phase counterpart satisfies:
\begin{equation}
\begin{split}
 \partial_t & \bar{\mathcal{P}}\argc{\eta,t}
 = \int dy \, \frac{\delta}{\delta \eta (y)} \arga{- \mathcal{G}_{\eta} \argc{\eta;t,y} \bar{\mathcal{P}}\argc{\eta,t}} \\ 
 & - \frac{D}{2} \int dy \, dy' \cdot R''_{\xi} (y-y') \, \frac{\delta^2 \bar{\mathcal{P}}\argc{\eta,t}}{\delta \eta(y) \delta \eta(y')}
\end{split}
 \label{eq-FPequa-eta}
\end{equation}
The detailed notations $F_V (t,y)$, $\bar{F}_V (t,y)$ and $\eta_V (t,y)$ have been simplified to the random functions $F(y)$, $\bar{F}(y)$ ad $\eta(y)$ of respective distributions  $\bar{\mathcal{P}} \argc{F,t}$,  $\bar{\mathcal{P}} \argc{\bar{F},t}$ and  $\bar{\mathcal{P}} \argc{\eta,t}$ at a fixed `time' $t$,
though their random nature is initially stemming from the microscopic random potential $V$.

\subsection{Steady-state solution at ${\xi=0}$}
\label{A-FokkerPlanck-equation-xi0}

For an uncorrelated disorder, the disorder correlator reduces to a normalized Dirac $\delta$-function
$R_{\xi=0}(y) = \delta(y)$ and the Gaussian distribution
\begin{equation}
 \bar{\mathcal{P}}_{\text{st}} \argc{F}
 \propto \exp \arga{- \frac{\lambda}{2} \int dy \argc{\partial_y F(y)}^2}
 \label{equa-steadystate-F}
\end{equation}
is a steady-state solution of \eqref{eq-FPequa-FV}, \textit{\textit{i.e.}} $\partial_t  \bar{\mathcal{P}}_{\text{st}} \argc{F}=0$, provided that ${\lambda^{-1}=\frac{cD}{T}}$ (see Ref.~\cite{huse_henley_fisher_1985_PhysRevLett55_2924}).
This condition comes solely from the counterbalance of the diffusive term $\frac{T}{2c} \partial_y^2 F(y)$ and the stochastic term $\frac{D}{2} \frac{\delta \bar{\mathcal{P}}\argc{F}}{\delta F(y)}$.
As for the contribution of the non-linear KPZ term $-\frac{1}{2c} \argc{\partial_y F(y)}^2$, it disappears~\cite{krug_1997_AdvPhys_46_139} under the boundary condition ${F'(y)\vert_{y \pm \infty}=0}$.

There is no `time' dependence in the FP equation \eqref{eq-FP-KPZ-1}, but the Gaussian PDF \eqref{equa-steadystate-F} is a steady-state solution only at ${t=\infty}$ for our physical definition of the pseudo free-energy \eqref{eq-def-free-energy-1} which satisfies
\begin{equation}
\partial_y F_V (t,y) = \partial_y F_{\text{th}}(t,y) + \partial_y \bar{F}_V (t,y) = c\frac{y}{t} + \eta_V (t,y)
\end{equation}
and it then becomes equivalent to the two normalized Gaussian PDF:
\begin{eqnarray}
 \bar{\mathcal{P}}^{0}_{\text{G}} \argc{\bar{F}}
 & = & \frac{1}{C_1(\lambda)} \exp \arga{- \frac{\lambda}{2} \int dy \argc{\partial_y \bar{F}(y)}^2}
 \label{eq-steadystate-Fbar} \\
 \bar{\mathcal{P}}^{0}_{\text{G}} \argc{\eta}
 & = & \frac{1}{C_2(\lambda)} \exp \arga{- \frac{\lambda}{2}  \int dy \argc{\eta (y)}^2}
 \label{eq-steadystate-eta}
\end{eqnarray}
with $C_2(\lambda)$ defined by $\int \mathcal{D}\eta(y) \cdot  \bar{\mathcal{P}}^{0}_{\text{G}} \argc{\eta} =1$ and similarly for $C_1(\lambda)$.
Choosing $\lambda^{-1}=\frac{cD}{T}$ and the boundary condition ${\bar{F}'(y)\vert_{y \pm \infty}=\eta(y)\vert_{y \pm \infty}=0}$, the FP equation eventually yields for the Gaussian distributions:
\begin{eqnarray}
 & \partial_t \bar{\mathcal{P}}^{0}_{\text{G}} \argc{\bar{F}}
 =& \frac{1}{t} \bar{\mathcal{P}}^{0}_{\text{G}} \argc{\bar{F}} \cdot \log \arga{C_1 (\lambda) \bar{\mathcal{P}}^{0}_{\text{G}} \argc{\bar{F}}}
  \label{eq-FP-Gauss-Fbar}
 \\ & \partial_t \bar{\mathcal{P}}^{0}_{\text{G}} \argc{\eta}
 =& \frac{1}{t} \bar{\mathcal{P}}^{0}_{\text{G}} \argc{\eta}
 \cdot \arga{\int dy \, \delta (0) + \log \arga{C_1 (\lambda) \bar{\mathcal{P}}^{0}_{\text{G}} \argc{\eta}}}
  \label{eq-FP-Gauss-eta} \nonumber
\end{eqnarray}
the equivalent equation for $\partial_t \bar{\mathcal{P}}^{0}_{\text{G}} \argc{\eta}$ being apparently ill-defined in the functional derivative framework, due to an additive divergent constant $\delta(0)$.
So the Gaussian distributions \eqref{eq-steadystate-Fbar} and \eqref{eq-steadystate-eta} (with ${\xi=0}$) become a steady-state solution only at infinite `time' in order to cancel the whole contribution \eqref{eq-FP-Gauss-Fbar}, since the value of $\lambda$ has been fixed as a `time'-independent constant.

Actually the chosen boundary condition is the only one physically possible at all `times':
\begin{equation}
 \partial_y \bar{F}_V (t,y) \vert_{y \pm \infty} = \eta_V (t,y)\vert_{y \pm \infty}  = 0
 \label{eq-BC-yinfy}
\end{equation}
since by construction $\bar{\mathcal{P}}^{0}_{\text{G}} \argc{\bar{F}}$ and $\bar{\mathcal{P}}^{0}_{\text{G}} \argc{\eta}$ penalize the functions $\bar{F}(y)$ and $\eta(y)$ whose fluctuations extend too much in the $y$-direction. The boundary condition at ${y = \pm \infty}$ is thus free from that point of view but should also be compatible with ${\overline{\eta_V(t,y)}=0}$ \eqref{eq-mean-etaV}.

\subsection{Steady-state solution of the linearized FP equation at ${\xi>0}$}
\label{A-FokkerPlanck-equation-xinonzero}

The Gaussian PDF \eqref{eq-steadystate-Fbar}-\eqref{eq-steadystate-eta} can be generalized with the introduction of the correlator $\bar{R}^{-1}(t,y)$ whose functional inverse is defined by \eqref{eq-def-corr-etaeta}:
\begin{align}
 \bar{\mathcal{P}}_{\text{G}} \argc{\bar{F},t}
 & \propto e^{- \frac{1}{2} \int dy \, dy' \bar{F}'(y) \bar{R}^{-1}(t,\valabs{y-y'}) \bar{F}'(y') }
 \label{eq-GaussianPDF-Fbar} \\
 \bar{\mathcal{P}}_{\text{G}} \argc{\eta,t}
 & \propto e^{- \frac{1}{2} \int dy \, dy'  \eta(y) \bar{R}^{-1}(t,\valabs{y-y'}) \eta(y') }
 \label{eq-GaussianPDF-eta}
\end{align}
with the proper `time'-dependent normalization of PDF such that
$\int \mathcal{D}\bar{F}(y) \cdot  \bar{\mathcal{P}}_{\text{G}} \argc{\bar{F},t}
= \int \mathcal{D}\eta(y) \cdot  \bar{\mathcal{P}}_{\text{G}} \argc{\eta,t} =1$.

These expressions actually describe correctly the distributions corresponding to the \textit{linearized} problem, namely to the equations~(\ref{eq-FP-KPZ-1}-\ref{eq-FP-eta-2}) where the KPZ quadratic contributions to the functionals $\mathcal G$ are set to 0. Indeed, the solution in the fields ${F_V}$, ${\bar F_V}$ and ${\eta_V}$ of those equations is linear in the disorder potential $V(t,y)$, whose distribution is Gaussian, implying that the distributions of the fields are themselves Gaussian. We refer the reader to~Appendix~\ref{A-short-time-dynamics-Fbar-generic} for motivations to study the linearized problem and for a solution leading to the two-point correlator ${\bar R^\text{lin}(t,y)}$ given in~(\ref{eq-decomposition-Rbarlin-1}-\ref{eq-decomposition-Rbarlin-2}), and actually providing the full `time'-dependent distributions through~(\ref{eq-GaussianPDF-Fbar}-\ref{eq-GaussianPDF-eta}) with ${\bar{R}=\bar R^\text{lin}}$.

At $\xi=0$ the solution of the full problems (\ref{eq-FP-KPZ-1}-\ref{eq-FP-eta-2}) is not Gaussian at finite `time', as known from the
exact solutions~\cite{calabrese_2010_EPL90_20002,amir_arXiv:1003.0443,sasamoto_2010_NuclPhysB_834_523,dotsenko_2010_EPL90_20003} but
becomes Gaussian at infinite `time', as detailed in the previous subsection.
At $\xi>0$ this last property does however not hold anymore. Indeed, anticipating slightly on Sec.~\ref{section-Dtildeinfty-T-xi}, if the steady distribution of $\eta$ were Gaussian then the three-point function ${\bar R_3(t,0)}$ would be zero by parity in the field $\eta$, and from~\eqref{eq:evoltRbart0bis} the infinite-`time' limit of the height ${\bar R(t,0)}$ of the correlator would be the same in the linearized and in the original problem, which is not true.

Another way to illustrate this fact is to try to solve the FP equation~\eqref{eq-FPequa-FV} inserting the Gaussian Ansatz~\eqref{eq-GaussianPDF-Fbar}. 
Using the fast decay of $R(y)$ and $R'(y)$ at large $y$ for the vanishing of boundary terms, we obtain from the right hand side of~\eqref{eq-FPequa-FV}
\begin{small}
\begin{equation}
\begin{split}
 & \frac{\partial_t \bar{\mathcal{P}}_{\text{G}} \argc{\bar{F},t}}{\bar{\mathcal{P}}_{\text{G}} \argc{\bar{F},t} }
 = \frac{1}{2} \int \! dy \, d\tilde{y} \bar{F}''(\tilde{y}) \bar{R}^{-1}(t,\valabs{\tilde{y}-y}) \\
 & \quad \times \arga{D \int dy' \mathcal{F}(t,\valabs{y-y'}) \bar{F}''(y') - 2 \bar{\mathcal{G}}\argc{\bar{F};t,y} } \\
 & \quad \quad \quad \quad \quad \quad - \frac{T}{2c} \int dy \, dy' \cdot \delta(y-y') \delta''(y-y')
\end{split}
\label{eq-GaussFP-general}
\end{equation}
\end{small}
with the definition:
\begin{equation}
 \mathcal{F}(t,y_1-y_2) \equiv \int dy_3 R_\xi(\valabs{y_1-y_3}) \bar{R}^{-1} (t,\valabs{y_2-y_3})
 \label{eq-def-convolutionF}
\end{equation}
On the other hand, the left hand side of~\eqref{eq-FPequa-FV} yields, differentiating with respect to `time':
\begin{small}
\begin{equation}
 \frac{\partial_t \bar{\mathcal{P}}_{\text{G}} \argc{\bar{F},t}}{\bar{\mathcal{P}}_{\text{G}} \argc{\bar{F},t} }
 =  - \frac{1}{2} \int \! dy \, dy' \bar{F}'(y) \partial_t \bar{R}^{-1}(t,\valabs{y-y'}) \bar{F}'(y')
\label{eq-GaussFP-general_time-derivative}
\end{equation}
\end{small}
Considering first the linearized case, the identification of~\eqref{eq-GaussFP-general} and~\eqref{eq-GaussFP-general_time-derivative}
yields, upon appropriate integrations by part, an equation of the form
\begin{equation}
  \int dy\,dy' \bar{F}'(y') M(t;y',y) \bar{F}'(y) =0
  \label{eq-FP-Gauss-functional}
\end{equation}
where $M(t;y',y)=M(t;y'-y)$ is a translation-invariant symmetric functional operator which combines $\bar R$ and $R$.
Since~\eqref{eq-FP-Gauss-functional} is valid for any function $\bar F'$ decaying fast enough at infinity,
solving this equation amounts to canceling the operator $M(t;y',y)$. After some manipulations aiming at casting the functional equation ${M(t;y',y)=0}$ into a diagonal form, the linearized form of the `flow' equation~\eqref{eq-noclose-Rbar} on ${\bar{R}(t,y)}$ is precisely recovered, namely
\begin{align}
 \partial_t \bar{R}(t,y)
 =	& \frac{T}{c} \partial_y^2 \bar{R}(t,y) - \frac{1}{t} \arga{\bar{R}(t,y) + \partial_y \argc{y \bar{R}(t,y)}} \nonumber \\ 
 	&  \quad - D R_{\xi}''(y) 
\label{eq-noclose-Rbar_linearized}
\end{align}
In the process, the divergent part on the last line of~\eqref{eq-GaussFP-general} was discarded.
This flow equation was obtained in Appendix~\ref{A-flow_CbarRbar} using It\=o's lemma, including the non-linearized case, and without having singular terms to discard (which we attribute to an artefact of functional calculus in the computation above).
The infinite-`time' steady-state solution thus verifies: ${\frac{T}{c} \partial_y^2 \bar{R}(\infty,y)- D R_{\xi}''(y)}=0$ which implies directly the expected result ${\bar{R}(\infty,y)= \frac{cD}{T} R_{\xi}(y)}$. It ensures that ${\mathcal F (t,y)\to (\frac{cD}{T})^{-1}\delta(y)}$ in~\eqref{eq-def-convolutionF} as $t$ goes to infinity, which actually prevents the divergent term to appear in~\eqref{eq-GaussFP-general}. The finite-`time' solution is studied in Appendix~\ref{A-short-time-dynamics-Fbar-generic}.

Taking however the corresponding steady-state distribution~\eqref{eq-GaussianPDF-Fbar} with ${\bar R(t,y)=\frac{cD}{T} R_{\xi}(y)}$ as trial steady solution for the full equation~\eqref{eq-FPequa-FV} yields a remaining term, arising from the non-linearity, cubic in ${\bar F}$ (\textit{i.e.} not of the form~\eqref{eq-FP-Gauss-functional}), which vanishes only at $\xi=0$.


\section{Scaling laws for a temperature-independent elastic weight} \label{A-saddle_scalings-maths}

The model defined in~Sec.~\ref{section-def-DES} depends on the four independent parameters $\arga{c,D,T,\xi}$, with the elastic constant $c$ being fixed independently from the temperature $T$ in the parametrization describing the elastic interface. In the language of the directed polymer, the elastic weight
${e^{-\frac 1T\int_0^{t_1} dt \, \frac{c}{2} \argp{\partial_t y(t)}^2 }}$
of a trajectory consequently depends explicitly on $T$,
contrarily to an alternative convention often used in the mathematics litterature which amounts to choose ${c=T}$.
The consequences on the scaling arguments of~Sec.~\ref{section-DPtoymodel-C} and the {low-$T$} saddle-point arguments of Sec.~\ref{section-DPtoymodel-D} are discussed in this appendix.

One is interested in the generic scaling of the prefactor of the roughness $B(t;c,D,T,\xi)$ in the random-manifold regime (of roughness exponent $\zeta_{\text{RM}}=\frac 23$):
\begin{equation}
 B(t;c,D,T,\xi)
 \stackrel{t\to\infty}{\sim}
 A_{\text{RM}}(c,D,T,\xi) t^{2\zeta_{\text{RM}}}
\end{equation}
The scaling in temperature of the prefactor is described by the \textit{thorn} exponent $\tho$ defined by $A(c,D,T,\xi) \sim T^{2\tho} $.
We have derived in section~\ref{section-DPtoymodel-C} from a scaling analysis that, depending on the temperature regime with respect to ${T_c=(\xi c D)^{1/3}}$, the expressions of the prefactor ${A_{\text{RM}}(c,D,T,\xi)}$ are
\begin{equation}
 A_{\text{RM}}
\stackrel{(T \gg T_c)}{=}
 \argp{\frac{D}{cT}}^{\frac 23}
\qquad
 A_{\text{RM}}
\stackrel{(T \ll T_c)}{=}
 \argp{\frac{D^2}{c^4\xi}}^{\frac 29}
\end{equation}
While
$ \tho_{\text{RM}}^{T \gg T_c}=-\frac 13 $
at high temperatures, the existence of the \textit{microscopic} length $\xi>0$ alters the value of the thorn exponent to
$ \tho_{\text{RM}}^{T \ll T_c}=0 $
at low temperatures, even though this exponent describes \textit{large-scale} properties of the polymer.

In this appendix, we determine how those exponents change when taking the particular convention $c=T$, often chosen in the mathematics community --~prompting us to denote by a subscript `$\mathfrak m$' the observables
defined with this convention \textit{e.g.}
\begin{equation}
 B^{\mathfrak m}(t;D,T,\xi)
 \stackrel{t\to\infty}{\sim} A_{\text{RM}}^{\mathfrak m}(D,T,\xi) t^{2\zeta_{\text{RM}}}
\end{equation}
Physically, the choice ${c=T}$ amounts to render the elastic weight temperature-independent, \eqref{eq-def-unnorm-Boltzmann-Wv} becoming
\begin{equation}
\!  W^{\mathfrak m}_V(t_1,y_1) = \!\int_{y(0)=0}^{y(t_1)=y_1}
  \!\!\!\!\!\!\!\!\!\mathcal Dy\,
  e^{\mbox{\scriptsize$\displaystyle -\!\int_{0}^{t_1} \!\!\!dt\Big[ \frac{(\partial_ty)^2}2+\frac 1T V(t,y(t))\Big]$}}
  \label{eq-Wy1t1_math}
\end{equation}
Here $\frac 1T$ only tunes the relative importance of disorder with respect to elasticity.  The parametrization $c=T$ also arises in the continuum limit of the discrete simple (solid-on-solid) SOS directed polymer model~\cite{bustingorry_2010_PhysRevB82_140201,agoritsas-2012-FHHpenta} and is thus of interest to analyze numerical results of this system.
There are other possible parametrizations depending on the physical model described by the KPZ equation. Another example is provided in Sec.~\ref{section-discussion-exp-C}, for which the regime where the disorder correlations matter is a high-velocity regime.

Before handling the different limits with respect to $T$ of~\eqref{eq-Wy1t1_math} in a functional integral saddle-point approach similar to that of section~\ref{section-DPtoymodel-D}, we first recall some results on the asymptotics of integrals with one variable.
The aim is to determine a (logarithmic) equivalent at large $p$ of integrals of the form
$
 I(p) = \int dy\ f(y) e^{-pg(y)}    
$.
The following result holds: if $g(y)$ has a unique, finite, minimum value reached in $y^\star $, then
\begin{equation}
 I(p) \stackrel{p\to\infty}{\sim} f(y^\star) e^{-pg(y^\star)}   \label{eq-saddle_1variable}
\end{equation}
Here $y^\star$ is the point (or one point) where the minimum of $g(y)$ is reached, and is thus by definition independent of $p$.
Powerlaw corrections in $p$ may arise from the integration of fluctuations around $y^\star$, but they disappear \textit{e.g.} in ratios of the following form (see~\eqref{eq-explicitBsaddle_lowT} and \eqref{eq-rescaling_B_zeta23} for the DP):
\begin{equation}
 \frac{\int dy\ f(y) e^{-pg(y)}}{\int dy\ e^{-pg(y)}} \stackrel{p\to\infty}{\sim} 
 \frac{f(y^\star) e^{-pg(y^\star)}}{e^{-pg(y^\star)}} = f(y^\star)
\end{equation}
The existence of the finite minimum is crucial, as illustrated from the derivation of Stirling's formula for the equivalent of the factorial. Starting from
\begin{equation}
  p! = \int_{\mathbb R^+} dy\ e^{-y} y^p = \int_{\mathbb R^+} dy\ e^{-y} e^{p \log y} \label{eq-factorialintegral}
\end{equation}
one may be tempted to apply~\eqref{eq-saddle_1variable} with
\begin{equation}
  f(y)= e^{-y}  \qquad g(y)= - \log y   \label{eq-def_fg_saddle_factorial}
\end{equation}
which, assuming blindly that $ g(y)$ reaches a \textit{finite} minimum in $y^\star$ would yield the wrong result ${p! \ \stackrel{p\to\infty}{\sim}\ e^{-y^\star} e^{p \log y^\star}}$.
The loophole here is that ${g(y)= - \log y}$ reaches no finite minimum on $\mathbb R^+$. On this simple example, the clue is to rescale $y$ by a factor $p$ ($y=p\bar y$) and write instead of~\eqref{eq-factorialintegral}
\begin{equation}
  p! = p^{p+1} \int_{\mathbb R^+} d\bar y \ e^{-p \bar y} \bar y^p =  p^{p+1} \int_{\mathbb R^+} d\bar y\ e^{-p(\bar y- \log \bar y)}
  \label{eq-rescaling_saddle_factorial}
\end{equation}
now with $f(\bar y)=1$ and $g(\bar y)= \bar y - \log \bar y$ which is minimal in $\bar y^\star=1$ one obtains correctly
\footnote{Again, the equivalent is logarithmic:
$ \log \frac{p!}{p^{p+1}} \stackrel{p\to\infty}{\sim} - p $. The real equivalent
$p! \stackrel{p\to\infty}{\sim} \sqrt{2\pi p}\,(p/e)^p$ is obtained after integration of fluctuations around the saddle.}
$ p! \stackrel{p\to\infty}{\sim} p^{p+1} e^{-p}$.
Note that, coming back to the initial variable $y$, we see that the optimum $y^\star$ of $g(y)$ in~\eqref{eq-def_fg_saddle_factorial} was not finite but diverging to infinity as $y^\star=p\bar y^\star=p$ for $p\to\infty$.
In other words, the rescaling $y=p\bar y$ in~\eqref{eq-rescaling_saddle_factorial} allows to find the optimal
$y$ at the correct scale in the large parameter~$p$.

Consider first the well-controlled {high-temperature} regime. Since the rescaling of section~\ref{section-DPtoymodel-C} is at $c=T=1$, it is compatible with the mathematician's convention and one can export directly \eqref{eq-rescalingsB-for-highT}-\eqref{eq-rescaledB-for-highT} imposing ${c=T}$
\begin{eqnarray}
&  B^{\mathfrak m}(t;D,T,\xi)= 
  \xi^{\mathfrak m}_{\text{th}}(T)^2 B \big(\frac{t}{t^{\mathfrak m}_*(T)};1,1,\frac{\xi}{\xi^{\mathfrak m}_{\text{th}}(T)}\big)
  \label{eq-rescalingB_math_for-lowT}
&\\
&t^{\mathfrak m}_*(T)
 = \frac{T^4}{D^2}
 \, , \;
 \xi^{\mathfrak m}_{\text{th}}(T)
 = \frac{T^2}{D}&
\end{eqnarray}
The regime $\xi\ll\xi^{\mathfrak m}_{\text{th}}(T)$, or equivalently ${T\gg T_c^{\mathfrak m}}$, with ${T_c^{\mathfrak m}=\sqrt{\xi D}}$ describes the high-temperature limit and consists in replacing
$\frac{\xi}{\xi^{\mathfrak m}_{\text{th}}(T)}$ by $0$ in~\eqref{eq-rescalingB_math_for-lowT}.
In the large-time limit ($t\gg t^{\mathfrak m}_*(T)$), this yields
$
 A^{\mathfrak m}_{\text{RM}}(D,T,\xi)|_{(T \gg T^{\mathfrak m}_c)}{=}
 (DT^{-2})^{\frac 23}
$
and the high-temperature thorn exponent is thus
$ \tho^{\mathfrak m}_{\text{RM}}\stackrel{(T \gg T_c)}=-\frac 23 $.

The {low-temperature} regime is however less direct to handle, since the rescaling~(\ref{eq-rescalingsB-for-lowT}-\ref{eq-rescaledB-for-lowT}) of section~\ref{section-DPtoymodel-C} is not at $c=T$ and thus cannot
be directly exported to the mathematician's convention.
Anyway as first choice the rescaling ${a=\xi, b=\xi^2}$ allows us to rescale at $\xi=1$ and to respect the mathematician's
convention: the elastic term is unchanged ($c=T$) in the weight
\begin{equation}
\!  W^{\mathfrak m}_V(t_1,y_1)   \stackrel{(d)}{=} 
 \!\int_{\bar{y}(0)=0}^{\bar{y}(t_1)=y_1/\xi}
  \!\!\!\!\!\!\!\!\!\mathcal D\bar{y} \,
  e^{\mbox{\scriptsize$\displaystyle -\!\int_{0}^{t_1/\xi^2} \!\!\!\!dt\Big[ \frac{(\partial_t \bar{y})^2}2+\frac{T_c^{\mathfrak m}}{T} V_1(t,\bar{y}(t))\Big]$}}
\label{eq-Wrescaledmath}
\end{equation}
where
$ V_1(t,y(t))\equiv V\big(t, y(t)\,\big)\big|_{D=1,\xi=1} $,
from which one reads
\begin{equation}
  B^{\mathfrak m}(t;D,T,\xi) = \xi^2 B^{\mathfrak m}\big(\tfrac t{\xi^2};1,\tfrac T{T_c^{\mathfrak m}},1\big)  
  \label{eq-rescalinglowTmathnaive}
\end{equation}
However, the limit $T\to 0$ cannot be taken by candidly replacing $\tfrac T{T_c^{\mathfrak m}}$ by~$0$
in~\eqref{eq-rescalinglowTmathnaive}. This would lead to $A^{\mathfrak{m}}_{\text{RM}}|_{(T \ll T_c^{\mathfrak m})}{=} \xi^{-2/3}$ and yield a corresponding zero thorn exponent, but this appears incorrect as we now discuss.  Indeed, the term $\frac{T_c^{\mathfrak m}}{T}$ in the Hamiltonian in~\eqref{eq-Wrescaledmath} appears only in front of the disorder term, and not in front of both contributions as in~\eqref{eq-explicitBsaddle_lowT}. In terms of path integrals
\begin{align}
\!\!   B^{\mathfrak m}&(t_1;D,T,\xi) =
\nonumber\\
& \xi^2\, \overline{
\frac
{\displaystyle 
\int_{y(0)=0}
\!\!\!\!\!\!\!\!\!\!\!\!\!\!\mathcal Dy\  y(\tfrac{t_1}{\xi})^2 \,
  e^{\mbox{\scriptsize$\displaystyle-\!\int_{0}^{\frac{t_1}{\xi}}\!\! dt\, \tfrac 12 {(\partial_ty)^2} $}}
  e^{\mbox{\scriptsize$\displaystyle-\frac{T_c}{T}\!\!\int_{0}^{\frac{t_1}{\xi}}\! dt\,V_1(t,y(t))$}}
}
{\displaystyle 
\int_{y(0)=0}
\!\!\!\!\!\!\!\!\!\!\!\!\mathcal Dy \
  e^{\mbox{\scriptsize$\displaystyle-\!\int_{0}^{\frac{t_1}{\xi}}\!\! dt\, \tfrac 12 {(\partial_ty)^2} $}}
  e^{\mbox{\scriptsize$\displaystyle-\frac{T_c}{T}\!\!\int_{0}^{\frac{t_1}{\xi}}\! dt\,V_1(t,y(t))$}}
}  }
\label{eq-explicitBsaddle_lowT_math}
\end{align}
the large prefactor $\frac{T_c}{T}$ actually selects the path which minimizes the disorder contribution along the polymer trajectory, and not the full Hamiltonian as in~\eqref{eq-explicitBsaddle_lowT}, and this path
is `too anomalous' (no elastic constraint enforces it to stay in a bounded region as $T\to 0$).
To contend with this singular limit, instead of starting with~\eqref{eq-rescalinglowTmathnaive}, one may better work in the physicists convention starting from the scaling construction
\eqref{eq-rescalingsB-for-lowT}-\eqref{eq-rescaledB-for-lowT}, where the low $T$ behavior is controlled.
One checks that the only possible rescaling of~\eqref{eq-rescalinglowTmathnaive} into a physicists roughness~${B(t;c,D,T,\xi)}$ satisfying ${c=D=\xi=1}$ is:
\begin{equation}
 a= \xi \, , \; \widetilde{E} = (\xi T D)^{1/3} \, , \; b = t_{**}^{\mathfrak m}(T)=\argp{\frac{\xi^5 T^2}{D}}^{1/3}
\end{equation}
\begin{align}
  B^{\mathfrak m}(t;D,T,\xi) &  \equiv B(t;T,D,T,\xi) \nonumber \\
  &= \xi^2 B \argp{\tfrac{t}{t_{**}^{\mathfrak m}(T)};1,1,\frac T{(\xi\, T D)^{1/3}},1}
  \label{eq-fromBmathtoBPhys} 
\end{align}
This rescaling is similar in spirit to the rescaling $y=p\bar y$ in~\eqref{eq-rescaling_saddle_factorial} for the saddle-point asymptotics study of $p!$\,: it allows to find the optimal path $y(t)$ at the correct scale in the large parameter $\frac{T^{\mathfrak m}_c}{T}$ encountered in~\eqref{eq-rescalinglowTmathnaive}.
The limit ${T\to 0}$ in the mathematician's ${B^{\mathfrak m}(t;D,T,\xi)}$ coincides with the limit ${T\to 0}$ in the physicist's roughness of~\eqref{eq-fromBmathtoBPhys} since
$  \frac T{(\xi \,T D)^{\frac 13}}\stackrel{T\to 0}{\longrightarrow} 0$.
One reads from~\eqref{eq-fromBmathtoBPhys} at asymptotically large time, according to the known results ${\zeta_{\text{RM}}=2/3}$:
\begin{align}
  B^{\mathfrak m}(t;D,T,\xi) &  \ \stackrel{t\to \infty}{\sim} \
  \xi^2 A_{\text{RM}}\Big(1,1,\tfrac T{(\xi\, T D)^{1/3}},1\Big) \argc{\tfrac{t}{t_{**}^{\mathfrak m}(T)}}^{2\zeta_{\text{RM}}}  
\end{align}
In the low-temperature regime ($T\ll T_c^{\mathfrak m}$) the physicist's $A_{\text{RM}}\Big(1,1,\tfrac T{(\xi\, T D)^{1/3}},1\Big)$
remains finite and goes to a $T$-independent finite constant in the limit $T\to 0$, as discussed previously in~Sec.~\ref{section-DPtoymodel-D}.
This finally yields
\begin{equation}
 A^{\mathfrak m}_{\text{RM}}(D,T,\xi)
\stackrel{(T \ll T_c^{\mathfrak m})}{=}
 \Big(\frac{D^2}{T^4\xi}\Big)^{\frac 29}
\end{equation}
At low temperature, the mathematician's thorn exponent $\tho$ is thus \textit{non-zero}:
$ \tho_{\text{RM}}^{\mathfrak m}|_{(T \ll T_c^{\mathfrak m})}{=} -
\frac 49 $.



%

\end{document}